\documentclass{article}

\normalbaselines

\usepackage[utf8]{inputenc}

\usepackage{booktabs}

\usepackage{amsmath}
\usepackage{amssymb}
\usepackage{stmaryrd}
\usepackage{xstring}

\usepackage{url}

\usepackage{ltl}

\usepackage{graphicx}
\usepackage{tikz}
\usepackage{color}

\usepackage{enumerate}

\makeglossary
\newcommand{\IF}{\mbox{{\bf if}\ }}
\newcommand{\FI}{\mbox{{\bf fi}}}
\newcommand{\DO}{\mbox{{\bf do}\ }}
\newcommand{\OD}{\mbox{{\bf od}}}
\newcommand{\WHILE}{\mbox{{\bf while}\ }}
\newcommand{\END}{\mbox{{\bf end}}}
\newcommand{\THEN}{\mbox{\ {\bf then}\ }}
\newcommand{\ELSE}{\mbox{\ {\bf else}\ }}
\newcommand{\EELSE}{\mbox{{\bf else}\ }}

\newcommand{\T}{\mbox{{\bf true}}}

\newcommand{\ES}{\mbox{$\emptyset$}}

\newcommand{\ra}{\mbox{$\:\rightarrow\:$}}

\newcommand{\lra}{\mbox{$\:\leftrightarrow\:$}}

\newcommand{\A}{\mbox{$\ \wedge\ $}}

\newcommand{\fa}{\mbox{$\forall$}}
\newcommand{\te}{\mbox{$\exists$}}

\newcommand{\LL}{\mbox{$\ldots$}}
\newcommand{\LLn}{\mbox{$1,\ldots,n$}}

\newcommand{\qed}{\Box}

\newcommand{\IFP}{\mbox{$\IF B_1 \ra S_1 \qed \LL \qed B_n \ra S_n\ \FI$}}
\newcommand{\DOP}{\mbox{$\DO B_1 \ra S_1 \qed \LL \qed B_n \ra S_n\ \OD$}}

\newcommand{\IFPa}{\mbox{$\IF \qed^n_{i=1}\ B_i \ra S_i\ \FI$}}
\newcommand{\DOPa}{\mbox{$\DO \qed^n_{i=1}\ B_i \ra S_i\ \OD$}}

\newcommand{\ITE}[3]{\mbox{$\IF {#1} \THEN {#2} \ELSE {#3}\ \FI$}}

\newcommand{\WD}[1]{\mbox{$\WHILE {#1}\ \DO$}}
\newcommand{\WDD}[2]{\mbox{$\WHILE {#1}\ \DO {#2}\ \OD$}}

\newcommand{\ATE}[2]{\mbox{${\bf await}\ {#1} \THEN {#2}\ \END$}}
\newcommand{\ATOM}[1]{\mbox{$\langle {#1} \rangle$}}

\newcommand{\HT}[3]{\mbox{$\{{#1}\}\ {#2}\ \{{#3}\}$}}

\newcommand{\C}[1]{\mbox{$\{{#1}\}$}}         % curly braces

\newcommand{\NI}{\noindent} % With the new Springer sty this seems the better choice of the two
\newcommand{\HB}{\qed}

\newcommand{\III}{\vspace{3 mm}}
\newcommand{\PP}{\mbox{$[S_1 \| \LL \| S_n]$}}

\newcommand{\SENDER}    {\mbox{$S\!E\!N\!D\!E\!R$}}

\newcommand{\RECEIVER}  {\mbox{$R\!EC\!EIV\!E\!R$}}
\newcommand{\FILTER}    {\mbox{$F\!I\!LT\!E\!R$}}

\newcommand{\BLANK}     {\mbox{`\ '}}
\newcommand{\AST}       {\mbox{`$*$'}}

\newenvironment{mytabbing}{\begin{tabbing}\hspace*{\leftskip}\=\+\kill}{\end{tabbing}}

\def\nlni{\par\ifvmode\removelastskip\fi\vskip\baselineskip\noindent}

\newcommand{\BEGIN}{\mbox{{\bf begin}}}
\newcommand{\block}[1]{\mbox{$\BEGIN \ {#1}\ \END$}}

\newcommand{\object}{\mbox{{\bf object}}}

\newcommand{\nulll}{\mbox{{\bf null}}}

\newcommand{\this}{\mbox{{\bf self}}}
\newcommand{\new}{\mbox{{\bf new}}}

\newenvironment{newtabbing}{\begin{tabbing}\hspace*{\leftskip}\=\+\kill}{\end{tabbing}}

\newenvironment{Example}{\begin{example}}{\end{example}}
\newenvironment{Lemma}{\begin{lemma}}{\end{lemma}}
\newenvironment{Def}{\begin{definition}}{\end{definition}}
\newenvironment{Note}{\begin{note}}{\end{note}}
\newenvironment{Cor}{\begin{corollary}}{\end{corollary}}

\newenvironment{Theorem-HB}{\begin{theorem}}{\HB\end{theorem}}
\newenvironment{Example-HB}{\begin{Example}}{\HB\end{Example}}
\newenvironment{Lemma-HB}{\begin{Lemma}}{\HB\end{Lemma}}
\newenvironment{Def-HB}{\begin{Def}}{\HB\end{Def}}
\newenvironment{Note-HB}{\begin{Note}}{\HB\end{Note}}
\newenvironment{Cor-HB}{\begin{Cor}}{\HB\end{Cor}}
\newenvironment{Warn-HB}{\begin{Warn}}{\HB\end{Warn}}

 \newcommand{\MSI}[1]{\mbox{${\cal M}_I[\![{#1}]\!]$}}

\usepackage{tikz}
\usetikzlibrary{positioning,shapes.geometric}
\DeclareRobustCommand{\fac}[1]{
\def\arraystretch{0}%
\setlength\arraycolsep{1pt}%
\setlength\arrayrulewidth{.2pt}%
\begin{array}[b]{|@{}c}
\strut\mkern2mu%
#1%
\\[\arraycolsep]%
\hline
\end{array}%
\mkern3mu%
}

\title{Fifty Years of Hoare's Logic}

\author{ Krzysztof R. Apt \\
{CWI, Amsterdam, The Netherlands} \\
{MIMUW, University of Warsaw, Warsaw, Poland} \\[2mm]
Ernst-R\"{u}diger Olderog \\
{University of Oldenburg, Oldenburg, Germany}
}

\date{}
\begin{document}

\maketitle

%%%%%%%%%%%%%%%%%%%%%%%%%%%%%%%%%%%%%%%%%%%%%%
\tableofcontents
\begin{abstract} 
  We present a history of Hoare's logic.
\end{abstract}

\section{Introduction}
\label{sec:introduction}

Hoare's logic is a formalism allowing us to reason about program
correctness.  It was introduced fifty years ago in the seminal article
\cite{Hoa69} of Tony Hoare that focused on a small class of
\textbf{while} programs, and was soon extended by him in \cite{Hoa71}
to programs allowing local variables and recursive procedures.  This
approach became the most influential method of verifying programs,
mainly because its syntax-oriented style made it possible to extend it
to almost any type of programs.  Also, thanks to parallel developments
in program semantics, this approach leans itself naturally to a
rigorous analysis based on the methods of mathematical logic.  Since
then, several books appeared that discuss Hoare's logic, or at least
have a chapter on it: \cite{Bak80,LS87,TZ88,Fra92,Win93,AFPS11,BA12},
to name a few. 

More than thirty years ago two surveys of Hoare's logic appeared,
\cite{Apt81b}, concerned with deterministic programs, and
\cite{Apt84}, concerned with nondeterministic programs.  At the
beginning of the nineties an extensive survey \cite{Cou90} was published
that also included an account of verification of parallel programs and
a discussion of other approaches to program verification.

A systematic exposition of Hoare's logics for deterministic,
nondeterministic and parallel programs appeared in our book
\cite{AO91}. The last edition of it, \cite{ABO09}, written jointly
with F.S.~de Boer, extended the presentation to recursive procedures
and object-oriented programs.  In this paper we occasionally rely on
the material presented in this book, notably to structure the
presentation.

Given that the literature on the subject is really vast, we would like
to clarify the scope of this survey.
Our objective is to trace the main developments in Hoare's logic that
originated with the original article \cite{Hoa69} and resulted in proof
systems dealing with several classes of programs. To realize this
natural research programme various difficulties, notably concerning the logical
status of the proposed proof systems, had to be resolved. We clarify here 
these and other complications and explain how they were addressed.

In particular, we discuss the problem of completeness, the issues
concerning local variables, parameter mechanisms, and auxiliary rules.
We also discuss various alternatives to Hoare's logic that were
proposed in the literature.  Given the scope of this exposition we
discuss only briefly (in the final section) the use of automated
reasoning in the context of Hoare's logic. This important sequel to
the research reported here deserves a separate survey.

According to Google Scholar, the original article \cite{Hoa69} has
been cited more than 7000 times.  This forced us to make some
selection in the presented material.  Some omissions, such as the
treatment of the nowadays less often used \textbf{goto} statement or
coroutines, were dictated by our effort to trace and explain the
developments that withstood the test of time.

Further, we do not introduce any program semantics. Consequently, we
do not prove any soundness or completeness results. Instead, we
focus on a systematic account of the established results combined with an
explanation of the reasons why some concepts were introduced, and on a
discussion of some, occasionally subtle, ways Hoare's logic differs
from customary logics.

We begin the exposition by discussing in the next section the
contributions to program verification by Alan Turing and Robert Floyd
that preceded those of Hoare.  Then, in Section \ref{sec:hoare}, we
discuss Hoare's initial contributions that focused on the
\textbf{while} programs and programs with recursive procedures, though
we extend the exposition by an account of program termination.  Next,
we discuss in Section \ref{sec:sound} the soundness and completeness
of the discussed proof systems. An essential difference between
Hoare's logic and first-order logic has to do with the features
specific to programming languages, such as subscripted variables,
local variables, and parameter mechanisms. We discuss these matters in
Section \ref{sec:parameters}.  This provides a natural starting point
for an account of verification of programs with arbitrary procedures,
notably procedures that allow procedures as parameters. This forms the
subject of Section \ref{sec:procedures}.

In Section \ref{sec:fairness} we discuss verification of
nondeterministic programs, the corresponding issue of fairness, and
verification of probabilistic programs. Then, in Section
\ref{sec:parallel} we focus on the verification of parallel and
distributed programs.  Next, in Section \ref{sec:oo}, we provide an
account of verification of object-oriented programs.  In the final two
sections, \ref{sec:alternative} and \ref{sec:concluding}, we shed light
on alternative approaches to program verification,
discuss briefly the use of automated verification in relation to Hoare's logic,
and attempt to explain and assess the impact of Hoare's logic.

\section{Precursors}
\label{sec:precursors}

\subsection{Turing}

\label{subsec:Turing}

The concern about correctness of computer programs is as old as
computers themselves.  Tracing the history of program verification is
not the focus of this survey, so we only discuss here two key
contributions preceding those of Tony Hoare.  Readers interested in
the early history of this subject are referred to the paper
\cite{Jones03} by C.B.~Jones.

In 1949, Alan Turing gave a presentation entitled ``Checking a Large
Routine'' at a conference in Cambridge U.K.~at the occasion of the
launching of the computer EDSAC (Electronic Delay Storage Automatic
Calculator), published as \cite{Tur49}.  F.L.~Morris and C.B.~Jones
recovered \cite{MorJon84} the 
original typescript of Turing's presentation and made it available for
a wider audience, thereby correcting several typing errors.
\begin{figure}%
  \centering
  
 \scalebox{1.8}{
\begin{tikzpicture}
\node[below right]{\includegraphics[width=2.9cm,clip,trim=230 60 50 60,angle=90]{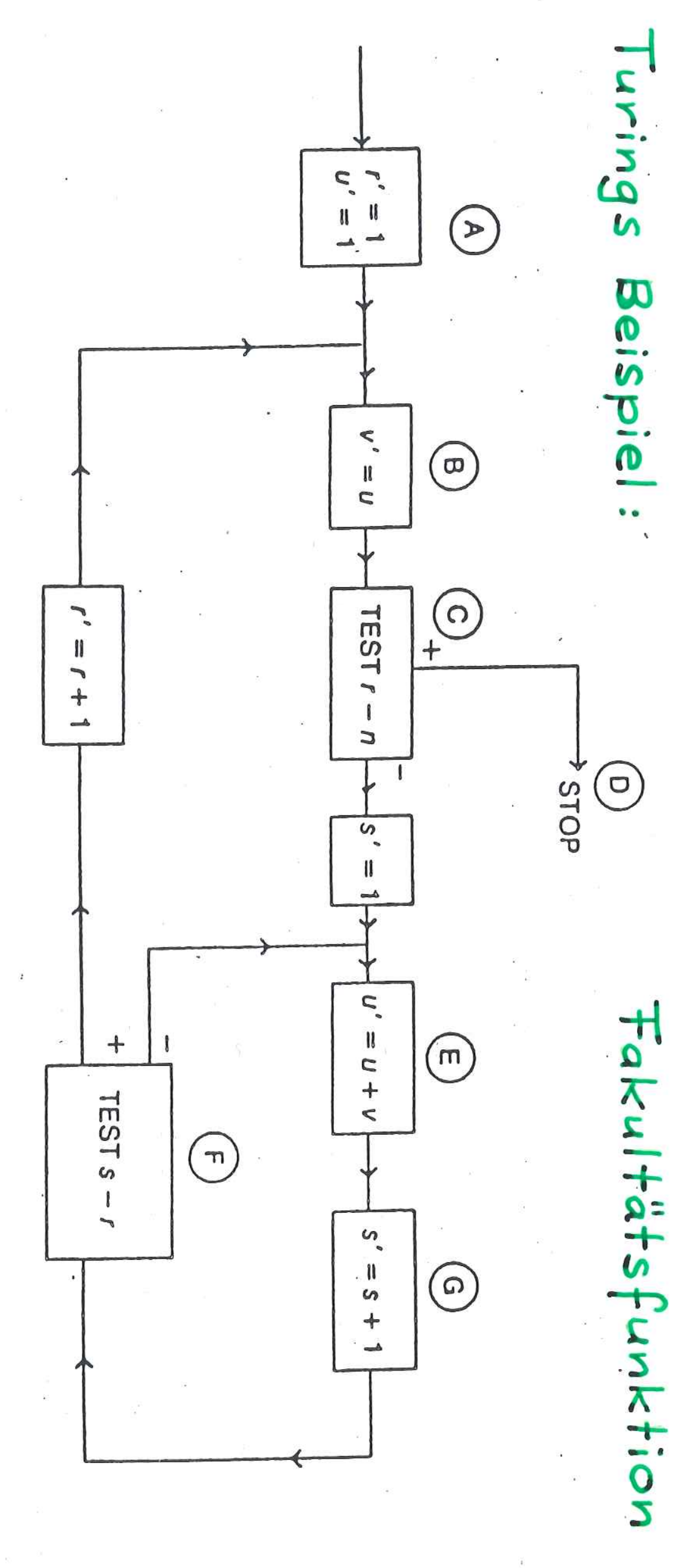}};
\fill[white](0.1,-0.1) rectangle (2.5,-0.6);
\fill[white](4,-0.1) rectangle (6.7,-0.5);
\end{tikzpicture}
 }

\caption{Turing's flowchart, 
               reconstructed by F.L.~Morris and C.B.~Jones~\cite{MorJon84}.}
\label{Turing-example}

\end{figure}
Turing started by asking

\begin{quote}
``How can one check a routine in the sense of making sure that it is right?''
\end{quote}
and proposed that

\begin{quote}
``...
the programmer should make a number of definite assertions 
which can be checked individually,	 
and from which the correctness of the whole programme easily follows.''
\end{quote}

Turing demonstrated his ideas for a flowchart program with nested
loops computing the factorial $n!$ of a given natural number $n$,
where multiplication is achieved by repeated addition; see
Figure~\ref{Turing-example}.  Note that the effect of a command in the
flowchart is represented by an equation like $u' = u+v$, where the primed
variable~$u'$ denotes the value of the variable $u$ after the execution of the
command. Today, this notation is still in use in logical
representations of computation steps like in the specification
language Z (see, e.g., \cite{Spivey92}) and bounded model checking.

\begin{figure}
  \centering
  
\includegraphics[width=4.3cm,clip,trim=280 60 50 50,angle=-90]{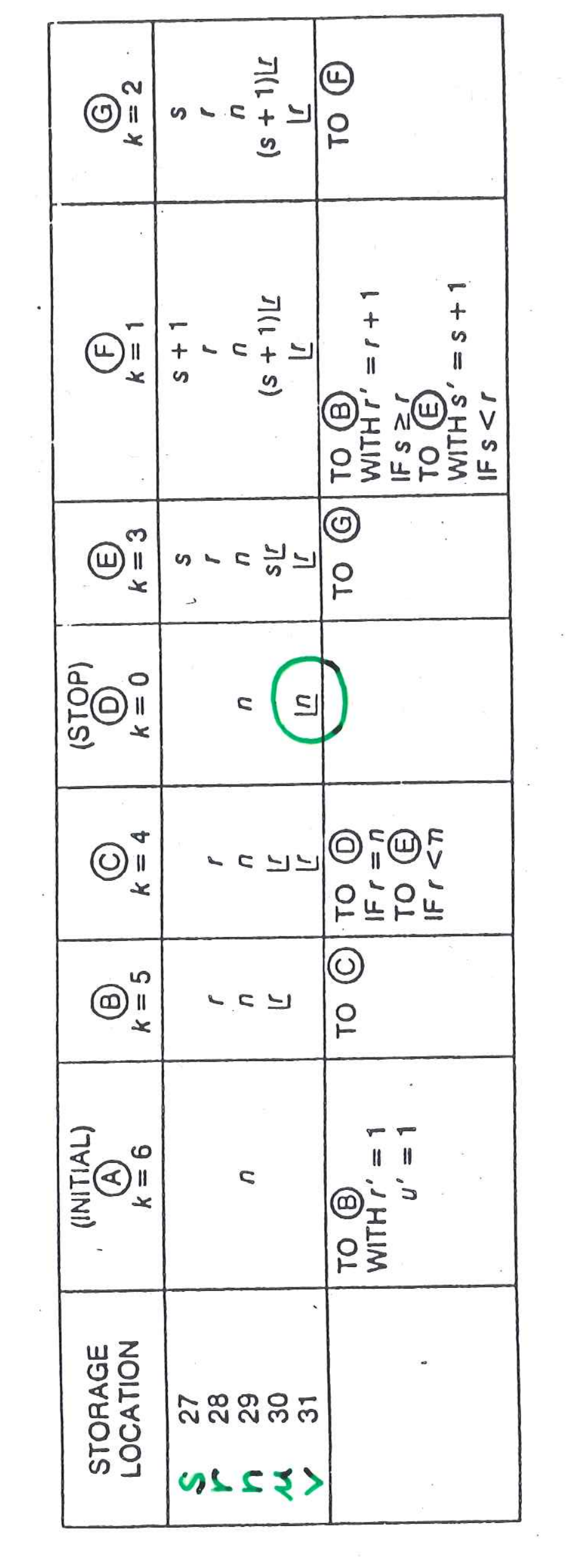}

\caption{Turing's assertions, 
              reconstructed by F.L.~Morris and C.B.~Jones~\cite{MorJon84}.
             Turing writes $\fac{n}$ for the factorial of $n$.
             }
\label{assertions}

\end{figure}

Turing referred already to \emph{assertions}.  In the example he
presented them in the form of a table referring to the numbers of the
locations storing the variables $s,r,n,u,v$, see
Figure~\ref{assertions}.  From today's viewpoint these assertions are
admittedly very specific and difficult to read.

Turing was not only concerned with delivering correct
values, but also with termination. He wrote

\begin{quote}
``Finally the checker has to verify that the process comes to an end.
Here again he should be assisted by the programmer giving a further definite assertion to be verified. 
This may take the form of a quantity which is asserted to decrease continually and vanish when the machine stops.''
\end{quote}
This refers already to the concept of a termination function.
Turing stated a global termination function for the example program,
i.e., an integer expression yielding a non-negative value
that decreases with every step of the program.

Summarizing, Turing introduced the notions of assertions and
termination functions, but did not state loop invariants and local
termination functions for the two loops of the program.  
Still, as we show in the Appendix, his approach can be represented 
within the framework of Hoare's logic.

\subsection{Floyd}

\label{subsec:Floud}

Robert Floyd was the first to propose a fully formal method for
proving the correctness of flowchart programs known as
\emph{inductive assertions method} in \cite{Flo67a}.
(Floyd actually mentions that his work is ``based on ideas of Perlis
and Gorn, and may have made their earliest appearance in an
unpublished paper by Gorn.'')
Here the assertions are logical formulas in terms of the variables appearing in
the flowcharts. The begin of the flowchart is annotated with an
assertion stating the assumptions under which the flowchart is
supposed to work. The end of the flowchart is annotated with an
assertion specifying the desired result.  To verify that these
input-output annotations are correct, each loop of the flowchart needs
to be annotated at some selected point with an assertion that should
hold whenever the control reaches this point.  The assertion should
thus be an \emph{invariant} at this selected point.  Floyd states rules how
to verify this by completing the flowchart so that there is at least
one assertion between any two subsequent statements. The rules explain
how to modify a given assertion when passing a test statement and when
passing an assignment statement.  When two assertions are adjacent to
the same arc then the logical implication has to hold in the direction
of the arc.

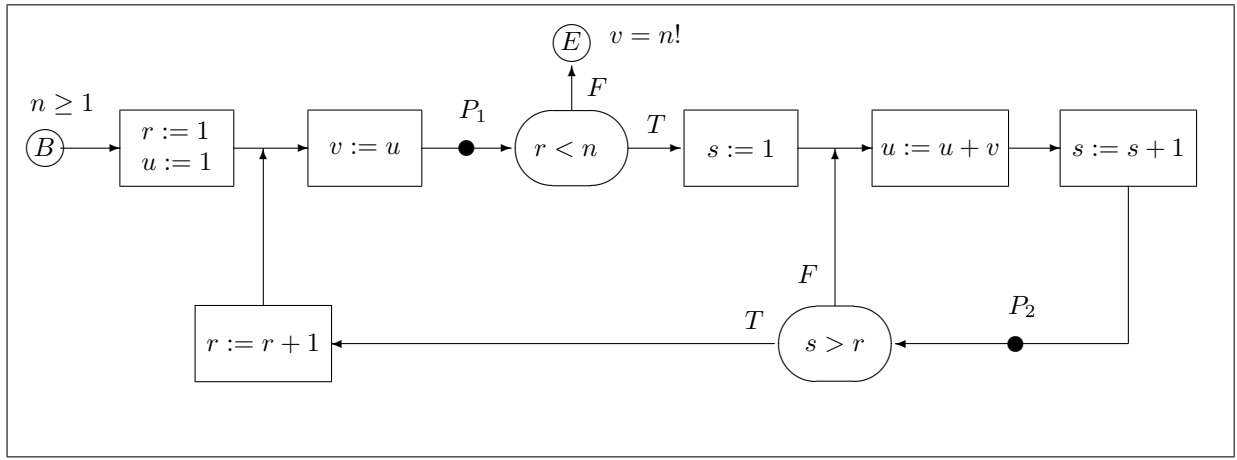
\begin{figure}
  \centering
  
\unitlength1cm

\begin{picture}(16.3,6)

\put(0,0){\framebox(16.3,6){}}

\put(0.5,4.1){\circle{.5}} \put(0.35,4){$B$} 
\put(0.3,4.6){$n \ge 1$}

\put(1.5,3.6){\framebox(1.5,1){$\begin{array}{l}
                                  r:=1 \\
                                  u:=1
                                 \end{array}$}}

 \put(7.5,5.5){\circle{.5}} \put(7.35,5.4){$E$} 
 \put(8,5.5){$v = n!$}
 
 \put(4,3.6){\framebox(1.5,1){$v:=u$}}
 
 \put(7.5,4.1){\oval(1.5,1)} \put(7,4){$r<n$}
  \put(7.7,4.8){$F$}
  \put(8.5,4.3){$T$}
 
 \put(9,3.6){\framebox(1.5,1){$s:=1$}}
 
 \put(11.5,3.6){\framebox(1.8,1){$u:=u+v$}}
 
 \put(14,3.6){\framebox(1.8,1){$s:=s+1$}}

 \put(2.5,1){\framebox(1.8,1){$r:=r+1$}}
 
 \put(11,1.5){\oval(1.5,1)} \put(10.6,1.4){$s>r$}
  \put(9.8,1.7){$T$}
  \put(10.5,2.3){$F$}

 \put(.7,4.1){\vector(1,0){.8}}     % from B to r:=1; u:=1
 \put(3,4.1){\vector(1,0){1}}       % from r:=1; u:=1 to v:=u
 \put(5.5,4.1){\vector(1,0){1.2}}    % from v:=u to r<n
 
 \put(6.1,4.1){\circle*{0.2}}  \put(6,4.5){$P_1$}
 
 \put(8.25,4.1){\vector(1,0){0.7}}  % from r<n to s:=1
  \put(7.5,4.6){\vector(0,1){0.6}}    % from r<n to E
  
 \put(10.5,4.1){\vector(1,0){1}}    % from s:=1 to u:=u+v
 \put(13.3,4.1){\vector(1,0){.7}}   % from u:=u+v to s:=s+1
 
 \put(14.9,3.6){\line(0,-1){2.1}}   % from s:=s+1 to s>r
  \put(14.9,1.5){\vector(-1,0){3.1}}
  \put(13.4,1.5){\circle*{0.2}} \put(13.3,1.9){$P_2$}
 
 \put(3.4,2){\vector(0,1){2.1}}     % from s>r upwards
 \put(10.2,1.5){\vector(-1,0){5.9}} % from s>r to r:=r+1
  
  \put(11,2){\vector(0,1){2.1}}     % from r:=r*1 upwards

\end{picture}

\caption{Turing's example as a flowchart with annotations
according to Floyd's method, where $P_1$ and $P_2$ are the following invariants:
$P_1 \equiv v=r! \land u=r! \land 1 \le r \le n$  and 
$P_2 \equiv v=r! \land u = s \cdot v \land 1 \le s \le r+1 \le n$.
See also \cite{deBakker75}.}
\label{Turing-annotated}

\end{figure}

In Figure~\ref{Turing-annotated} we show Turing's example as a
flowchart with annotations according to Floyd's method. At the begin
$B$ of the flowchart the annotation $n \ge 1$ states the assumption of
the computation, at the end $E$ the annotation $v=n!$ specifies the
desired result that should hold whenever the computation reaches $E$.
To verify that this annotation is correct, in every loop a point needs
to be selected and annotated with an \emph{invariant}. In this
example, we select the bullet points and annotate them
with the assertions
\[
  P_1 \equiv v=r! \land u=r! \land 1 \le r \le n
\]
and 
\[
 P_2 \equiv v=r! \land u = s \cdot v \land 1 \le s \le r+1 \le n.
\]

While the inductive assertions method is a natural approach to program
verification, it is limited to programs that can be represented as
flowcharts. This obvious deficiency was overcome by Hoare by proposing
an approach that is applicable to programs written in the customary
textual form.

\section{Hoare's Contributions}
\label{sec:hoare}

\subsection{Reasoning about \textbf{while} programs}
\label{subsec:hoare1}

To reason about programs Hoare introduced in \cite{Hoa69}
a new notation
\[
P \ \{S\} \ Q,
\]
with the interpretation
\begin{quote}
``If the assertion $P$ is true before initiation of a program $S$,
then the assertion $Q$ will be true on its completion.''
\end{quote}
Nowadays one rather writes
\[
\HT{P}{S}{Q}
\]
so that additional assertions can be freely inserted in the program
text, by putting the $\{ \cdot \}$ brackets around them. Such a
possibility will turn out to be especially important when reasoning
about parallel programs. In what follows we shall use the latter
notation.  In this context $P$ is referred to as a \emph{precondition}
and $Q$ as a \emph{postcondition}.

Subsequently Hoare introduced an axiom to reason about the
assignment statement and the proof rules to reason about the program
composition and the \textbf{while} statement.  He also
introduced two consequence rules, now combined into one, that allow
one to strengthen the precondition and to weaken the postcondition.
He then used these axioms and rules to establish correctness of the following
simple program, let us call it \texttt{DIV}, that finds ``the quotient
$q$ and a remainder $r$ obtained on dividing $x$ by $y$'':
\[ 
r:=x;\ q:=0;\ \WDD{y \leq r}{r:=r-y;\ q:=1+q}. 
\]
All variables are assumed to range over the nonnegative integers.

In what follows we review these steps.
We view  assertions as formulas in some first-order language and assume from the
reader some rudimentary knowledge of basic concepts of logic.
The assignment axiom has the form:
\III

\NI
ASSIGNMENT
\[ \HT{P[x:=t]}{x:=t}{P}, \]
where $x$ is a variable, $t$ is an expression, and 
$P[x:=t]$ is the result of substituting $t$ for all free occurrences of $x$ in $P$.
\III

The already mentioned consequence rule has the following form:
\III

\NI
CONSEQUENCE
\[
 \frac{ P \ra P_1, \HT{P_1}{S}{Q_1}, Q_1 \ra Q        }
        { \HT{P}{S}{Q}  }
\]
\vspace{1mm}

Here it is assumed that the mentioned implications can be established
in some further unspecified proof system exclusively concerned with the assertions.
(Hoare just referred to $\ra$ as `logical implication'.)

The final two rules were:
\III

\NI
COMPOSITION
\[ \frac{ \HT{P}{S_1}{R}, \HT{R}{S_2}{Q}                }
        { \HT{P}{S_1;\ S_2}{Q}                          }\]
and
\III

\NI
WHILE
\[ \frac{ \HT{P \A B}{S}{P} } { \HT{P}{\WDD{B}{S}}{P \A \neg B}}\]

Nowadays one refers to the assertion $P$ that satisfies the premise of
this rule as a \emph{loop invariant}.

Hoare's correctness proof of the \texttt{DIV} program is presented in
Figure \ref{fig:1}. (Hoare wrote the postcondition of the conclusion
of the WHILE rule as $\neg B \A P$ and this is how it is recorded in
Figure \ref{fig:1}.)  It yields the desired conclusion that $q$ is the
quotient and $r$ the remainder resulting from dividing $x$ by $y$.
The crucial step in this proof is line 10 that clarifies the role
played by the assertion $x = r + y\cdot q$. This line establishes that
$x = r + y\cdot q$ is a loop invariant of the considered
\textbf{while} statement and its discovery is essential for the proof
to succeed.

\begin{figure}
  \centering
  
  {\small
\begin{tabular}{lll}
Line    & Formal proof &   Justification \\
number  \hspace{-12mm} &  &  \\[2mm]
1            & $\T \ra x = x + y\cdot 0$ & logic \\
2            & $\HT{x = x + y\cdot 0}{r:=x}{x =r + y \cdot 0}$ & ASSIGNMENT \\
3            & $\HT{x =r + y \cdot 0}{q:=0}{x =r + y \cdot q}$ & ASSIGNMENT \\
4            & $\HT{\T}{r:=x}{x =r + y \cdot 0}$ & CONSEQUENCE (1,2) \\
5            & $\HT{\T}{r:=x;\ q:=0}{x =r + y \cdot q}$ & COMPOSITION (4,3) \\
6            & $x = r + y\cdot q \land y \leq r \ra x = (r-y) + y \cdot (1+q) $ & logic \\
7            & $\HT{x = (r-y) + y \cdot (1+q)}{r:=r-y}{x = r + y \cdot (1+q)}$ & ASSIGNMENT \\
8            & $\HT{x = r + y \cdot (1+q)}{q:=1+q}{x = r + y \cdot q}$ & ASSIGNMENT \\
9            & $\HT{x = (r-y) + y \cdot (1+q)}{r:=r-y; \ q:=1+q}{x = r + y \cdot q}$ & COMPOSITION (7,8) \\
10            & $\HT{x = r + y\cdot q \land y \leq r}{r:=r-y; \ q:=1+q}{x = r + y \cdot q}$ & CONSEQUENCE (6,9) \\
11            & $\HT{x = r + y\cdot q }{\WDD{y \leq r}{r:=r-y; \ q:=1+q}}{\neg y \leq r \land x = r + y \cdot q}$ \hspace{-3mm} & WHILE (10) \\
 12            & $\{\T\}\ r:=x;\ q:=0; \ \WDD{y \leq r}{r:=r-y; \ q:=1+q} $ & \\
               & \hspace{74mm} $\{\neg y \leq r \land x = r + y \cdot q\}$ & COMPOSITION (5,11) \\[2mm]
\end{tabular}
The arguments in the right column  of the rules refer to the line numbers to which they
were applied and `logic' indicates that the relevant formulas are true (Hoare referred to 
specific axioms of Peano arithmetic).
\caption{Correctness proof of the \texttt{DIV} program \label{fig:1}}}
\end{figure}

As pointed out in \cite{JR10} the assignment axiom was originally
proposed in \cite{Kin69}, the PhD thesis of J.~King.  From \cite{Flo67a}
one can distill a more complex assignment axiom
\III

\NI
ASSIGNMENT I
\[ 
 \HT{P}{x:=t}{\te y:(P[x:=y] \A x=t[x:=y])}. 
\]
that reasons ``forward'' starting from the precondition $P$.
\III

The striking simplicity of the ASSIGNMENT axiom reveals a close
relation between the assignment statement and the substitution
operation.  This is achieved, in contrast to Floyd's approach, by
reasoning `backwards', so starting from the postcondition $P$.  The
adoption of this axiom by Hoare probably influenced a couple of years
later Edsger W.~Dijkstra to propose the weakest precondition semantics
that adopted this reasoning `backward' to all program statements. We
shall discuss this alternative approach to program verification in
Section \ref{sec:alternative}.  From the mathematical point of view,
Hoare's proof rules and axioms form an unusual mix: the assignment
axiom adopts `backward' reasoning, while the proof rules
embrace `forward' reasoning, in the sense that they maintain the view that
a program is a transformation of a precondition to a postcondition.

Historically, Hoare's approach was preceded by P.~Naur's paper
\cite{Nau66} in which a simple {\sc Algol 60} program that finds the
greatest element in an array of numbers was verified using
`snapshots', which are in fact assertions. However, Naur's reasoning
is not backed up by any proof rules. His paper closes with this
remark: ``Similar concepts have been developed independently by
Robert W. Floyd (unpublished paper, communicated privately).'' that
clearly refers to Floyd's paper discussed in the previous subsection.
  
Hoare's paper turned out to be the beginning of a far reaching change in
reasoning about programs, resulting from moving from flowcharts to
programs expressed in the customary textual form. This opened the way
to reasoning about programs that cannot be readily expressed as
flowcharts, for example, recursive procedures or programs with
variable declarations.  Also it made it possible to adopt a
syntax-directed reasoning about programs by using their structure as 
guidance in organizing the proof.

A related, implicit, feature of the proof system proposed by Hoare is
that it encourages program development by allowing one to first
specify the desired preconditions and postconditions of a program
component and subsequently to look for a program fragment for which
the corresponding correctness statement can be established.  Hoare
took a lead in this novel view of program correctness by publishing in
\cite{Hoa71a} a correctness proof of the \texttt{FIND} program, the
purpose of which is to find the $f$th largest element of an array
$A[1:N]$ by rearranging its elements so that upon termination
\[
  A[1], A[2], \dots, A[f-1] \leq A[f] \leq A[f+1], \dots, A[N].
\]

The program is very subtle ---it uses a triply nested \textbf{while}
loop--- and as a result its correctness proof is highly
nontrivial. The proof is not carried out in the proof system of
\cite{Hoa69} but from the way it is written it is clear that it can be
done so. In fact, Hoare refers in a number of places to
\emph{invariants} that he defines as formulas that remain true
throughout the execution of the program independently of the values of
the program variables.

A similar in style contribution is \cite{Hoa72a}, in which a
correctness proof was given of a program encoding the sieve of
Eratosthenes.  The difference was that the program was developed
together with its correctness proof and presented using non-recursive
procedures and classes, drawing on the contemporary works of
E.W.~Dijkstra on structured programming and O.J.~Dahl on the
object-oriented programming language SIMULA 67, which appeared as
chapters in \cite{DDH72}.  These two contributions of Hoare,
\cite{Hoa71a} and \cite{Hoa72a}, showed that his original logic could
be seen not only as a tool to verify programs but also as a guide to
design correct programs. These ideas were further developed by
Dijkstra, notably in his book \cite{Dij76}.

\subsection{Reasoning about recursive procedures}
\label{subsec:hoare2}

Let us continue with another milestone in the history of Hoare's
logic. Foley and Hoare established correctness of the
program \texttt{Quicksort}~\cite{FH71}, originally proposed by Hoare in
\cite{Hoa61b}. Foley and Hoare stated: 
\begin{quote}
``The purpose of the program \texttt{Quicksort} is to sort the elements 
$a[m]$ to $a[n]$ of an array into ascending order, while leaving untouched 
those below $a[m]$ and above $a[n]$.''
\end{quote}
The main difficulty was that
\texttt{Quicksort} uses recursion. (Actually it was the first
non-trivial example of a successful use of recursion in imperative programming.) 
This required appropriate proof rules 
that were introduced by Hoare in \cite{Hoa71}.

In what follows, given a program $S$ we denote by $change(S)$ the set
of variables that are subject to change in it.
In general, this set depends on the initial state of the program, but
one can define syntactically a superset of it that takes care
correctly of local variables and multiple procedure declarations. In
what follows we mean by $change(S)$ this syntactically defined
superset.

Further, we use $\mathbf{proc} \: p(\mathbf{x:v}): S$ to denote the
declaration of a procedure $p$ with the body $S$ and two disjoint
lists of distinct formal parameters: $\mathbf{x}$ is the list of all
global variables of $S$ which are subject to change by $S$, i.e.,
$\{\mathbf{x}\} = change(S)$, and $\mathbf{v}$ is the list of all
other global variables of $S$ (read-only variables).  (Hoare actually
used a slightly different notation that is now obsolete.)

\emph{Legal} procedure calls are of the form $\mathbf{call} \: p(\mathbf{a:e})$, where
\begin{itemize}

\item $\mathbf{a}$ is a list of distinct variables of the same length
  as $\mathbf{x}$ that are substituted for $\mathbf{x}$,
  
\item $\mathbf{e}$ is a list of expressions not containing any
  variable of $\mathbf{a}$, of the same length as $\mathbf{v}$, that are
  substituted for $\mathbf{v}$.
\end{itemize}

The following proof rule dealt with a `generic'
procedure call $\mathbf{call} \: p(\mathbf{x:v})$:
\III

\NI
RECURSION
\[
\begin{array}{l}
\HT{P}{\mathbf{call} \: p(\mathbf{x:v})}{Q} \vdash \HT{P}{S}{Q}                    \\
[-\medskipamount]
\hrulefill                                                      \\
\HT{P}{\mathbf{call} \: p(\mathbf{x:v})}{Q} 
\end{array}
\]
where the procedure $p$ is declared by $\mathbf{proc} \: p(\mathbf{x:v}): S$.
\III

(Hoare actually included the procedure declaration as an additional
premise of the rule.)  What is the intuition behind this rule?  Hoare
states in \cite{Hoa71} that it permits
\begin{quote}
``the use of the desired conclusion as a hypothesis in the proof of the body itself.''
\end{quote}

More specifically, the symbol $\vdash$ in the premise denotes the
provability relation. So this rule is actually a metarule.  According
to \cite{FH71} the premise of this rule
\begin{quote}
``permits
$\HT{P}{\mathbf{call} \: p(\mathbf{x:v})}{Q}$ to be
assumed as a hypothesis in the proof of $\HT{P}{S}{Q}$.''
\end{quote}
This proof is supposed to be carried out using the remaining axioms
and proof rules.  The conclusion of the rule then coincides with this
hypothesis.

To transfer a result established by the recursion rule
to any other procedure call with actual parameters, say the lists
$\mathbf{a}$ and $\mathbf{e}$, the following substitution rule was introduced:
\III

\NI
SUBSTITUTION 
\[
\frac{\HT{P}{\mathbf{call} \: p(\mathbf{x:v})}{Q}}
{\HT{P[\mathbf{k} := \mathbf{k'}, \mathbf{x} := \mathbf{a}, \mathbf{v} := \mathbf{e}]}
    {\mathbf{call} \: p(\mathbf{a:e})}
    {Q[\mathbf{k} := \mathbf{k'}, \mathbf{x} := \mathbf{a}, \mathbf{v} := \mathbf{e}]}}
\]
where the following holds for the substitutions applied to $P$ and $Q$:

\begin{itemize}
\item $\mathbf{k}$ is a list of free variables of $P$ or $Q$ that do not occur 
  in $\mathbf{x}$ or $\mathbf{v}$, but which occur in $\mathbf{a}$ or $\mathbf{e}$,
  
\item $\mathbf{k'}$ is a list of fresh variables of the same length as
  $\mathbf{k}$ that are substituted for $\mathbf{k}$. So the variables
  in $\mathbf{k'}$ do not appear globally in the procedure body $S$,
  are not free in $P$ or $Q$, and do not occur in $\mathbf{a}$ or
  $\mathbf{e}$,

\item $\mathbf{a}$ and $\mathbf{e}$ are such that the call
$\mathbf{call} \: p(\mathbf{a:e})$ is legal.

\end{itemize}

\noindent
So the substitution
$[\mathbf{x} := \mathbf{a}, \mathbf{v} := \mathbf{e}]$ of the formal
parameters by the actual ones is carried out together with an appropriate renaming
$[\mathbf{k} := \mathbf{k'}]$ of the `potentially conflicting'
variables in $P$ and $Q$.

Hoare noted that the above two rules are not sufficient to reason
about recursive procedures. To have a more powerful proof method, he
introduced the following rule, where $\mathit{free}(P)$ stands for the
set of free variables in an assertion $P$ and similarly with
$\mathit{free}(P,Q)$.  Further, for a list $\mathbf{z}$ of variables,
$\{\mathbf{z}\}$ denotes the \emph{set} of all variables occurring in
$\mathbf{z}$.  Similarly, $\{\mathbf{a,e}\}$ denotes the set of all
variables occurring in the lists $\mathbf{a}$ and $\mathbf{e}$.
\III

\NI
ADAPTATION
\[
\frac{\HT{P}{\mathbf{call} \: p(\mathbf{a:e})}{Q}}
{\HT{\exists\, \mathbf{z}\; (P \land \forall \mathbf{a}\; (Q \ra R))}
     {\mathbf{call} \: p(\mathbf{a:e})}{R}}
\]
where 
$\mathbf{z}$
is a list of variables with
$\{\mathbf{z}\} = \mathit{free}(P,Q) \setminus (\mathit{free}(R) \cup \{\mathbf{a,e}\})$.
\III

The precondition of the conclusion of this rule looks complicated.
What does it express? Hoare explained in \cite{Hoa71}:

\begin{quote}
``If $R$ is the desired result of executing a procedure call,
$\mathbf{call} \: p(\mathbf{a:e})$,
and $\HT{P}{\mathbf{call} \: p(\mathbf{a:e})}{Q}$ is already given,
what is the weakest precondition $W$ such that  
$\HT{W}{\mathbf{call} \: p(\mathbf{a:e})}{R}$ 
is universally valid?
It turns out that this precondition is 
$\exists\, \mathbf{z}\; (P \land \forall \mathbf{a}\; (Q \ra R))$.''
\end{quote}

\NI
We shall discuss this rule further in Subsection \ref{subsec:adaptation}.

To deal with  the declarations of local variables Hoare introduced the
following rule:
\III

\NI
DECLARATION
\[
\frac{\HT{P}{S[x:=y]}{Q}}
{\HT{P}{\mathbf{begin \; var}\; x; \: S \; \mathbf{end}}{Q}}
\]
where $y \not\in \mathit{free}(P,Q)$ and $y$ does not appear in $S$ 
unless the variables $x$ and $y$ are the same.
\III

Additionally, a proof rule, originally proposed in \cite{Lau71},
to reason about the conditional \textbf{if-then} statement was used.
We present here instead a more common rule that deals with 
the \textbf{if-then-else} statement:
\III

\NI
CONDITIONAL
\[ \frac{ \HT{p \A B}{S_1}{q}, \HT{p \A \neg B}{S_2}{q}         }
  { \HT{p}{\ITE{B}{S_1}{S_2}}{q}                          }
\]
      
The correctness proof of \texttt{Quicksort} by Foley and Hoare in
\cite{FH71} was carried out using the above proof rules, originally
presented in \cite{Hoa71}.  The authors formulated two correctness
criteria that should hold upon termination of \texttt{Quicksort}:

\begin{itemize}

\item \emph{Sorted}: the output array should be sorted within
the given bounds $m$ and $n$.

\item \emph{Perm}: the output array should be a permutation of the original
input array within the given bounds $m$ and $n$ but untouched outside
these bounds.

\end{itemize}

The proof established these properties simultaneously, using
appropriate assertions.
In \cite{ABO09} a detailed modular correctness proof of
\texttt{Quicksort} was given.  Modular means here that the property
\emph{Perm} was proved first and next, based on this result, the
property \emph{Sorted}.

\subsection{Reasoning about termination}
\label{subsec:termination}

So far we did not discuss within Hoare's logic the subject of program
termination.  Nowadays, one talks of \emph{partial correctness}, which
refers to the conditional statement `if the program terminates
starting from a given precondition, then it satisfies the desired
postcondition' and this is precisely what Hoare's proof system allows
one to accomplish. A more demanding property is \emph{total
  correctness}, which stipulates that all program computations
starting from a given precondition terminate and satisfy the desired
postcondition. To formalize these notions we need to refer to the
program semantics. We shall discuss it in the next section.

All approaches to proving program termination within Hoare's logic
formalize Floyd's \cite{Flo67a} observation that
\begin{quote}
  ``Proofs of termination are dealt with by showing that each step of
  a program decreases some entity which cannot decrease
  indefinitely.''
\end{quote}
The challenge is to incorporate such a reasoning into Hoare's
framework in a simple way.  The first extension of Hoare's proof
system to total correctness was proposed in \cite{MP74}, but the
proposed strengthening of the WHILE rule was somewhat elaborate.  In
\cite{Har79} the appropriate rule took a simpler form:
\III

\NI
WHILE I
\[ 
\frac{P(n+1) \ra B, \ \HT{P(n+1)}{S}{P(n)}, \ P(0) \ra \neg B } 
{ \HT{\te n \: P(n)}{\WDD{B}{S}}{P(0)} }
\]
where $P(n)$ is an assertion with a free variable $n$ that does not
appear in $S$ and ranges over natural numbers.
\III

Still, a disadvantage of this rule is that it requires to find a
parameterized loop invariant $P(n)$ such that the value of $n$
decreases exactly by 1 with each loop iteration.  Such a precise
information is not needed to establish termination and sometimes is
difficult to come up with.

Additionally, it is often inconvenient to reason about partial
correctness and termination at the same time. These concerns were
addressed in the following proof rule introduced in \cite{OG76a} that
adds two new premises to the original WHILE rule: \III

\NI
WHILE II
\[
\begin{array}{l}
\HT{P \A B}{S}{P},                      \\
\HT{P \A B \A t=z}{S}{t<z},             \\
P \ra t \geq 0                         \\
[-\medskipamount]
\hrulefill                              \\
\HT{P}{\WDD{B}{S}}{P \A \neg\ B}
\end{array}
\]
where $t$ is an integer expression, called a \emph{termination
  function} (sometimes called a \emph{bound function} or a
\emph{variant}), and $z$ is an integer variable that does not appear
in $P,B,t$ or $S$.
\III

This proof rule corresponds to Dijkstra's modification of his weakest
precondition semantics proposed in \cite{EWD:EWD573} and reproduced
as \cite{EWD:EWD573pub}.  Returning to the \texttt{DIV} program from
Subsection \ref{subsec:hoare1} note that it does not terminate when
$y = 0$. To prove its termination one needs to assume that initially
$x \geq 0 \land y>0$ and use a stronger loop invariant, namely
$P' \equiv r \geq 0 \land y>0 \land x = r + y\cdot q$. The termination
function is particularly simple here: it is just $r$. The relevant
claims,
\[
\HT{P' \A y \leq r \A r=z}{r:=r-y;\ q:=1+q}{r<z}
\]
and
\[
P' \ra r \geq 0,
\]
are straightforward to prove.

Hoare did discuss program termination in the already discussed
\cite{Hoa71a}, where he showed termination of the \texttt{FIND}
program. Since this property is not captured by his proof system from
\cite{Hoa69}, he used informal arguments.  According to the above
terminology, he established total correctness of the program
\texttt{FIND}. Hoare noticed that the termination proof required new
invariants in addition to those needed for proving partial
correctness.  However, he did not introduce the concept of a
termination function with a corresponding proof rule for total
correctness of \textbf{while} programs.

To deal with the total correctness of the recursive procedures the following
analogue of the WHILE I rule was proposed independently in
\cite{Cla76} and \cite{Sok77}:
\III

\NI
RECURSION I
\[
\frac{\neg P(0), 
\HT{P(n)}{\mathbf{call} \:p}{Q} \vdash \HT{P(n+1)}{S}{Q}}
{\HT{\te n P(n)}{\mathbf{call} \:p}{Q} }
\]
given the procedure declaration $\mathbf{proc} \: p: S$,
and where $P(n)$ is an assertion with a free variable $n$ that does not
appear in $S$ and ranges over natural numbers.

This rule shares with the WHILE I rule the same disadvantage
concerning the parameterized assertion $P(n)$.  This matter was
addressed in \cite{ABO09}, where the following analogue of the WHILE II
rule was used for recursive procedures:
\III

\NI
RECURSION II
\[
\begin{array}{l}
  \HT{P\wedge t< z}{\mathbf{call} \: p}{Q} \vdash \HT{P \wedge t= z}{S}{Q} \\
  P \ra t \geq 0                         \\
[-\medskipamount]
\hrulefill                                                      \\
\HT{P}{\mathbf{call} \: p}{Q} 
\end{array}
\]
given the procedure declaration
$\mathbf{proc} \: p: S$, where $t$ is a termination function, 
$z$ is an integer
variable that does not occur in $P, t, Q$ and $S$ and is treated in
the proofs as a constant.
\III

The last restriction means that in the proof in the first premise no
proof rules are apply quantification or substitutions to the variable
$z$.  Examples of such rules are introduced in Subsection
\ref{subsec:adaptation}.  We shall discuss the need for such a
restriction in Subsection \ref{subsec:complete-recur-proc}.

In the presentation of the correctness of the recursive procedure
\texttt{Quicksort} in \cite{FH71} only few
remarks were spent on termination.  In \cite{ABO09} termination of
this program was proved using an extension of the
RECURSION II rule to procedures with the call-by-value parameter
mechanism.  

\section{Soundness and Completeness Matters}

\label{sec:sound}

\subsection{Preliminaries}
\label{subsec:preliminaries}

In mathematical logic a standard way to judge the adequacy of a proof
system is by means of the soundness and completeness concepts. It is
then legitimate to address these matters for the proof systems introduced
in the previous section. This requires some care since the CONSEQUENCE
rule also uses implications between assertions as premises, the WHILE II rule
refers to integer variables and expressions, whereas the RECURSION rules
refer in their premises to the provability relation.
For these considerations one needs to define some semantics with
respect to which the introduced axioms and proof rules can be
assessed.

To proceed in a systematic way we need to recall some basic
notions from mathematical logic.  Assume a first-order language
$\cal L$.  An \emph{interpretation} $I$ for $\cal L$ consists of

\begin{itemize}
\item a non-empty domain $D$,
  
\item an assignment to each $n$-ary function symbol in $\cal L$ of 
  an $n$-ary function over $D$,

\item an assignment to each $n$-ary predicate symbol $\cal L$ of an
  $n$-ary relation over $D$.

\end{itemize}

Given an interpretation $I$, a \emph{state} $\sigma$ is a function
from the set of variables to the domain $D$.  The definition of
$I$ disregards our assumption that all variables are typed.  However,
it is easy to amend it by replacing the domain $D$ by the set of typed
domains and by stipulating that each variable ranges over the domain
associated with its type.

The next step is to define, given an interpretation $I$ and a state
$\sigma$, the value $\sigma(t)$ of an expression and when $\sigma$
\emph{satisfies} a formula $\phi$ of $\cal L$, written as
$\sigma \models_I \phi$. We omit both definitions. We then say that a
formula $\phi$ is \emph{true} (or \emph{valid}) in $I$, written as
$\models_I \phi$, if for all states $\sigma$ we have
$\sigma \models_I \phi$.

The consecutive step is to define semantics of the underlying
programming concepts.  This can be done in a number of ways. The
common denominator of all approaches is the above concept of a
state. Assume that all function and predicate symbols
of the considered programming languages belong to the assumed
first-order language.  (In fact, this assumption is already implicit
in the way assertions are formed in the ASSIGNMENT axiom and the WHILE
rule.)  However, some adjustments are needed to deal with more complex
expressions that are customary in programming languages.  In
particular, we need to extend the domain of the states to array
variables. We stipulate that to such variables a state assigns a
function from the domain of the array to the range type. Using such a
function we can then assign values to expressions that use array
variables.  As the complexity of the considered programming language
grows, its expressions become more involved and the concept of the
state gets more complex. At this stage we limit ourselves to the
notion of a state that assigns values to all simple variables (i.e,
variables of a simple type) and array variables.

The final step is to define \emph{semantics of the programs}.
Several approaches were proposed in the literature. Their discussion
and comparison is beyond the scope of this paper.  For the sake of the
subsequent discussion we assume a semantics of the programs that
allows us to define \emph{computations} of each considered program,
which are identified here with the sequences of states that can be
generated by it.

In Hoare's logic the types of the variables in the considered programs, for
instance in the program \texttt{DIV} studied in Figure~\ref{fig:1}, are usually
omitted and one simply assumes that all variables are typed and that
the considered programs are correctly typed.  

\subsection{Soundness}
\label{subsec:soundness}

Let us return now to assertions and programs.  Suppose that all
assertions are formulas in a given first-order language $\cal L$ and
that all considered programs use function and predicate symbols of
$\cal L$. Each interpretation $I$ for $\cal L$ then determines the set
of states and thus allows us for each program to define the set of its
computations over $I$.  This in turn allows us to introduce the
following notions.

Let $I$ be an interpretation.
We say that the correctness formula $\HT{P}{S}{Q}$ is
true in $I$ in the sense of \emph{partial correctness} if
the following holds:
\begin{center}
  every terminating computation of $S$ over $I$ that starts \\
  in a state that satisfies $P$ ends in a state that satisfies $Q$.
\end{center}

Further, we say that the correctness formula $\HT{P}{S}{Q}$ is
true in $I$ in the sense of \emph{total correctness} if
the following holds:
\begin{center}
  every computation of $S$ over $I$ that starts in a state that \\
  satisfies $P$ terminates and ends in a state that satisfies $Q$.
\end{center}
  
Denote now by ${\cal H}$ the original proof system of Hoare presented
in Subsection \ref{subsec:hoare1} and by ${\cal HT}$ the proof system
obtained from ${\cal H}$ by replacing the WHILE rule by the WHILE II
rule.  The following two results capture the crucial properties of
these proof systems.
\III

\NI \textbf{Soundness Theorem 1}
Consider a proof of the correctness formula $\HT{P}{S}{Q}$ in the
proof system ${\cal H}$ that can use in the applications of the
CONSEQUENCE rule assertions from a set $\cal A$. Consider an
interpretation $I$ in which all assertions from $\cal A$ are
true. Then $\HT{P}{S}{Q}$ is true in $I$ in the sense of partial
correctness.
\III

This property of the proof system ${\cal H}$ is called soundness in
the sense of \emph{partial correctness}.  It was first established in
\cite{HL74} w.r.t.~the relational semantics in which programs were
represented as binary relations on the sets of states.
Returning to the original proof of \cite{Hoa69} presented in Figure \ref{fig:1},
the appropriate set $\cal A$ of the assertions consists of the formulas listed in lines 1 and 6.
Since they are true in the standard interpretation $I$ of Peano arithmetic, we conclude
by the above theorem that the established correctness formula listed in line 12
of Figure \ref{fig:1} is true in $I$ in the sense of partial correctness.

The following counterpart of of the above theorem justifies the reasoning about
termination. It is, however, important to read it in conjunction with
the qualifications that follow.
\III

\NI \textbf{Soundness Theorem 2}
Consider a proof of the correctness formula $\HT{P}{S}{Q}$ in the
proof system ${\cal HT}$ that can use in the applications of the
CONSEQUENCE rule assertions from a set $\cal A$. Consider an
interpretation $I$ in which all assertions from $\cal A$ are
true. Then $\HT{P}{S}{Q}$ is true in $I$ in the sense of total
correctness.
\III

This property of the proof system ${\cal HT}$ is called soundness in
the sense of \emph{total correctness}.  The first proof was given in
\cite{Har79} and referred to the proof system in which instead of the
WHILE II rule the WHILE I rule was used.  In this rule the assertion
$P(n)$ refers to a free variable $n$ that ranges over natural
numbers. To guarantee the correct interpretation of such assertions
one needs to ensure that in each state such a variable $n$ is
interpreted as a variable of type `natural number'. In \cite{Har79}
this is achieved by considering assertion languages that extend the
language of Peano arithmetics and by limiting one's attention to
\emph{arithmetic interpretations}.  These are interpretations that
extend the standard model for arithmetic.  

In the case of the WHILE II rule similar considerations are needed to
ensure the correct interpretation of the integer expression $t$ and
the integer variable $z$.  The corresponding result was given in
\cite{AO91} and reproduced in the subsequent two editions of the book.
As in \cite{AO91} all variables are assumed to be typed, $t$ and $z$
are correctly interpreted and the need for the arithmetic
interpretations disappears.

\subsection{Completeness}
\label{subsec:completeness}

The completeness of the ${\cal H}$ and ${\cal HT}$ proof systems aims
at establishing some form of converse of the Soundness Theorems.  It
is a subtle matter and requires a careful analysis. Let us start with
the proof system ${\cal H}$.  It is incomplete for an obvious reason.
Consider for instance the correctness formula
$\HT{\mathbf{true}}{x:= 0}{x \neq 1}$.  By the ASSIGNMENT axiom we get
$\HT{0 \neq 1}{x:= 0}{x \neq 1}$.  To conclude the proof we need to
establish the obvious implication $\mathbf{true} \to 0 \neq 1$ and
apply the CONSEQUENCE rule. However, we have no proof rules and axioms
that allow us to derive this implication.

A way out is to augment ${\cal H}$ by a proof system allowing us to
prove all true implications between the assertions. Unfortunately, in
general such proof systems do not exist. This is a consequence of two
results in mathematical logic. The first one states that the set of
theorems in a proof system with recursive sets of axioms and finitary
rules is recursively enumerable.  The second one is Tarski's
undefinability theorem of 1936, see \cite{Tar36}. It implies that the
set of formulas of Peano arithmetic that are true in the standard
model of arithmetic is not arithmetically definable, so in particular
not recursively enumerable. This means that completeness of the proof
system ${\cal H}$ cannot be established even if we add to it a proof
system concerned with the assertions.

A natural solution is to try to establish completeness \emph{relative}
to the set of true assertions, that is to use the set of true
assertions as an `oracle' that can be freely consulted in the
correctness proof.  However, even then a problem arises because the
assertion language can fail to be sufficiently expressive.  Namely
\cite{Wan78} exhibited a true correctness formula that cannot be proved
because the necessary intermediate assertions cannot be expressed in
the considered assertion language.  Simpler examples of such
assertion languages were provided in \cite{BT82}.

A solution to these complications was proposed by S.A.~Cook in \cite{Coo78}. To
explain it we need to introduce some additional notions.  We call a set
of states $\Sigma$ \emph{definable} in an interpretation $I$ iff
for some formula $\phi$ we have
$\Sigma = \{\sigma \mid \sigma \models_I \phi\}$.

Next, we assign to each program $S$ its \emph{meaning}
$\MSI{S}$ relative to $I$, defined by
\[
  \MSI{S}(\sigma) = \{\tau \mid \mbox{there exists a computation of $S$ over $I$
        that starts in $\sigma$ and terminates in $\tau$}\}.
\]
At this moment the set $\MSI{S}(\sigma)$ has at most one element, which
will not be anymore the case when nondeterministic or parallel programs are
considered.

Then given an assertion $P$ and a program $S$ we define
\[
  sp_I(P,S) = \{\tau \mid \te \sigma (\sigma \models_I P \land \tau \in \MSI{S}(\sigma))\}.
\]
So $sp_I(P,S)$ is the set of states that can be reached by executing
$S$ over $I$ starting in a state satisfying $P$; `$sp$' stands for the
\emph{strongest postcondition}.

We then say that the language $\cal L$ is \emph{expressive} relative
to an interpretation $I$ and a class of programs $\cal S$ if for every
assertion $P$ and program $S \in \cal S$ the set of states $sp_I(P,S)$
is definable.  Finally, given a first-order language $\cal L$, a proof
system $PS$ for a class of programs $\cal S$ is called \emph{complete in the
  sense of Cook} if for every interpretation $I$ such that $\cal L$ is
expressive relative to $I$ and $\cal S$ the following holds:
\begin{center}
  every correctness formula true in $I$ in the sense of partial \\
  correctness can be proved in $PS$ assuming all true formulas in $I$.
\end{center}
In other words, completeness
in the sense of Cook is a restricted form of relative completeness
mentioned above, in which we limit ourselves to the class of
interpretations w.r.t.~which the underlying language $\cal L$ is expressive.
The following result is a special case of a theorem established in \cite{Coo78}.
\III

\NI \textbf{Completeness Theorem} The proof system $\cal H$ for partial correctness of 
the \textbf{while} programs is complete in the sense of Cook.
\III

The main difficulty in the proof, that proceeds by induction on the
program structure, consists in finding the loop invariants. A simpler
argument was provided in \cite{Cla79}, where a dual definition of
expressiveness was used.  Instead of the strongest postcondition it
relied on the so-called \emph{weakest liberal precondition}, which,
given an interpretation $I$, assertion $Q$ and a program $S$, is
defined by
\[
  wlp_I(S,Q) = \{\sigma \mid \fa \tau (\tau \in \MSI{S}(\sigma) \to \tau \models_I Q)\}.
\]
So $wlp_I(S,Q)$ is the set of states from which all terminating
computations of $S$ over $I$ end in a state satisfying $Q$.  The
qualification `liberal' refers to the fact that termination is not
guaranteed. The assumption that the set of states $wlp_I(Q,S)$ is
definable makes it possible to find a very simple loop
invariant. Namely, assuming that $\HT{P}{\WDD{B}{S}}{Q}$ is true in an
interpretation $I$ such that $\cal L$ is expressive relative to it in
this revised sense, it turns out that a loop invariant is simply an
assertion $R$ defining $wlp_I(\WDD{B}{S},Q)$. Additionally, both $P \to R$ and
$R \land \neg B \to Q$ are true in $I$, which allows one to establish
$\HT{P}{\WDD{B}{S}}{Q}$ by the WHILE and CONSEQUENCE rules.

The above completeness result led to a number of works that tried to
clarify it in a logical setting.  In particular, J.A.~Bergstra and 
J.V.~Tucker noticed in \cite{BT82a} that expressiveness is not a necessary
condition for a Hoare's logic to be complete. Indeed, if one takes as the
assertion language the language of Peano arithmetic, then there exist
interpretations $I$ for which it is not expressive, while all
correctness formulas that are true in $I$ in the sense of partial
correctness can be proved in the proof system~$\cal H$.

We should also mention here a proposal put forward in \cite{BG87} by
A.~Blass and Y.~Gurevich.  They suggested to use a different assertion
language than first-order logic (or its multi-sorted variants dealing
with subscripted variables or typed variables). The proposed assertion
language is a fragment of the second-order logic, called
\emph{existential fixed-point logic} (EFL).  EFL extends a fragment of
first-order logic, in which negation is applied only to atomic
formulas and the universal quantifier is absent, by a fixed-point
operator.  The authors showed that EFL is sufficient for proving
relative completeness of the proof system $\cal H$ without any
expressiveness assumption. The reason is that both the strongest
postconditions and the weakest liberal preconditions of the
\textbf{while} programs (also in presence of recursive parameterless
procedures) are definable in EFL.

Different insights on the relative completeness can be obtained by
just limiting the assertion language to propositions.  The resulting
propositional fragment of Hoare's logic of \textbf{while} programs
(PHL)  can then be studied in a purely algebraic way, by means of an
equational reasoning.  D.~Kozen showed in \cite{Koz00} that an
extension of Kleene's algebra of regular expressions with tests (KAT)
subsumes PHL. In this setting the $\WDD{B}{S}$ program is represented
by the expression $(BS)^* \bar{B}$ and the correctness formula
$\HT{P}{S}{Q}$ by the equation $PS\bar{Q} = 0$, where $\: \bar{}\: $
is the complementation operator of KAT.  After this translation the
proof rules of the proof system $\cal H$ become theorems of
KAT. Further, after defining validity of implications between
conjunctions of expressions using a relational semantics, a
counterpart of the completeness result can be established for PHL.
This shows that reasoning within propositional Hoare's logic can be
represented by purely equational reasoning.  This work was followed by
\cite{KT01}, where it was clarified how the notion of completeness in
the sense of Cook can be captured by adding the weakest liberal
precondition to the assertion language and an appropriate
axiomatization of it to the axioms of KAT.

An even more general analysis of soundness and relative completeness
of the proof system $\cal H$ was presented in \cite{AMMO09} within the
framework of category theory, using a simple theory of pre-ordered
sets. Such a general approach allowed the authors to derive new sound
and relatively complete Hoare's logics for the runtime analysis and for
verification of linear dynamical systems.

Consider now the proof system $\cal HT$ for total correctness of
\textbf{while} programs. To establish its completeness in the
appropriate sense we encounter the same complications as in the case
of $\cal H$, but additionally we have to deal with the problem of
definability of the termination functions used in the WHILE II rule.
In \cite{Har79} completeness was established for the assertion
languages that extend the language of Peano arithmetic and for
arithmetic interpretations defined in the previous subsection, but the
paper considered the WHILE I rule in which the termination functions
are absent.  To be able to express in the assertion language the
parameterized loop invariants $P(n)$ the definition of an arithmetic
interpretation stipulates that there is a formula in the assertion
language that, when interpreted, encodes finite sequences of the
domain elements by one element.

In \cite{AO91} and the subsequent two editions of the book relative
completeness of $\cal HT$ was established. To this end, it was assumed
that the underlying assertion language is \emph{expressive}, which
meant that for every \textbf{while} loop $S$ there exists an integer
expression $t$ such that whenever $S$ terminates when started in a
state $\sigma$, then the value $\sigma(t)$ is the number of loop
iterations.  In the adopted setup the assumption that all variables
are typed automatically ensures that the considered interpretations
included the standard model of Peano arithmetic and that $\sigma(t)$
is a natural number.

\section{Fine-tuning the Approach}

\label{sec:parameters}

The matters discussed until now gloss over certain issues that have to
do with the adjustments of the preconditions and postconditions,
various uses of variables, and procedure parameters. In this section
we discuss closely these matters, as they reveal some differences
between customary logics and Hoare's logic and show the subtleties of
reasoning about various uses of variables in the context of programs.

\subsection{Adaptation rules}

\label{subsec:adaptation}

In Hoare's logic we see two types of rules. First, for each
programming construct there is at least one axiom or rule dealing with
its correctness. Together, they make possible a syntax-directed
reasoning about program correctness.  Second, there are proof rules
where the same program $S$ is considered in the premise and the
conclusion. These rules allow us to \emph{adapt} an already
established correctness formula $\HT{P_1}{S}{Q_1}$ about $S$ to
another proof context.  Most prominent is the CONSEQUENCE rule that
allows us to strengthen the precondition $P_1$ to a precondition $P$
with $P \ra P_1$ and to weaken the postcondition $Q_1$ to a
postcondition $Q$ with $Q_1 \ra Q$, thus arriving at the conclusion
$\HT{P}{S}{Q}$.  Another one is Hoare's ADAPTATION rule dealing with
procedure calls.  Hoare stated in \cite{Hoa71} that in the absence of
recursion, i.e., in his proof system for \textbf{while} programs, his
ADAPTATION rule is a derived rule.  So the power of this rule is only
noticeable in the context of recursion.

Other rules can be conceived that are concerned with the same program
in the premise and conclusion. For example, the following rules were
introduced in \cite{Har79} and \cite{ABO09}. Here and elsewhere we
denote the set of variables of a program $S$ by $var(S)$.

\III

\NI
INVARIANCE 
\[ \frac{ \HT{R}{S}{Q}           }
        { \HT{P \A R}{S}{P \A Q} } 
\]
where $\mathit{free}(P) \cap change(S)=\ES$.  
\III

\NI
$\te$-INTRODUCTION 
\[ \frac{ \HT{P}{S}{Q}          }
        { \HT{\te\, x: P}{S}{Q}    }
\]
where $x \not \in var(S) \cup \mathit{free}(Q)$.
\III

\NI
SUBSTITUTION I 
\[ \frac{ \HT{P}{S}{Q} }  
        { \HT{P[\mathbf{z}:=\mathbf{t}]}{S}{Q[\mathbf{z}:=\mathbf{t}]} } 
\]
where $(\{\mathbf{z}\}\cap var(S)) \cup (var(\mathbf{t}) \cap change(S))=\ES$.
\III

We shall return to these rules shortly. But first, following
\cite{Old83}, let us discuss the ADAPTATION rule in 
the more general setting of programs.
We say that a program $S$ is \emph{based on} a finite set $X$ of variables
if $var(S) \subseteq X$ holds. 
Now we can recast Hoare's ADAPTATION rule as follows:
\III

\NI
ADAPTATION I
\[
 \frac{ \HT{P}{S}{Q}       }
      { \HT{\exists\, \mathbf{z} (P \land \forall \mathbf{x} (Q \ra R))}{S}{R}  }
\]
where $\mathbf{x}$ and $\mathbf{z}$ are lists of variables, $S$ is
based on $\mathbf{\{x\}}$, and
$\{\mathbf{z}\} = \mathit{free}(P,Q) \setminus (\mathit{free}(R) \cup
\{\mathbf{x}\})$.
\III

Following Hoare, the precondition in the conclusion of this rule
intends to express the weakest precondition $W$ such that
$\HT{W}{S}{R}$ holds (in the sense of partial correctness), assuming
that $R$ is the desired result of executing $S$ and $\HT{P}{S}{Q}$ is
already established.  This intention can be phrased as follows: 
find the weakest assertion $W$ such that $\HT{W}{S}{R}$ holds 
for \emph{all} programs based on $\{\mathbf{x}\}$ that satisfy
$\HT{P}{S}{Q}$.  In \cite{Old83} this precondition was calculated as
follows:
\[
  W \equiv \forall \mathbf{y}\, 
           ( \forall \mathbf{u}\, (P \ra Q[\mathbf{x}:=\mathbf{y}])
                                     \ra R[\mathbf{x}:=\mathbf{y}]),
\]
where $\mathbf{y}$ is a list of fresh variables of the same length
as $\mathbf{x}$ with $\{\mathbf{y}\} \cap \mathit{free}(P,Q,R) = \emptyset$
and $\{\mathbf{u}\} = \mathit{free}(P,Q) \setminus \{\mathbf{x}\}$.

Comparing $W$ with the precondition used in the conclusion of the ADAPTATION I rule
shows that
the implication
\[
  \exists\, \mathbf{z} (P \land \forall \mathbf{x} (Q \ra R)) \ra W
\]
holds but the converse is false.  Thus  Hoare's precondition
is sound but is stronger than necessary.
This suggests the following variant of the rule:
\III

\NI
ADAPTATION II
\[
 \frac{ \HT{P}{S}{Q}       }
      { \HT{W}{S}{R}  }
\]
where $S$ is based on $\{\mathbf{x}\}$ and $W$ is the precondition calculated above.
\III

\NI
To show the weakness of the Hoare's ADAPTATION I rule in the context of recursive
procedures, consider the following example essentially taken from \cite{Old83}.
Let $I$ be the standard interpretation of integers and
consider the procedure declaration
\[
 \mathbf{proc} \: p(\mathbf{x:}): 
  \mathbf{if}\ x>1\ 
  \mathbf{then}\ x:=x-1;\ \mathbf{call} \: p(\mathbf{x:});\ x:= x+1\ \mathbf{fi}, 
\]
where $x$ is an integer variable.
Then the procedure call satisfies the following correctness formula
\[
 \HT{x=y}{\mathbf{call} \: p(\mathbf{x:})}{x=y},
\]
where $y$ is a fresh variable $x$. It expresses the fact that the
procedure call leaves the variable $x$ unchanged.  For the
postcondition $x=y+1$ the ADAPTATION I rule (or, equivalently, the
original ADAPTATION rule) permits us to conclude only
\[
 \HT{\mathbf{false}}{\mathbf{call} \: p(\mathbf{x:})}{x=y+1},
\]
whereas the weakest liberal precondition 
$wlp_I(\mathbf{call} \: p(\mathbf{x:}),x=y+1)$
is expressed by the formula $x=y+1$.
This precondition is correctly calculated by the ADAPTATION II rule, which yields
\[
 \HT{x=y+1}{\mathbf{call} \: p(\mathbf{x:})}{x=y+1}.
\]

\NI

To compare the power of different adaptation rules in a systematic way, 
S.~de Gouw and J.~Rot
\cite{GR16} used the following notion due to \cite{Kle99}. A set
$\mathcal{R}$ of proof rules for a class $\mathcal{S}$ of programs is
called \emph{adaptation complete} if for all assertions $P,Q,P',Q'$
and finite sets $X$ of variables
\begin{itemize}
\item 
whenever for all programs $S \in \mathcal{S}$ based on $X$
the truth of $\HT{P}{S}{Q}$ implies the truth of $\HT{P'}{S}{Q'}$
in the sense of partial correctness,

\item then for all program $S \in \mathcal{S}$ based on $X$
there is a derivation of $\HT{P'}{S}{Q'}$ from $\HT{P}{S}{Q}$
using only rules of $\mathcal{R}$, written as
$\HT{P}{S}{Q} \vdash_{\mathcal{R}} \HT{P'}{S}{Q'}$.
\end{itemize}
By the result of \cite{Old83}, the set
$\mathcal{R}_\text{O} = \{\text{ADAPTATION II}, \text{CONSEQUENCE}\}$
is adaptation complete. Further, $\mathcal{R}_\text{O}$ enjoys two
properties, as noted in \cite{GR16}:

\begin{enumerate}
\item 
  Other adaptation rules, like 
$\exists$-INTRODUCTION and slightly weakened versions of the INVARIANCE and SUBSTITUTION I rules
are derivable from~$\mathcal{R}_\text{O}$. 

\item 
Any derivation in $\mathcal{R}_\text{O}$ can be replaced 
by a single application of each
of the two rules in $\mathcal{R}_\text{O}$.

\end{enumerate}

What about Hoare's adaptation rule?  Let
$\mathcal{R}_\text{H} = \{\text{ADAPTATION I}, \text{CONSEQUENCE}\}$.
From the above example it follows that this set is \emph{not}
adaptation complete.  Nevertheless, $\mathcal{R}_\text{H}$ enjoys
property~1, but not property~2 of $\mathcal{R}_\text{O}$.  \III

The paper \cite{Old83} also investigated three other adaptation
rules proposed in the literature.  An adaptation rule introduced in
\cite{GriesLevin80} turned out to be sound but \emph{not} adaptation complete
when grouped together with the CONSEQUENCE rule. In turn an
adaptation rule for the programming language {\sc Euclid} given in
\cite{LondonGHLMP78} is not even sound, while an
adaptation rule introduced in \cite{CartwrightOppen81} is both sound and
adaptation complete when grouped together with the CONSEQUENCE rule.

\subsection{Subscripted and local variables}
\label{subsec:local}

\NI
{\bf Subscripted variables}
In both \cite{Hoa71a} and \cite{FH71} programs with arrays were
studied. To reason about assignments to the subscripted variables the
ASSIGNMENT axiom was used, implicitly assuming that the definition of
substitution to subscripted variables is obvious.
This is indeed the case when the array subscripts are simple
expressions, for example a constant or a simple variable, which was
indeed the case for both programs analyzed there. However, in the case
of more complex subscripts difficulties may arise, as the following
example discussed in \cite{Bak80} shows.  In the case of an assignment
to a simple variable any correctness formula $\HT{P}{x:=t}{x = t}$,
where $t$ is a constant is true. However, the correctness formula
\[
\HT{a[1] = 2 \A a[2] = 2}{a[a[2]]:=1}{a[a[2]]=1}
\]
is false. Indeed, given the precondition the execution of the
assignment $a[a[2]]:=1$ amounts to executing the assignment $a[2]:=1$
after which the expression $a[a[2]]$ evaluates to 2 and not 1.  This
suggests that the ASSIGNMENT axiom cannot be used for arbitrary
subscripted variables.

This complication was clarified and solved in a systematic way in
\cite{Bak80}, by extending the definition of substitution to an
arbitrary subscripted variable.  The crucial step in the inductive
definition of the substitution $s[u:=t]$ deals with the case when
$s \equiv a[s_1]$ and $u \equiv a[u_1]$, for which one defines
\[
  s[u:=t] \equiv \IF{s_1[u:=t]=u_1} \THEN{t} \ \EELSE{a[s_1[u:=t]]} \ \FI.
\]
So in the \textbf{if} case one checks whether after performing the
substitution $[u:=t]$ on $s_1$ the subscripts $s_1$ and $u_1$ are
\emph{aliases} ---and substitutes in that case
$a[s_1]$ by $t$--- while in the \textbf{else} case one applies the
substitution $[u:=t]$ inductively to the subscript $s_1$ of $a[s_1]$.

J.W.~de Bakker showed that with this extended definition of substitution
the ASSIGNMENT axiom remains sound for subscripted variables.
Different axioms for assignment to subscripted variables were given in
\cite{HW73,Gri78,Apt81b}.
In the first two references in the corresponding axiom
an array is treated as a partial function and 
an assignment to a subscripted variable is
interpreted as a change in the whole function.

\smallskip

\NI
{\bf Local variables}
Consider now the case of local variables.  They can be viewed as a counterpart of bound
variables in logical formulas. However, the situation is more complicated
because of the dynamic character of variables in programming languages and the
presence of procedures.

We discussed already completeness in the sense of Cook of the proof
system $\cal H$ given in \cite{Coo78}. Cook actually considered an
extension of the proof system ${\cal H}$ by axioms and proof rules for
a small programming language that allows variable declarations and
nonrecursive procedures and proved its completeness in the above
sense.  However, the semantics of the block statement made the
corresponding completeness result invalid.  It is useful to discuss
this matter more closely.

Local variables were already dealt with in the DECLARATION rule
mentioned in Subsection \ref{subsec:hoare2}.
In \cite{Coo78} a different rule was used in which the substitution was applied
to the assertions and not to the programs, namely:
\III

\NI
BLOCK
\[
\frac{\HT{P[x := y]}{S}{Q[x:=y]}}
{\HT{P}{\mathbf{begin \; var \;} x; \: S \; \mathbf{end}}{Q}}
\]
where $y \not\in \mathit{free}(P,Q) \cup var(S)$.
\III

More precisely, a slightly adjusted version of this rule was used so
that one could reason about variable declarations in the context of
non-recursive procedures.  But even without this adjustment a possible
problem arises.  Consider the program
\[
  \mathbf{begin \; var}\; x; \: x := 1 \; \mathbf{end};
    \mathbf{begin \; var}\; y; \: z := y \; \mathbf{end}.
\]

In many programming languages it would yield an error because the
right-hand side of the second assignment refers to a value of the
uninstantiated variable $y$.  However, according to the semantics
proposed in \cite{Coo78} such assignments were allowed. Local
variables were modelled using a stack in which the last used value was
kept on the stack and implicitly assigned to the next local
variable. As a result the correctness formula
\[
\HT{\mathbf{true}}{\mathbf{begin \; var}\; x; \: x := 1 \; \mathbf{end};
    \mathbf{begin \; var}\; x; \: y := x \; \mathbf{end}}{y = 1}
\]
was true according to the semantics though there is no way to prove it.

\cite{Coo81} provided a corrigendum in which two possible fixes were
suggested. One was to modify the semantics so that the proposed proof
system is still complete. This can be achieved by assigning to each
newly declared variable a register that has not been used before and
modifying the notion of a state accordingly.

Another one was to require all newly declared variables to be
initialized to some fixed value, say $\omega$. This option was first used
in \cite{Gor75},
where a variant of the BLOCK rule was used in which the initialization
was modelled on the proof-theoretic level by the conjunct
$x = \omega$ added to the precondition of the premise.
To correct the proof of the relative completeness result given in
\cite{Coo78}, one should then
use an initialization of the local variables, both in the semantics and in the proof
theory, so use the variant of the BLOCK rule given in \cite{Gor75}.

Yet another option is to require all newly declared
variables to be explicitly initialized to some arbitrary expression.
This approach was taken in \cite{ABO09}, where the following more
general version of the corresponding rule was used that allowed a
declaration of a list of new variables:
\III

\NI
BLOCK I
\[
\frac{\HT{P}{\mathbf{x} := \mathbf{t}; S}{Q}}
{\HT{P}{\block{\mathbf{var} \  \mathbf{x} := \mathbf{t}; S}}{Q}}
\]
where $\C{\mathbf{x}} \cap free(Q) = \ES$.
\III

Here $\mathbf{x}:=\mathbf{t}$, where $\mathbf{x}$ is a list of
different variables and $\mathbf{t}$ a corresponding list of
expressions, is a \emph{parallel assignment}, introduced in
\cite{Dij75} and further discussed in Section
\ref{sec:alternative}.
\III

It is natural to postulate in this rule that the variables
listed in $\mathbf{x}$ do not appear in the expressions from
$\mathbf{t}$. However, this is a syntactic condition concerning the
program formation that is not needed to reason about partial
correctness. Further, as we shall soon see, putting no restrictions on
$\mathbf{x}$ and $\mathbf{t}$ turns out to be useful for modelling
parameter passing in a subtle situation when some formal parameters
happen to coincide with the global variables that are used in actual
parameters.

The observant reader will notice that in the discussed rules
substitution is used differently.  In the DECLARATION rule the
substitution is applied to the programs, in the BLOCK rule it is
applied to the assertions, while ---interestingly--- in the BLOCK I
rule it is not used at all.  The resulting proof systems yield
different results when applied to programs that use procedures. To
illustrate the problem consider the parameterless procedure declared
by $\mathbf{proc} \: p: z:= x$, the program
\[
  S_0 \equiv \mathbf{begin \; var \;} x; \ x:=0; \ \mathbf{call} \: p \; \mathbf{end},
\]
and the correctness formula
\begin{equation}
  \HT{x=1}{S_0}{z=1}.
  \label{equ:1}
\end{equation}

To reason about the procedure call $\mathbf{call} \: p$ we add to the proof system
$\cal H$ the following degenerated version of the RECURSION rule:
\III

\NI
COPY
\[
\frac{\HT{P}{S}{Q}}
{\HT{P}{\mathbf{call} \: p}{Q}}
\]
assuming the declaration of a parameterless non-recursive procedure $\mathbf{proc} \: p: S$.
\III

In our case it allows us to derive $\HT{x=1}{\mathbf{call} \: p}{z=1}$ from
$\HT{x=1}{z:=x}{z=1}$.
This in turn allows us to derive 
\[
  \HT{x=1}{y:=0; \ \mathbf{call} \: p}{z=1}.
\]
Now, applying the DECLARATION rule we get (\ref{equ:1}). 

However, using the BLOCK rule we get a different conclusion. Namely, we first establish
\[
    \HT{y=1}{x:=0; \ \mathbf{call} \: p}{z=0},
\]
from which
\begin{equation}
\HT{x=1}{S_0}{z=0}.
  \label{equ:2}
\end{equation}
follows.

Finally, if we use the BLOCK I rule,
and therefore consider a slightly modified
program
\[
  S' \equiv \mathbf{begin \; var \;} x:=0; \ \mathbf{call} \: p \; \mathbf{end},
\]
we get 
\begin{equation}
  \label{equ:blockI}
\HT{x=1}{S'}{z=0}.  
\end{equation}

These differences have to do with the way local variables are
interpreted in the presence of procedures.  According to the
\emph{static} scope policy the procedures should be evaluated in the
environment in which they were declared, while according to the
\emph{dynamic} scope policy they should be evaluated in the environment in
which they were called.  So according to the static scope, which is
adopted in most imperative languages, we should conclude
(\ref{equ:1}) and not (\ref{equ:2}) or (\ref{equ:blockI}).

We conclude that the proof systems studied in \cite{Coo78} and
\cite{Gor75} dealt with dynamic scope and not static scope. The same
is in principle the case for \cite{ABO09}, but one requires there that
the local variables are first renamed so that they differ from global
variables.  The effect is that then static and dynamic scopes
coincide.  
In the above example one thus considers the statement
\[
  S_1 \equiv \mathbf{begin \; var \;} y; \ y:=0; \ \mathbf{call} \: p \; \mathbf{end}
\]
instead of $S_0$.  Then we get $\HT{x=1}{S_1}{z=1}$, as desired.

\subsection{Parameter mechanisms and procedure calls}

The \emph{call-by-name} parameter mechanism was originally proposed in {\sc
  Algol 60}. It was used in \cite{Hoa71} and \cite{Coo78} and adopted
in all subsequently discussed papers on procedures, unless stated
otherwise.  It boils down to a simultaneous substitution of the actual
parameters for the formal ones, so naturally it was modelled
in the SUBSTITUTION rule by a straightforward substitution.

However, a most commonly used parameter mechanism is
\emph{call-by-value}.  According to its semantics the actual
parameters are evaluated first and subsequently their values assigned
to the formal parameters.  Some other parameter mechanisms were
occasionally used. For example, the programming language {\sc Pascal}
(see \cite{JW75}) also allows the \emph{call-by-variable} mechanism
(also called \emph{call-by-reference}), which is a mixture of
call-by-name and call-by-value. The actual parameter has to be a
variable. In case it is a subscripted variable, its index is evaluated
first and the resulting subscripted variable is substituted for the
formal parameter.

In \cite{AdB77} it was proposed to model these two parameter
mechanisms of {\sc Pascal} by means of an appropriate syntactic transformation
of the procedure body, that was called `syntactic application'.  In
what follows we use in the procedure declaration the qualification
$\mathbf{val}$ to indicate call-by-value and $\mathbf{var}$ to
indicate call-by-variable.  Given a procedure declaration
$\mathbf{proc} \: p(\mathbf{val} \; x, \mathbf{var} \; y): S$, so with
$x$ called by value and $y$ called by variable, the call $p(t,v)$,
where $t$ is an expression and $v$ a, possibly subscripted, variable,
was modelled by the syntactic application $S[t,v]$ defined by
\[
\begin{array}{rll}
  S[t,z] & \equiv & \mathbf{begin \; var \;} u; \ u:=t; \ S[x:=u, y:=z ] \; \mathbf{end}, \\
    S[t,a[s]] & \equiv & \mathbf{begin \; var \;} u_1, u_2; \ u_1:=t; \ u_2:=s; \ S[x:=u_1, y:=a[u_2] ] \; \mathbf{end},
\end{array}
\]
where $z$ is a simple variable and $u, u_1, u_2$ do not appear in $s, t, z$ or $S$.

This leads to the following generalization of the COPY rule
from the previous subsection:
\III

\NI
CALL-BY-VALUE/CALL-BY-VARIABLE
\[ \frac{ \HT{P}{S[t,v]}{Q} }
  { \HT{P}{\mathbf{call} \: p(t,v)}{Q} }  
\]
where the non-recursive procedure $p$ is declared by
$\mathbf{proc} \: p(\mathbf{val} \; x, \mathbf{var} \; y): S$.
\III

In \cite{ABO09} this approach to call-by-value was simplified
by noting that no variable renaming is needed to model it. The
resulting rule, that needs to be used in conjunction with the BLOCK I
rule, became:
\III

\NI
CALL-BY-VALUE
\[
  \frac{\HT{P}{\block{\mathbf{var} \  \mathbf{x} := \mathbf{t}; S}}{Q}}
         { \HT{P}{\mathbf{call} \: p(\mathbf{t})}{Q}                          }
\]
where the non-recursive procedure $p$ is declared by
$\mathbf{proc} \: p(\mathbf{val} \; \mathbf{x}): S$.
\III

To see how this rule correctly handles a subtle situation when a formal
parameter coincides with a global variable used in an actual parameter,
consider a procedure declared by
\[
  \mathbf{proc} \: p(\mathbf{val} \; x): x:=x+2; \ y:= y+x.
\]
Using the BLOCK I rule we can then establish the correctness formula
\[
  \HT{y=0 \land x=1}{\mathbf{begin \; var \;} x; \ x:=x+1; \ x:=x+2; \ 
y:=y+x \; \mathbf{end}}{y=4},
\]
from which
\[
  \HT{y=0 \land x=1}{p(x+1)}{y=4}
\]
follows by the CALL-BY-VALUE rule. This agrees with the semantics of
the call-by-value parameter mechanism. (The stronger postcondition,
$y=4 \land x=1$ can be established using the axioms and proof rules
introduced in the next section.)  So the assignment $x := x+1$ refers
on the left-hand side to the formal parameter $x$ and on the
right-hand side to the actual parameter $x+1$ that contains the global
variable $x$.

In \cite{Apt81b} it was suggested that other parameter mechanisms can
also be modelled by means of an appropriate syntactic application, and
subsequently reasoned about within Hoare's logic.  An example is the
\emph{call-by-result} parameter mechanism of {\sc Algol W}, a
precursor of {\sc Pascal} (see \cite{WH66}).  According to it the
actual parameter is either a simple or a subscripted variable. Upon
termination of the call the value of the formal parameter is assigned
to the actual parameter. In the case the actual parameter is a
subscripted variable, its index is evaluated first.  This parameter
mechanism is used in conjunction with the call-by-value.

We shall return to this subject in the next section when discussing
completeness of the counterparts of such rules
for recursive procedures.

\section{Reasoning about Arbitrary Procedures}
\label{sec:procedures}

\subsection{Completeness results for recursive procedures}
\label{subsec:complete-recur-proc}

The relative completeness result established in \cite{Coo78} dealt
with the language considered in \cite{Hoa71} and Subsection
\ref{subsec:hoare2}, except that recursion was disallowed. To ensure
soundness in the sense of partial correctness Cook stipulated that for
the procedure calls $\mathbf{call} \: p(\mathbf{a:e})$ no variable in
$(\mathbf{a:e})$ different from formal parameters occurs globally in
the procedure body.

Cook's work was extended to the language in which recursive procedures
were allowed in the Master Thesis of G.A.~Gorelick, 
written under the supervision of Cook.  The details are only available
as a technical report \cite{Gor75}.  We present the essentials for the
case of a single recursive procedure
$\mathbf{proc} \: p(\mathbf{x:v}): S$, in line with the presentation
in Subsection \ref{subsec:hoare2}.

The conceptual contribution of Gorelick is the introduction of 
\emph{most general formulas}. He wrote:

\begin{quote}
``The completeness result for recursive programs is then obtained by exhibiting,
for each recursive procedure $p$, a ``most general formula'' $\alpha_p$
such that $\vdash \alpha_p$, and $\alpha_p \vdash \beta$ for all
true formulas $\beta$ about $p$.''
\end{quote}
Given a procedure declaration
$\mathbf{proc} \: p(\mathbf{x:v}):  S$,
a \emph{most general formula for} the procedure $p$ is a correctness formula 
\[
 \HT{\mathbf{c}=\mathbf{z}}{\mathbf{call} \: p(\mathbf{x:v})}{G},
\]
where $\mathbf{c}$ is the list of variables that appear in the formal
parameters $\mathbf{x}$ and $\mathbf{v}$ or have a global occurrence
in $S$, and $\mathbf{z}$ is a list a fresh variables (not occurring in
$\mathbf{x}, \mathbf{v}$, or $S$), of the same length as $\mathbf{c}$
that serves to \emph{freeze} the initial values of the variables in
$\mathbf{c}$ before they are changed by $S$.  The formula $G$ is taken
to express the strongest postcondition $sp_I(\mathbf{c}=\mathbf{z},S)$
introduced in Subsection \ref{subsec:completeness}.  Since the
variables in $\mathbf{c}$ and $\mathbf{z}$ may appear in $G$ as free
variables, $G$ describes the relationship between the initial values of
$\mathbf{c}$ represented by $\mathbf{z}$
and the final values of $\mathbf{c}$ computed by the
procedure body $S$ and represented by $\mathbf{c}$ itself.

The crucial properties of most general formulas are captured by the following lemmas
due to \cite{Gor75}.
\III

\NI \textbf{Lemma G1}
If $\HT{P}{\mathbf{call} \: p(\mathbf{x:v})}{Q}$ is true in $I$ 
in the sense of partial correctness
then it can be derived 
from $\HT{\mathbf{c}=\mathbf{z}}{\mathbf{call} \: p(\mathbf{x:v})}{G}$ and
the set of all true formulas in $I$ using ``suitable adaptation rules''.
\III

\NI \textbf{Lemma G2}
For each procedure call $\mathbf{call} \: p(\mathbf{c:v})$ the most general formula
$\HT{\mathbf{c}=\mathbf{z}}{\mathbf{call} \: p(\mathbf{x:v})}{G}$ can be derived 
from the set of all true formulas in $I$
using Lemma G1 and the RECURSION rule.
\III

\NI
The proof of Lemma G1 is based on the following axiom and adaptation rules,
proposed in \cite{Gor75} for the case where $S$ is a procedure call
$\mathbf{call} \: p(\mathbf{a:e})$:
\III

\NI
INVARIANCE 
\[ 
 \HT{P}{S}{P} 
\]
where $\mathit{free}(P) \cap var(S)=\ES$.

\III

\NI
CONJUNCTION
\[ 
  \frac{ \HT{P}{S}{Q}, \HT{P}{S}{R}     }
       { \HT{P}{S}{Q \A R}                } 
\]
\vspace{1mm}

\NI
VARIABLE SUBSTITUTION 
\[ 
  \frac{ \HT{P}{S}{Q} }  
       { \HT{P[\mathbf{z}:=\mathbf{t}]}{S}{Q[\mathbf{z}:=\mathbf{t}]} } 
\]
where 
\begin{itemize}
\item 
  $\{\mathbf{z}\}\cap var(S) =\ES$,

\item if for any component $t_i$ of the list $\mathbf{t}$,
$var(t_i) \cap var(S) \neq \ES$,
then the corresponding component $z_i$ of the list $\mathbf{z}$
satisfies $z_i \not\in \mathit{free}(Q)$.
\end{itemize}
\III

Using Lemmas G1 and G2, Gorelick established the following result.
(The result was actually established for a system of recursive procedures,
so the RECURSION rule was appropriately generalized.)

\III

\NI \textbf{Completeness Theorem} For programs with recursive
procedures as defined in Subsection~\ref{subsec:hoare2}, the proof system
$\mathcal{H}$ extended by the RECURSION and SUBSTITUTION rules of
Subsection \ref{subsec:hoare2} and the above axiom and adaptation
rules is complete in the sense of Cook.
\III

Both \cite{Coo78} and \cite{Gor75} followed \cite{Hoa71} and used the
call-by-name parameter mechanism.  The restrictions imposed on the
actual parameters of `legal' procedure calls were partly lifted in
\cite{CartwrightOppen81}.  Analogous research was carried out in
\cite{Bak80} for the case of the call-by-value and call-by-variable
parameter mechanisms in the presence of recursive procedures. To this
end, the CALL-BY-VALUE/CALL-BY-VARIABLE rule from the previous section
was modified to the following rule:
\III

\NI
RECURSION III
\[
\begin{array}{l}
\HT{P_i}{\mathbf{call} \: p(t_i,v_i)}{Q_i}, i \in \C{\LLn} \vdash \HT{P_i}{S[t_i,v_i]}{Q_i}, i \in \C{\LLn}                    \\
[-\medskipamount]
\hrulefill                                                      \\
\HT{P_1}{\mathbf{call} \: p(t_1,v_1)}{Q_1} 
\end{array}
\]
where the procedure $p$ is declared by
$\mathbf{proc} \: p(\mathbf{val} \; x, \mathbf{var} \; y): S$ and
$p(t_i,v_i)$,  $i \in \C{1, \LL, n}$,
are the procedure calls that appear in $p(t_1,v_1)$ and $S[t_1,v_1]$.

\III

In \cite{Bak80} soundness and relative completeness for a
corresponding proof system was proved.
However, the proof was established only for the special case of a
single recursive procedure, given the combinatorial explosion of the
cases concerned with the relation between the actual and formal
parameters. The main ideas of this proof were discussed in
\cite{Apt81b}.

Proof systems for recursive procedures with the call-by-value
parameter mechanism were studied in a couple of other publications.
In \cite{Ohe99} a sound and relatively complete proof system was
introduced for a programming language with local and global variables
and mutually recursive procedures with the call-by-value mechanism,
that allows for both static and dynamic scope. The proofs were not
provided but it was mentioned that they were certified in the Isabelle
theorem prover.  In this work each assertion was identified with its
meaning represented as a function from states to Boolean values.  This
approach simplifies the soundness and relative completeness proofs by
avoiding various complications concerned with possible variable
clashes (in particular, it obviates the use of the SUBSTITUTION
rules), but makes it cumbersome to verify specific programs and
deviates from the syntactic approach pursued in the literature on
Hoare's logic.

Further, in \cite{ABOG12} soundness and relative completeness of the
proof system introduced in \cite{ABO09} was established. The recursive
procedures with the call-by-value parameter mechanism were dealt with
by an appropriate modification of the CALL-BY-VALUE rule from the
previous section.

However, in all these works on the call-by-value and call-by-variable
parameter mechanisms correctness of each procedure call had to be
dealt with separately.  This is an obvious drawback, since the
resulting correctness proofs are then not linear in the length of the
program.  It would be preferable if could establish a desired property
for a `generic call' just once, from which the needed properties of
all specific calls would follow by some form of a substitution rule.
In \cite{ABO09} it was observed that this can be achieved for the
calls in which no actual parameter contains a global variable.
Recently, K.R.~Apt and F.S.~de Boer introduced in \cite{AB19} a
sound and relatively complete proof
system that provides a satisfactory solution to this problem, without
any restriction.  They used a generalization of the following
recursion rule to a system of mutually recursive procedures:
\III

\NI
RECURSION IV
\[
  \frac{\HT{P}{p(\mathbf{x})}{Q} \vdash \HT{P}{S}{Q}}
  {\HT{P}{p(\mathbf{x})}{Q}}
\]
where the procedure $p$ is declared by
$\mathbf{proc} \: p(\mathbf{val} \; \mathbf{x}): S$
and $\C{\mathbf{x}} \cap free(Q) = \ES$.
\III

It was combined with the following rule:
\III

\NI
PROCEDURE CALL
\[
\frac{\HT{P}{p(\mathbf{x})}{Q}}
{\HT{P[\mathbf{x}:=\mathbf{t}]}{p(\mathbf{t})}{Q}}
\]
where the side conditions are the same as in the previous rule.
\III

The resulting proof system uses the BLOCK I rule from the previous
section, so it is appropriate for the dynamic scope. However, the
authors pointed out that for a natural class of programs in which the
global variables that occur in procedure bodies differ from the local
variables, static and dynamic scopes coincide. Such a class of
programs was first considered in \cite{CartwrightOppen81}.

So far we discussed partial correctness of programs with recursive
procedures.  In \cite{Apt81b} it was stated without a proof that a
proof system corresponding to the one used in \cite{Gor75} is sound in
the sense of total correctness.  In this system the INVARIANCE axiom
was dropped (it is not sound for total correctness), the procedures
had no parameters and the RECURSION rule was replaced by the RECURSION
I rule.  However, it was discovered in \cite{AB90} that this claim is
false. The problem has to do with the fact that the counter variable
$n$ can be subject to quantifier elimination in the
$\exists$-INTRODUCTION rule, and to substitution in the SUBSTITUTION
rule.

For example, given the procedure declaration
$\mathbf{proc} \: p: p;p$ of an obviously nonterminating procedure
one can establish the premises $\neg P(0)$ and
$\HT{P(n)}{\mathbf{call} \:p}{\mathbf{true}} \vdash \HT{P(n+1)}{S}{\mathbf{true}}$ of
the above rule for $P(n) \equiv n>1$ and conclude
$\HT{\mathbf{true}}{\mathbf{call} \:p}{\mathbf{true}}$ by the RECURSION I and
CONSEQUENCES rules.

The solution proposed in \cite{AB90} was to stipulate that the counter
variables are treated as constants in the $\exists$-INTRODUCTION and
SUBSTITUTION rules.  This allowed the authors to prove both soundness
and relative completeness of the resulting proof system for total
correctness of recursive procedures without parameters w.r.t.~the
arithmetic interpretations introduced in Subsection \ref{subsec:soundness}.
The above remarks explain why we added in the
RECURSION II rule a qualification concerning the counter variable~$z$.

\subsection{Clarke's incompleteness result}
\label{subsec:incompleteness}

Programming languages like {\sc Algol 60} \cite{Nau63} and {\sc
  Pascal} \cite{JW75} contain procedure mechanisms that are
considerably more complex than what we discussed so far.  In his
seminal paper \cite{Cla79}, E.M.~Clarke identified a combination
of features for which it is impossible to obtain a Hoare-like proof
system that is sound and relatively complete (in the sense of Cook).
By a Hoare-like proof system we mean a set of syntax-directed proof
rules for correctness formulas about programs in the programming
language under consideration such that the application of each proof
rule is decidable. Moreover, the proof system allows one to prove
correctness formulas concerning programs from a given programming
language.

Clarke proved this incompleteness result for a block-structured
programming language $L_0$ which includes the following features that
are present in {\sc Algol~60} and {\sc Pascal}:

\begin{enumerate}

\item procedure names as parameters of procedure calls,

\item recursion,

\item static scope,

\item global variables in procedure bodies,

\item nested procedure declarations.

\end{enumerate}

\noindent
Clarke's incompleteness result is based on the following two lemmas
for Hoare-like proof systems $H$ and programming languages $L$.
\III

\NI \textbf{Lemma A}
If $H$ is sound and relatively complete for $L$, then
the divergence problem for $L$ is decidable for finite interpretations $I$.
\III

\NI \textbf{Lemma B}
If $L$ has a rich procedure concept including features 1--5, then
its divergence problem is undecidable for all finite
interpretations $I$ with at least two domain elements.
\III

\noindent
An interpretation $I$ is called \emph{finite} if its domain $D$ is
finite.  A program $S$ in $L$ \emph{diverges} for $I$ if $S$ never
terminates when started with any input values from $D$ for its
variables.  The \emph{divergence problem} for a programming language
$L$ for an interpretation $I$ is the problem of deciding whether an
arbitrary program $S$ of $L$ \emph{diverges}.

The proof of Lemma A rests on the following general observations.
A program $S$ of $L$ diverges for $I$ if and only if the correctness formula
$\HT{true}{S}{\mathit{false}}$ is true in $I$ in the sense of partial correctness.
Since $H$ is sound and relatively complete, the latter is true if and only if
$\HT{true}{S}{\mathit{false}}$ is provable in $H$, when all assertions in the 
proofs are interpreted in $I$.
Thus all diverging programs $S$ in $L$ can be recursively enumerated by
enumerating all proofs in $H$,
thereby deciding in $I$ the assertions used in the applications
of the CONSEQUENCE rule, which is possible because $I$ is finite.
Trivially, also all non-diverging programs $S$ in $L$ can be recursively enumerated:
simply start each program $S$ for all of its finitely many inputs from $I$
and enumerate it in case it halts for one of these inputs.
Decidability of the set of all diverging programs follows since both the set and its 
complement are recursively enumerable.

For the proof of Lemma B, Clarke shows that \emph{queue machines} (or
\emph{Post machines}), which have an undecidable halting problem, see, e.g., 
\cite{Manna74}, can be simulated by programs from $L_0$.  A queue
machine manipulates a queue and finitely many registers.  A machine
program is a finite sequence of labelled instructions of three types:

\begin{itemize}
\item \texttt{enqueue} $x$ that adds the value of register $x$ to the rear of the
  queue,

\item \texttt{dequeue} $x$ that removes the front entry from the queue and
  puts into register $x$, and

\item \texttt{if} x=y \texttt{then go to} $\ell$ that
branches to the instruction labelled with $\ell$ if the values of the
registers $x$ and $y$ agree.
\end{itemize}

Clarke \cite{Cla79} described his simulation idea 
of a queue machine with a program in $L$ as follows:
\begin{quote}
``The queue is represented by successive activations of a recursive procedure $sim$
with the queue entries being maintained as values of the variable $top$ which is local
to $sim$. Thus an addition to the rear of the queue may be accomplished by having
$sim$ call itself recursively. Deletions from the front of the queue are more
complicated. $sim$ also contains a local procedure $up$ which is passed as a
parameter during the recursive call which takes place when an entry is added to the 
rear of the queue. In deleting an entry from the front of the queue, this
parameter is used to return control to previous activations of $sim$
and inspect the values of $top$ local to those activations.
The first entry in the queue will be indicated by marking (e.g. negating)
the appropriate copy of $top$.''
\end{quote}

From this description it is clear that the simulation program exploits the 
features 1, 2 and 5. Feature 4 is also needed since a global variable 
\emph{program\_counter} is used in the body of the procedure $sim$.
Feature~3 (static scope) concerns the semantics of procedures 
and is needed to achieve the correct back pointers for the procedure $up$
in the runtime stack generated by the successive activations of the 
procedure $sim$.

Note that the procedure $sim$ has a formal parameter that is
instantiated with procedures, which in turn have no formal procedure
parameters on their own.  Here the concept of mode depth is helpful.
The \emph{mode depth} of a procedure is defined inductively.  A
procedure that may not take identifiers of procedures as parameters
has mode depth 1.  For any $k \ge 1$, a procedure that may take
identifiers of procedures of mode depth $k$ has mode depth $k+1$.
Thus for Clarke's simulation, procedures of mode depth $\le 2$
suffice.  This restriction is also obeyed in {\sc Pascal}~\cite{JW75}.
Procedures of an arbitrary finite mode depth, corresponding to arbitrary
\emph{higher types}, as in {\sc Algol 68}~\cite{Wij75}, or even with
\emph{self-application}, i.e., with procedure calls of the form
$\mathbf{call} \: p( \dots, p, \dots)$, as possible in {\sc Algol
  60}~\cite{Nau63}, are not needed.

\subsection{Clarke's language $L_4$}

Clarke also claimed that each language $L_i$ obtained
from $L_0$ by disallowing self-application and the feature
$i \in \{1,2,3,4,5\}$ of the above list has a sound and relatively
complete Hoare-like proof system \cite{Cla79}. However, he proved this only for
the case of $L_3$ in which dynamic scope replaces static scope. The
completeness argument rests on the fact that each program $S$ in $L_3$
has a finite range, where the \emph{range} of $S$ is the set of
different procedure calls that can be invoked in the computations of
$S$.  He claimed that a similar proof system could be obtained for $L_4$
in which global variables are disallowed. His argument was that
programs of $L_4$ can be transformed into schematically equivalent
ones in $L_5$, where nested procedure declarations are
disallowed. Such programs have a sound and relatively complete
Hoare-like proof system.

H.~Langmaack and E.R.~Olderog \cite{LO80} discovered that this argument is wrong.
They considered the \emph{formal execution trees} of programs
and showed that programs in $L_4$ may have trees with context-free path languages
whereas programs in  $L_5$ can generate only trees with regular path languages.
Therefore they posed the challenge of developing a sound and relatively
complete Hoare-like proof system for $L_4$.

As a first step, Langmaack proved in \cite{Lan82} that for all {\sc
  Algol}-like programs in $L_4$ the divergence problem is decidable
for finite interpretations.  This is a necessary condition for the
existence of a sound and relatively complete Hoare-like proof system
according to Lemma A.  Moreover, due to a theorem by R.J.~Lipton
\cite{Lip77}, this decidability result is also sufficient for the
existence of a sound and relatively complete `Hoare logic' for {\sc
  Algol}-like programs in $L_4$. However, Lipton's notion of a `Hoare
logic' is rather weak: it means that the set of correctness formulas
that are true in the sense of partial correctness is recursively
enumerable relative to the underlying interpretation. Lipton's theorem
does not yield any usable, syntax-oriented proof rules.

First concrete axiomatizations for $L_4$ programs appeared
independently in papers by Olderog \cite{Old84} and W.~Damm and B.~Josko
\cite{DJ83}.  The former studied the case of {\sc
  Pascal}-like programs, i.e. with mode depth $\le 2$.  The key idea
is the use of second-order relation variables in the assertion
language to stand for the behaviour of uninterpreted, symbolic
procedures in a new SEPARATION rule.  In order to determine what a
procedure call $\mathbf{call} \: p( \dots q \dots)$ of a procedure $p$
with an actual procedure parameter $q$ does, this rule \emph{first}
determines separately what $q$ does and what $p$ 
does with a symbolic procedure instead of $q$ represented by a formula
and \emph{then} composes both results using substitution.
Applications of the SEPARATION rule have the effect of a systematic
transformation of an initially non-regular formal execution tree into
a regular tree at the cost of introducing higher-order tree
combinators.  The resulting regularity enables a suitably complete
Hoare-like proof system.

The paper \cite{DJ83} handled the more general case
of {\sc Algol}-like programs with arbitrary finite modes
by using higher-order predicate variables and 
unevaluated substitutions to such variables.
Common to both papers is that they deviated from the standard notion
of relative completeness in that they used a higher-oder assertion
language and an appropriate notion of expressiveness.

The status of the language $L_4$ was finally clarified in 1989 in an
over 90 pages long paper by S.M.~German, E.M.~Clarke and J.Y.~Halpern
\cite{GCH89} in which a Hoare-like proof system for $L_4$ was provided
that is sound and relatively complete in the (original) sense of
Cook. The completeness proof heavily relies on an appropriate
arithmetic encoding of procedure declarations so that the assertion
language remains first-order.

\subsection{The characterization problem} 

Exploring the border between relative completeness and incompleteness
(in the sense of Cook) of Hoare's logics has been the focus of
considerable research in the 1980s.  Clarke called it the
``Characterization Problem for Hoare Logics'' in his survey article
\cite{Cla85}.  The key question is what is meant by a `Hoare logic'.
Several researchers proved results for a weak interpretation where a
Hoare logic is just the set of correctness formulas
$\HT{P}{S}{\mathit{Q}}$ that are true in the sense of partial
correctness.  Here the problem is for which classes of programs $S$,
assertions $P, Q$, and underlying interpretations $I$ this set is
recursively enumerable or even decidable.

We cited already Lipton's result \cite{Lip77}.  Since its proof was
only sketched, Langmaack provided in \cite{Lan79} a rigorous proof for
the setting of {\sc Algol}-like programs $S$.  Both Lipton and
Langmaack restrict themselves to quantifier-free assertions $P, Q$.
Clarke, German and Halpern established \cite{CGH83} an ``effective
axiomatization of Hoare logics'' extending Lipton's result to
first-order formulas $P, Q$.  However, these results do not yield any
usable, syntax-directed proof rules that are in the spirit of
Hoare-like proof systems that appeared since the publication of
\cite{Hoa69}. Therefore Clarke stated in \cite{Cla85}:
\begin{quote}
``Certainly the most important research problem is to develop
a version of the characterization theorem that provides some insight as to when 
a syntax-directed proof system can be obtained.''
\end{quote}

Two such characterization theorems were established in \cite{Old81} and \cite{Old-STOC83}.
The first paper studied a language $L_{Algol}$ of {\sc Algol}-like programs
with the features 1, 2, 4 and 5 of \cite{Cla79} 
(see Subsection~\ref{subsec:incompleteness}).
Feature 3 (static scope) was not fixed but left as a parameter
in the form of a \emph{copy rule}.
Such a rule defines how during program execution a procedure call is
replaced by a copy of the procedure body.  In the copy certain
occurrences of local variables and procedures in the procedure body
are renamed to avoid name clashes.  By varying the copy rule,
different program semantics varying from dynamic to static scope were
defined.  The COPY rule from Subsection \ref{subsec:local} is a
trivial example of such a rule in which no renaming takes place.

Parameterized by a given copy rule $\mathcal{C}$, a Hoare-like proof system
$\mathcal{H}(\mathcal{C})$ for the partial correctness of 
{\sc Algol}-like programs was introduced.
A program $S$ was called $\mathcal{C}$-\emph{bounded} if applications of the copy
rule $\mathcal{C}$ do not lead to programs with ``procedural reference chains'' 
of arbitrary length.
A program $S$ has a \emph{finite} $\mathcal{C}$-\emph{index} if the relation of
``substitutional equivalence'' induces only finitely many equivalence classes in the
set of reachable procedure calls.
Due to this equivalence, 
the condition of finite $\mathcal{C}$-\emph{index} is more liberal than the conditions of
\emph{finite recursive cycle} and \emph{finite range} used 
by Gorelick \cite{Gor75} and Clarke \cite{Cla79}, respectively,
in their completeness proofs.
The following result holds.
\III

\NI \textbf{Theorem} 
Suppose the assertion language is expressive relative to the considered interpretation
$I$ and the copy rule $\mathcal{C}$. Then the following 
are equivalent for {\sc Algol}-like programs $S$ and assertions $P, Q$:

\begin{enumerate}

\item $\HT{P}{S}{Q}$ can be proved in the proof system $\mathcal{H}(\mathcal{C})$
      assuming all true formulas in $I$.

\item $S$ has a finite $\mathcal{C}$-index and $\HT{P}{S}{Q}$ is true in $I$
      in the sense of partial correctness.
      
\item $S$ is $\mathcal{C}$-bounded and $\HT{P}{S}{Q}$ is true in $I$
      in the sense of partial correctness.

\end{enumerate}

As corollaries to this theorem various completeness results were obtained 
in \cite{Old81}.
For the naive copy rule $\mathcal{C}_n$, which models ``dynamic scoping'',
the proof system $\mathcal{H}(\mathcal{C}_n)$ is sound and relatively complete for 
the full set $L_{Algol}$ because all programs in $L_{Algol}$ are
$\mathcal{C}_n$-bounded. This corresponds to Clarke's language~$L_3$.
For the Algol 60 copy rule $\mathcal{C}_{60}$, which models ``static scoping'',
the proof system $\mathcal{H}(\mathcal{C}_{60})$ is sound and relatively complete 
for the following sublanguages of $L_{Algol}$ because all programs in these
sublanguages are $\mathcal{C}_{60}$-bounded:

\begin{itemize}

\item $L_{pnes}$ -- all programs without procedure nesting, 
                    corresponding to Clarke's language $L_5$,

\item $L_{par}$ -- all programs with parameterless procedures only,

\item $L_{pp}$ -- all programs without procedures as parameters,
                  corresponding to Clarke's language $L_1$,

\item $L_{rp}$ -- all programs in which formally recursive procedures 
                  are disallowed to have procedure identifiers as formal parameters,
                  a superset of  Clarke's language $L_2$,

\item $L_{\mathit{gf}}$ -- all programs without global formal procedure parameters.

\end{itemize}

The paper \cite{Old-STOC83} studied a set  $L_{Pas}$ 
of programs with {\sc Pascal}-like procedures $S$, i.e., with
mode-depth $\le 2$.
Here a sublanguage $L \subseteq L_{Pas}$ is called
\emph{admissible} if it is closed under certain program
transformations that leave the procedure structure invariant.  In
particular, these transformations allow for the introduction of global
variables.  A tree $T$ is \emph{regular} if the set of paths in $T$ is
a regular language in the Chomsky hierarchy or, equivalently, if $T$
has only finitely many different patterns of subtrees.  The
\emph{formal call tree} of a program $S$ describes in which order the
procedures of $S$ are called, where a branching appears when from one
procedure several immediate successors can be called.  
The following theorem refers to a particular Hoare-like proof system $H_0$
that was introduced in \cite{Old-STOC83}.
\III

\NI \textbf{Theorem} 
For every admissible language $L \subseteq L_{Pas}$ the following 
are equivalent:

\begin{enumerate}

\item There exists a sound and relatively complete
      Hoare logic for $L$ in the sense of Lipton.
      
\item The divergence problem for $L$ is decidable for
      finite interpretations.

\item All programs in $L$ have regular formal call trees.

\item The Hoare-like proof system $H_0$ is sound and relatively
      complete for $L$.

\end{enumerate}

Since $L_{Pas}$ contains programs with a non-regular formal call tree,
the theorem implies that there is no sound and relatively complete
Hoare logic for $L_{Pas}$ itself, a fact that follows from Clarke's incompleteness
result.
Note that this theorem does not contradict the results obtained for $L_4$
because a sublanguage $L \subseteq L_{Pas}$  without global variables
is not admissible.

\section{Nondeterministic and Probabilistic Programs}
\label{sec:fairness}

\subsection{Reasoning about nondeterminism}
\label{subsec:nondeterminism}

In the context of programming languages \emph{nondeterminism} stands
for the phenomenon that a program can yield more than one answer.  In
the sixties and early seventies a couple of simple programming
constructs were proposed that introduced nondeterminism. In particular,
in \cite{Lau71} the nondeterministic statement  $S_1 \ \mathbf{or} \ S_2$
was considered with the meaning:
execute either $S_1$ or $S_2$, 
and the following proof rule:
\III

\NI
OR
\[ \frac{ \HT{P}{S_1}{Q}, \HT{P}{S_2}{Q}                }
        { \HT{P}{S_1 \ \mathbf{or} \ S_2}{Q}                          }\]

\vspace{1mm}
      
But the undoubtedly most successful and elegant proposal is the language of
\emph{guarded commands} introduced by E.W.~Dijkstra in \cite{Dij75}
and in the book form in \cite{Dij76}.  Dijkstra's original motivation
for introducing this language was to simplify programs by delaying
some arbitrary choices to the implementation level and to restore
symmetry that is not present in the \textbf{if-then-else}
statement. This led to a specific proposal for nondeterminism.

For the sake of what follows it is sufficient to consider
two crucial statements of Dijkstra's language:

\begin{itemize}
\item 
the \emph{alternative command}
\[\IFP, \]
\item 
\emph{repetitive command}
\[\DOP, \]
\end{itemize}
where each $B_i$ is a Boolean expression, called a \emph{guard}, and
each $S_i$ is a program statement.  The symbol $\Box$ represents a
nondeterministic choice.  

The alternative command is executed by selecting a guard $B_i$ that
evaluates to \textbf{true} and executing the associated statement
$S_i$.  If more than one guard $B_i$ evaluates to true any of the
corresponding statements $S_i$ may be executed next.  If all guards
evaluate to \textbf{false}, the execution of the alternative command
results in a \emph{failure}.
The repetitive command is executed in a similar way, with two
differences. First, after termination of a selected statement $S_i$
the repetitive command is executed again. Second, if all guards
evaluate to \textbf{false}, the execution of the repetitive command
simply terminates.  So it is a natural generalization of the
\textbf{while} statement.

As an illustration of the use of the alternative command consider the
customary program for computing the maximum of two numbers using the
conditional statement:
\begin{newtabbing}
\qquad
\ITE{x \geq y}{max:=x}{max:=y}.
\end{newtabbing}
A solution using the alternative command is
symmetric in the variables $x$ and $y$ and also involves nondeterminism:
\[
\IF x \geq y \ra max:=x\ \Box\ y \geq x \ra max:=y \ \FI.
\]

As an illustration of the use of the repetitive command consider
the customary {\bf while} program for computing the {\it greatest common
  divisor} ({\it gcd}\/) of two natural numbers, initially stored in
the variables $x$ and $y$:
\begin{newtabbing}
\qquad
\= \WD{x \neq y}                                        \\
\> \qquad \ITE{x>y}{x:=x-y}{y:=y-x}                     \\
\> \OD.
\end{newtabbing}

\NI
Using the repetitive command the same algorithm
can be written as
\[\DO x>y \ra x:=x-y \ \qed \ y>x \ra y:=y-x\ \OD. \]
Both programs terminate with the {\it gcd} stored in the variables $x$
and $y$ but the second program is more readable and, unlike the first
one, is symmetric in the variables $x$ and $y$.

To reason about the guarded command language Dijkstra introduced in
\cite{Dij75} the \emph{weakest precondition calculus} that we
briefly discuss in Section \ref{sec:alternative}.
This approach allows one to reason about total correctness.
It is easy to conceive Hoare-style proof rules that deal with partial correctness
of the alternative and repetitive commands (they were proposed first in \cite{Bak80}).
For total correctness some care has to be exercised to deal 
with the absence of failures, which is a new concept in this framework.
The appropriate rule, introduced in \cite{Apt84}, takes the following form:
\III

\NI
ALTERNATIVE COMMAND II
\[
\begin{array}{l}
P \ra \bigvee_{i=1}^n\ B_i,                                \\
\HT{P \A B_i}{S_i}{Q}, i \in \C{\LLn}                   \\
[-\medskipamount]
\hrulefill                                              \\
\HT{P}{\IFPa}{Q}
\end{array}
\]

To obtain a proof system for total correctness of guarded commands
it suffices now to deal with the
termination of the repetitive command. 
The following natural generalization of the WHILE II
rule was used in \cite{ABO09}:
\III

\NI
REPETITIVE COMMAND II
\[
\begin{array}{l}
\HT{P \A B_i}{S_i}{P}, i \in \C{\LLn},                  \\
\HT{P \A B_i \A t=z}{S_i}{t<z}, i \in \C{\LLn},         \\
P \ra t \geq 0                                                  \\
[-\medskipamount]
\hrulefill                                                      \\
\HT{P}{\DOPa}{P \A \bigwedge_{i=1}^{n} \neg B_i}
\end{array}
\]
\NI where $t$ is a termination function and $z$ is an
integer variable not occurring in $P, t, B_i$ or $S_i$ for
$i \in \C{\LLn}$.

\subsection{Reasoning about fairness}
\label{subsec:fairness}

\emph{Fairness} is a concept that arises in presence of any
form of nondeterministic choice.  Suppose that we repeatedly have some
choice among a fixed set of alternatives, for instance of going left
or going right. If we repeatedly select the alternative of `going
left', then we systematically ignore the other alternative. In such case
we can argue that the adopted selection procedure is not fair with
respect to the other alternative, `going right'. To exclude such unfair
selection procedures we need to focus on infinite `runs' of selections
of alternatives and make precise when an alternative can be selected.

These matters can be discussed in a more precise way using the guarded
commands language.  Consider the following program, where $k$ is a
fixed natural number:
\[
x:=1;\ \DO x>0 \ra x:=x+1 \ \qed \ x>k \ra x:=0\ \OD. 
\] 
It does not always terminate, since we can repeatedly select the first
guard.  The resulting computation is considered \emph{unfair} since
from some moment on the second guard is always enabled (i.e.,
evaluates to \textbf{true}), but never selected.
Once the second guard is selected when
it is enabled, the program terminates. More formally, we say that the
above program terminates under the \emph{fairness} assumption, which
stipulates that each guard that is from some moment on continuously
enabled is infinitely often selected.

As another example consider the following program, where
$odd(x)$ is a test with the expected meaning:
\[
x:=1;\ \DO x>0 \ra x:=x+1 \ \qed \ odd(x) \ra x:=0\
  \OD. 
\] 
Also this program does not always terminate. The only infinite
computation repeatedly ignores the second guard. However, in contrast
to the previous example, in this infinite computation at no moment the
second guard becomes continuously enabled. So even under the fairness
assumption this program does not terminate.

On the other hand in this infinite computation the second guard is
infinitely often enabled. We say that this computation is
\emph{strongly unfair}.  If the second guard is selected when it is
enabled, the program terminates.  In this case we say that the above
program terminates under the \emph{strong fairness} assumption, which
stipulates that each guard that is infinitely often enabled is
infinitely often selected. (To stress the difference the first notion of
fairness is usually called \emph{weak fairness}.)

Both forms of fair termination can be established by means of appropriate
proof rules. In what follows we explain the \emph{transformational approach} 
proposed in \cite{AO81} and in a journal form in \cite{AO83}.  To prove
termination of a repetitive command $S$ under the assumption of weak
(or strong) fairness (with respect to a precondition $P$) one can
transform it into a repetitive command $T(S)$ the computations of
which coincide with the weakly (or strongly) fair computations of $S$.
To this end, we need a \emph{random assignment} $x :=?$ that assigns
nondeterministically to the variable $x$ an arbitrary natural
number. To understand why this command naturally arises when
considering fairness, note that under both assumptions of fairness the
program
\[b:= \mathbf{true}; \ x:=0; \ \DO b \ra x:=x+1 \ \qed \ b \ra b:=\mathbf{false}\ \OD \]
mentioned in \cite{Dij76} is equivalent to
\[b:= \mathbf{false}; \ x:=?.\]

In what follows we limit ourselves to the presentation of strong
fairness and to explain the idea we focus on a repetitive command with
just two guards:
\[
S \equiv \DO B_1 \ra S_1 \: \Box \: B_2 \ra S_2 \ \OD.
\]
The transformed program $T(S)$ uses two auxiliary variables $z_1$ and $z_2$
ranging over natural numbers and has the following form:
\begin{mytabbing}
\qquad\qquad
$T(S) \equiv$ \= $z_1 := ?;\ z_2 := ?;$ \\
              \> \DO \= \\
       \>     \> $B_1 \land z_1 \leq z_2 \ra S_1; \ z_1 := ?; \ \IF B_2 \THEN z_2 := z_2 - 1 \FI$ \\
             \> $\Box$ \\
       \>     \> $B_2 \land z_2 \leq z_1    \ra S_2; \ z_2 := ?; \ \IF B_1 \THEN z_1 := z_1 - 1 \FI$ \\
              \> \OD.
\end{mytabbing}

Informally, each variable $z_i$ tracks the number of times the
corresponding guard is enabled and the augmented guards prevent that
an infinitely often enabled guard is never selected.

To reason about the strong fair termination of $S$ it is now
equivalent to reason about the termination of the program $T(S)$. 
To this end, we only need an axiom dealing with the random
assignment.  Such an axiom was proposed in \cite{Har79}:
\III

\NI
RANDOM ASSIGNMENT
\[ 
\HT{P}{x:=?}{P} 
\]
where $x$ does not appear free in $P$.
\III

But this indirect approach can be avoided by absorbing the
transformation $T(\cdot)$ into the proof of termination of the program
$T(S)$. This way one obtains a proof rule for establishing termination
under the strong fairness assumption that deals directly with the
original program $S$ and is similar in shape to the REPETITIVE
COMMAND II rule.  We omit the details, though we should mention that to
reason about fairness it is in general necessary that the termination
function takes values from an arbitrary well-founded ordering and not
just natural numbers (see \cite{AO83}, \cite{APS84},
and \cite{AP86}, where soundness and relative completeness of a
proof system for total correctness of \textbf{while}
programs augmented with random assignment was established.).

Different approaches to reason about fairness in the context of
nondeterministic programs were independently proposed in \cite{LPS81}
and \cite{GFMR81}, that appeared in a journal form as \cite{GFMR85}.
In \cite{APS84} the method presented here was extended to programs
with nested nondeterminism.  These approaches and proof rules were
discussed in book form in \cite{Fra86}, where also additional
versions of fairness were considered.

The transformational approach to fairness was originally developed in
\cite{AO81,AO83} for the Dijkstra's guarded command language in which
each repetitive command has a fixed finite number of alternatives.
The transformations can be seen as implementing a general fair
scheduler controlling finitely many processes.  These transformations
were extended in \cite{OlderogP10,HoenickeOP10} to deal with
\emph{dynamic control}, where processes can be created dynamically.
Then the overall number of processes can be infinite, but at each step
of an execution of the system the number of created processes is
finite.  J.~Hoenicke and A.~Podelski went in~\cite{HoenickeP15} one
step further and extended the transformations to deal with fairness
with an \emph{infinitary control}, where the number of created
processes can be infinite. Both dynamic and infinitary control were
expressed by repetitive commands with infinitely many alternatives.
However, these papers did not propose any proof rules derived from the
new transformations.

\subsection{Probabilistic programs}
\label{subsec:prob}

Probabilistic programs are sequential programs with the ability to
draw values at random from probabilistic distributions.  They have
attracted large attention in the research community due to many
applications, for example in security to describe randomized
encryptions, in machine learning to describe distribution functions,
and in randomized algorithms.  They have typically just a few lines of
code, but are hard to analyze, see, e.g., \cite{KatoenGJKO15}.  Properties of
interest for such programs include the \emph{expected runtime} and
\emph{almost sure termination}.

Most formal modelling takes place in the setting of an extended
Dijkstra's guarded command language, called 
\emph{probabilistic guarded command language}, abbreviated pGCL.
Here, both nondeterministic choice and probabilistic choice are admitted.
In particular, McIver and Morgan \cite{McIverM05} carried out their
research on probabilistic programs in this setting.
They extended the notion of weakest precondition to 
\emph{weakest pre-expectations}.

J.~den Hartog and E.P.~de Vink \cite{HartogV02} introduced a Hoare-like proof
system for partial correctness of probabilistic programs which are
defined as (determininistic) \textbf{while} programs extended by the
\emph{probabilistic choice} $S_1 \oplus_{\rho} S_2$ between the
statements $S_1$ and $S_2$.  This intention is that the statement
$S_1$ is chosen with the probability $\rho$
and  the statement $S_2$ with the probability $1-\rho$.
This necessitates the introduction of probabilistic predicates in the
assertion language.  In these predicates, the real-valued expression
$\mathbb{P}(R)$ yields the probability that the normal predicate $R$
holds.  For example, for an integer variable $x$ the correctness
formula
\[
  \HT{x=1}{x:=x+1 \oplus_{\frac{1}{3}} x:=x+2}
     {\mathbb{P}(x=2) = \frac{1}{3} \land \mathbb{P}(x=3) = \frac{2}{3}}
\]
holds.  The proof system of \cite{HartogV02} contains several new
rules that go beyond the ones concerning the \textbf{while}
programs. In particular, for the probabilistic choice the following
rule was proposed:
\III

\NI
PROBABILISTIC CHOICE
\[ \frac{ \HT{P}{S_1}{Q}, \HT{P}{S_2}{Q'}                }
        { \HT{P}{S_1  \oplus_{\rho}  S_2}{Q \oplus_{\rho} Q'}                          }\]

\vspace{1mm}

\NI
where the probabilistic choice operator $\oplus_{\rho}$ is also applied
to the probabilistic predicates $Q$ and $Q'$.
\III

Den Hartog and de Vink established soundness of their proof system.  They also
proved relative completeness for the subset of loop-free probabilistic
programs and a restricted set of predicates in the postcondition. The
proof is based on calculating the weakest precondition.

Some other developments concerning verification of probabilistic programs are
discussed in Subsection~\ref{subsec:relational}.

\section{Parallel and Distributed Programs}
\label{sec:parallel}

By a \emph{concurrent program} we mean a program that has a number of
components, the execution of which proceeds in parallel. If the
program components share some variables one usually refers to this
framework as shared memory parallelism. Following \cite{ABO09}, we
refer to the corresponding programs as \emph{parallel programs}.  If
the program components do not share variables but can communicate by
messages, one usually calls these components \emph{processes} and
refers to the corresponding concurrent programs as \emph{distributed
  programs}.  A most commonly used approach relies on asynchronous
message passing.  However, within the framework of program
verification, for reasons of simplicity, one rather focuses on
synchronous message passing, which means that the processes need to
synchronize their actions to make the communication possible.  
Again following \cite{ABO09}, we mean by \emph{distributed programs}
the ones with synchronous message passing, within the context
of a specific language proposal due to Hoare \cite{Hoa78}.

\subsection{Reasoning about parallel programs}
\label{subsec:parallel}

Verification of parallel programs calls for new insights. To discuss the matters
let us first introduce the syntax. By 
\[
\PP
\]
we mean a \emph{parallel composition} of $n$ sequential programs,
$S_1,\LL,S_n$, that may share variables. 

The main complication in reasoning about parallel programs is the
interference caused by the use of shared variables.  For example, both
\[\HT{x=0}{[x:=x+1;\ x:=x+1}{x=2} \]
and
\[\HT{x=0}{x:=2}{x=2} \]
hold but
\[\HT{x=0}{[x:=x+1;\ x:=x+1 \| x:=2]}{x=2 \lor x = 4} \]
does not since one possible computation consists of executing the
assignments in the following order $x:=x+1;\ x:=2;\ x:=x+1$, which
yields the final value 3.

The first limited approach to the verification of parallel programs
was proposed by Hoare in \cite{Hoa72d}, where he dealt with parallel composition
of disjoint program components. More formally, we say that the
component programs $S_1, \LL, S_n$ of $\PP$ are \emph{disjoint} if no
variable subject to change in one component appears in another
component.  This led to the following proof rule:
\III

\NI
DISJOINT PARALLELISM
\[ \frac{\raisebox{1mm}{$ \HT{P_i}{S_i}{Q_i}, i \in \C{\LLn}$}}
        {\raisebox{-1mm}{$ \HT{\bigwedge^n_{i=1}\ P_i}{\PP}{\bigwedge^n_{i=1}\ Q_i}$}}\]
\NI
where $\mathit{free}(P_i,Q_i) \cap change(S_j)=\ES$ for $i \neq j$.
\III

As noticed by Hoare in \cite{Hoa75}, parallel and sequential compositions of program
components coincide under the assumption of
disjointness. This brought him to suggest the following rule:
\III

\NI
SEQUENTIALIZATION
\[ \frac{\raisebox{1mm}{$ \HT{P}{S_1;\ \LL;\ S_n}{Q} $}}
        {\raisebox{-1mm}{$ \HT{P}{\PP}{Q} $}} \]
\vspace{1mm}
     
(In \cite{Hoa72d} and \cite{Hoa75} actually only parallel composition
of two components was studied).  In 1975 this limited approach to
program correctness was extended to arbitrary parallel programs. It
was based on a novel idea of \emph{interference freedom}, first
introduced in the PhD thesis of S.~Owicki \cite{Owi75}, and
subsequently published in two papers by her and supervisor
D.~Gries, \cite{OG76a} and \cite{OG76b}. This approach became known as
the \emph{Owicki-Gries} method.

In what follows we explain the ideas behind this extension of Hoare's
logic.  To deal with the problem of interference mentioned above,
Owicki and Gries suggested to compose not the correctness statements
about the component programs \emph{but} their proofs, presented in an
appropriate form.  This form relies on the fact that the correctness
proofs of the component programs are syntax-directed, that is they
follow in some sense the program structure. As a result the proof can
be recorded by retaining the used assertions in the program text. For
example, the application of the WHILE rule can be retained in the text
of the \textbf{while} statement by writing its conclusion as
\[
\HT{P}{\WDD{B}{\{P \A B\}\ S \ \{P\} }}{P \A \neg B},               
\]
while the conclusion of the CONSEQUENCE rule can be written as
\[
  \{P\} \HT{P_1}{S}{Q_1} \{Q\},
\]
which amounts to interpreting two consecutive assertions as an
implication.  The crucial point of such a proof representation, called
a \emph{proof outline}, is that each statement $S$ is preceded in it
by at least one assertion.  Subsequent assertions indicate implication
between them.

Here is an example of a proof outline for the correctness proof of the
\texttt{DIV} program that we considered in Subsection \ref{subsec:hoare1}:

\begin{mytabbing}
\qquad\qquad
\= \C{\T}                             \\
\> \C{x = x + y \cdot 0}                             \\
\> $r:=x;$                             \\
\> \C{x = r + y \cdot 0}                             \\
\> $q:=0;$                                   \\
\> \C{P}                                    \\
\> $\WD{y \leq r}$                                    \\
\> \qquad \C{P \A y \leq r}                           \\
\> \qquad \C{x = (r-y) + y \cdot (1+q)} \\
\> \qquad $r:=r-y;$ \\
\> \qquad \C{x = r + y \cdot (1+q)} \\
\> \qquad $q:=1+q$                     \\
\> \qquad \C{P} \\
\> \OD                                                  \\
\> \C{\neg y \leq r \A P},                                       \\
[\bigskipamount]
where                                                   \\
[\bigskipamount]
\> $P \equiv x = r + y\cdot q$.
\end{mytabbing}

In the Appendix we give another example of a proof outline by
representing in such a form Turing's proof discussed in Subsection
\ref{subsec:Turing}.  To understand the essence of the approach of
Owicki and Gries let us return for a moment to the original Hoare's
proof system. It is sound in the following stronger sense, where,
given a proof outline and a subprogram $T$, we denote by $pre(T)$
the assertion that directly precedes $T$.
For a proof see, e.g., \cite{ABO09}.
\III

\NI \textbf{Strong Soundness Theorem} Consider a
proof outline that represents the proof of a correctness formula
$\HT{P}{S}{Q}$.  Take a computation $\xi$ of $S$ that starts in a
state that satisfies $P$. Each time $\xi$ reaches a substatement $T$
of $S$, the precondition $pre(T)$ is satisfied.  Further, if $\xi$
terminates, its final state satisfies $Q$.
\III

Armed with this knowledge let us return to parallel programs. Suppose that
we established the correctness formulas $\HT{P_i}{S_i}{Q_i}$, where
$i \in \C{\LLn}$ and $S_1, \LL, S_n$ are the component programs of
the parallel program $\PP$. Let $\HT{P_i}{S_i^*}{Q_i}$ be the corresponding
proof outlines. We call them
\emph{interference-free}
if for all assignments $x:=s$ in $\PP$ and all assertions $R$ used in a proof
outline of another component
the correctness formula
\[ \HT{R \A pre(x:=s)}{x:=s}{R} \]
holds. Informally, the proof outlines are interference-free if the execution
of each assignment statement in the state that satisfies its
assertion does not invalidate the assertions used in the proof
outlines of other components.  Then the following proof rule allows us
to reason about parallel programs:
\III

\NI
PARALLELISM
\[
\begin{array}{c}
\mbox{The proof outlines\ } \HT{P_i}{S^*_i}{Q_i},  %            \\
i \in \C{\LLn},\ \mbox{are interference-free}                                   \\
[-\medskipamount]
\hrulefill                                                          \\
\HT{\bigwedge^n_{i=1}\ P_i}{\PP}{\bigwedge^{n}_{i=1} Q_i}
\end{array}
\]
\vspace{1mm}

Owicki and Gries noted that to reason about parallel programs they
needed \emph{auxiliary variables} (occasionally called by other
researchers \emph{ghost variables}), so variables that neither
influence the control flow nor the data flow of the program, but only
record some additional information about the program execution.
Formally, a set of simple variables $A$ is called a {\it set of
  auxiliary variables} of $S$ if each variable from $A$ occurs in $S$
only in assignments to the variables from $A$.  The appropriate proof
rule allowing one to delete them is:
\III

\NI
AUXILIARY VARIABLES
\[ \frac{ \HT{P}{S}{Q}          }
        { \HT{P}{S_A}{Q}        }\]
where for some set of auxiliary variables $A$ of $S$
the program $S_A$ results from $S$ by deleting all assignments
to the variables in $A$,  and no variable from $A$ appears free in $Q$.
\III

In presence of shared variables it is sometimes useful to restrict the
number of interference points. This can be done by allowing
\emph{atomic sections} (see \cite{Lip75} and \cite{Lam77}), which are
statements of the form $\ATOM{S}$, where $S$ does not contain any
loops or further atomic sections.  The idea is that such a statement
is executed without any interruption by another component of the
parallel program.  This suggests the following proof rule (see
\cite{Owi78}):
\III

\NI
ATOMIC SECTION
\[ \frac{ \HT{p}{S}{q}                                  }
        { \HT{p}{\ATOM{S}}{q}                           }
\]

To allow for synchronization between the component programs, Owicki and
Gries used a generalization of the atomic section statement, called
an \emph{await statement} $\ATE{B}{S}$, with the following
meaning: if $B$ evaluates to true, the statement $S$ is executed
as an atomic section.  The corresponding proof rule is:
\III

\NI
AWAIT
\[ \frac{ \HT{P \A B}{S}{Q}                             }
        { \HT{P}{\ATE{B}{S}}{Q}                         }
\]

The presence of synchronization statements leads to a possibility of a
\emph{deadlock}, an undesired situation in which some components of a
parallel program did not terminate while the nonterminated components
are all blocked.  Owicki and Gries noted that interference-free
outlines allow one to reason about absence of deadlock in a simple
way.  To this end, it suffices to identify the possible deadlock
situations and to show for each of them that the corresponding
conjunction of the $pre(T)$ assertions cannot be simultaneously
satisfied.

This approach to verification was successfully applied in \cite{OG76a}
to establish correctness of non-trivial parallel programs, in
particular two versions of the classical producer/consumer problem, a
parallel version of the \texttt{FIND} program mentioned in Subsection
\ref{subsec:hoare1}, and an implementation of Dijkstra's semaphores.
In \cite{OG76b} it was extended to parallel programs that use a more
efficient synchronization construct called conditional regions
introduced in \cite{Hoa72}, and in \cite{Owi78} to concurrent programs
with shared data classes.

\smallskip

\NI
{\bf Termination}
Let us discuss now the issue of termination of parallel
programs. Owicki and Gries proposed to establish it by using the WHILE
II rule instead of the WHILE rule and by postulating that no
component program increases the termination function of a \textbf{while}
statement used in another component program.  However, the authors of
\cite{ABO90} noticed that the termination of parallel programs is
subtler than it sounds and came up with an example of simple parallel
program that does not terminate, even though its termination can be
established using Owicki and Gries method.

The problem has to do with the fact that the assertions used in the
proof of the second premise of the WHILE II rule, so
$\HT{P \A B \A t=z}{S}{t<z}$, are not tested for interference freedom.
In \cite{ABO90} two ways of solving this problem are proposed. The
first is to retain in the proof outlines the correctness proofs of
both the first and second premise of this rule, so that all assertions
used in these proofs are subjected to the interference freedom
test. Another is by adding additional requirements to the definition
of interference freedom that ensure that the used termination functions
decrease along all syntactically possible paths through the
program. The latter approach suffices to validate the termination
proofs presented in \cite{OG76a}.

Finally, termination of parallel programs under the fairness
assumption was dealt with in \cite{OA88}, where the Owicki-Gries method
was combined with a transformational approach presented in Subsection
\ref{subsec:fairness}.

Independently of the Owicki-Gries method of \cite{OG76a}, L.~Lamport
proposed in \cite{Lam77} an essentially equivalent approach to
verification of parallel programs. The difference was that, as in
\cite{Flo67a}, he considered programs presented as flowcharts, now one
for each component program. As a result the interference-freedom test
referred to the assertions attached to the flowchart nodes and
translated into a requirement that these assertions are
\emph{monotone}, that is, that they are maintained by the actions of
the other components. Lamport used this approach to establish
correctness of a solution to the mutual exclusion problem. In the
paper also program properties were considered that had not been hitherto addressed.
We shall return to this matter in the last section of the
paper.

\smallskip

\NI
{\bf The rely-guarantee approach}
A drawback of the Owicki-Gries method is that due to the test of
interference freedom, verification of a parallel program
$S \equiv \PP$ is possible only if all components $S_1, \LL , S_n$ of
$S$ are explicitly given. Further, this method is not compositional,
where `compositionality' means that (partial) correctness of a
parallel program is derived directly from the correctness of its
components.

This prompted research on alternative formalisms for reasoning about
parallel programs. A discussion of these approaches can be found in
\cite[pp.~479-484]{RBH01}. Here we limit ourselves to an account of
one, perhaps most successful proposal, called the \emph{rely-guarantee
  approach}. It was introduced in the PhD thesis of C.B.~Jones,
\cite{Jon81}, the essence of which appeared in \cite{Jon83}. This
approach provides a compositional way reasoning about concurrent
programs by incorporating the interference-freedom test into the
proof.  This is achieved by using more informative correctness
formulas.

We follow here the presentation given in the book by W.P.~de Roever
et al.~\cite{RBH01}.  In the rely-guarantee approach one assumes that
a given program $S$ is executed in some environment and therefore uses
an extended specification format
\[
  \langle R,G\rangle:\HT{P}{S}{Q},
\]
where a correctness formula $\HT{P}{S}{Q}$ is extended by an 
\emph{interface specification} $\langle R,G\rangle$ of the environment 
consisting of a rely condition $R$ and a guarantee condition $G$. 
In contrast to the assertions $P$ and $Q$ in the correctness formula, 
$R$ and $G$ are \emph{predicates on transitions}, i.e., they
relate two states, the one before executing a transition and the one after it.
As in Turing's flowchart in Subsection~\ref{subsec:Turing},
primed versions of variables are used to refer to the state after executing the transition.
For example, $x' < x$ expresses that the value of the variable $x$ decreases.
The idea is that $R$ states assumptions that $S$ makes on the transitions
of its environment and that $G$ states the guarantees that $S$ provides to the environment.

Informally, $\langle R,G\rangle:\HT{P}{S}{Q}$
expresses that the correctness formula $\HT{P}{S}{Q}$ is true in the sense of partial
correctness if

\begin{itemize}

\item whenever at some moment during the computation of $S$ all
      past environmental transitions satisfy $R$

\item then all transitions of $S$ up to that moment satisfy $G$.

\end{itemize}

\NI
For parallel composition of programs $S_1$ and $S_2$ the following proof rule 
is presented in \cite{RBH01}:
\III

\NI
PARALLEL COMPOSITION
\[ 
\begin{array}{l}
   (R \lor G_1) \ra R_2, \\
   (R \lor G_2) \ra R_1, \\
   (G_1 \lor G_2) \ra G, \\[1mm]
   \langle R_i, G_i\rangle:\HT{P_i}{S_i}{Q_i}, i = 1,2, \\[-\medskipamount]
\hrulefill                                                      \\
\langle R,G\rangle: \HT{P_1 \land P_2}{[S_1 \| S_2]}{Q_1 \land Q_2} 
\end{array}
\]

In \cite[p.~453]{RBH01} this rule is explained as follows:

\begin{quote}

\begin{itemize}

\item every transition of $S_i$ (which is characterized by $G_i$) and
      every transition of the common environment of $S_1$ and $S_2$ 
      (characterized by $R$) is seen by $S_j$ with $i \neq j$ as an
      environment transition which has to satisfy $R_j$,
      
\item every transition by $S_1$ or $S_2$ is a transition of $[S_1 \| S_2]$,
      and therefore has to satisfy $G$, and

\item since the validity of $\langle R_i, G_i\rangle:\HT{P_i}{S_i}{Q_i}$
      implies $Q_i$ is invariant under $R_i$, the postcondition $Q_1 \land Q_2$
      holds upon termination of $[S_1 \| S_2]$;
      since $R$ implies $R_i$, $i=1,2$, this implies that $Q_1 \land Q_2$
      is invariant under $R$ after $[S_1 \| S_2]$ has terminated, too.

\end{itemize}

\end{quote}

Note that the rule is compositional for the customary correctness
formulas.  The crux of applying it is to find suitable rely-guarantee
conditions for which the implications in the premises hold.  In
\cite{Jon83} and \cite{RBH01} this approach was illustrated by
providing alternative correctness proofs of a solution to 
the mutual exclusion problem and of a
parallel version of the \texttt{FIND} program considered in
\cite{OG76a}.  In practice, this approach may be as difficult as
proving interference freedom.

\subsection{Reasoning about distributed programs}

Hoare introduced an elegant approach to distributed
programming based on synchronous communication~\cite{Hoa78}.  In an intentional
analogy to the title of Dijkstra's seminal paper \cite{Dij68} on
parallel programs, this proposal was called Communicating Sequential
Processes. In the paper Hoare introduced a simple programming language
for distributed programming, called since then CSP, in which
Dijkstra's guarded command language was extended by allowing
communication primitives for synchronous communication. This focus on
synchronous, as opposed to asynchronous, communication had a huge
impact on the theory of distributed programming and was also realized
in the programming language Occam \cite{INM84}.  The idea is that
synchronous communication is simpler to reason about as it obviates
the discussion of message ordering and buffers and their
management. Synchronous communication can be implemented by means of
asynchronous one, by adding additional processes that simulate message
buffers, so it can be viewed as an elegant abstraction.

In \cite{AFR80} a proof system was proposed to reason about a simple
class of CSP programs.  The crucial idea of this approach was an
introduction of a \emph{cooperation test} that corresponds to the
interference test of the Owicki-Gries method.

To explain the matters we first clarify the relevant aspects of CSP.
Each CSP program consists of a parallel composition of
processes, written as
\[
[{PR}_1 :: S_1 \| \LL \| {PR}_n :: S_n]
\]
Each ${PR}_i$ is a label of a process and $S_i$ is its program.  These
processes share no variables. They communicate by means of synchronous
communication that is achieved by means of two matching
\emph{input/output commands}, in short i/o commands.  An i/o command
has the form ${PR}_i ?x$ (an input command) or ${PR}_i !t$ (an output
command), where $x$ is a variable, and $t$ an expression.  The i/o
commands ${PR}_i ?x$ or ${PR}_j !t$ \emph{match} if $i \neq j$,
${PR}_i ?x$ appears in the program for process ${PR}_j$, ${PR}_j !t$
appears in the program for process ${PR}_i$, and the types of $x$ and
$t$ coincide.

When the control in the programs for processes ${PR}_i$ and ${PR}_j$
is just in front of the mentioned i/o commands and they match, they
can be executed jointly, with the effect that the value of $t$ is
assigned to $x$. So the effect of the joint execution of the commands
${PR}_i ?x$ or ${PR}_j !t$ is that of an assignment $x:=t$.

In the CSP language, Dijkstra's guarded commands are generalized by
allowing i/o commands to appear in the guards. So the guards can now
also be of the form $B; \alpha$, where $B$ is a Boolean expression and
$\alpha$ is an i/o command. If $B$ evaluates to \textbf{true} the i/o
command $\alpha$ of the generalized guard behaves the same way as the
usual i/o command, though it fails if it addresses a process that
terminated. If $B$ evaluates to \textbf{false} the generalized guard
fails. 

Further, Dijkstra's \textbf{do}-\textbf{od} notation is
replaced by using a star '*' and the `[' and `]' brackets, while the
\textbf{if}-\textbf{fi} notation is replaced by the `[' and `]'
brackets, that are also used to enclose the parallel composition.

As an example of a CSP program consider the following transmission problem,
taken from \cite{ABO09}, that is a simplified version of a similar problem 
discussed in \cite{Hoa78}.
We wish to transmit from the \SENDER\ process to the \RECEIVER\
process through a \FILTER\ process a sequence of characters in such a
way that \FILTER\ process deletes from the sequence all blank
characters.  The following CSP program 
\[
  [\,\SENDER\ :: S_1\, \|\, \FILTER\ :: S_2\, \|\, \RECEIVER\ :: S_3\,],
\]
where

\begin{mytabbing}
\qquad
\= $S_1 \equiv$ \=                                                \kill
\> $S_1 \equiv$  \> $i:=0$;  $* [\,i \neq M;\FILTER\,!\,a[i] \ra i:=i+1\,]$,   \\
[\medskipamount]
\> $S_2 \equiv$  \> $in:=0;\ out:=0;\ x:=\BLANK$;                   \\
\> \> *[\,\=$x \neq \AST;\SENDER\,?\,x \ra$                               \\
\> \>        \> \qquad \= \IF    \= $x=\BLANK \ra$ \= $skip$            \\
\> \>        \>        \> $\qed$ \> $x \neq \BLANK \ra$ \> $b[in]:=x;$  \\
\> \>        \>        \>        \>                     \> $in:=in+1$   \\
\> \>        \>        \> \FI                                           \\
\> \>        \> $\qed \ out \neq in;\RECEIVER\,!\,b[out] \ra out:=out+1$      \\
\> \> \phantom{*}],     \\                                                        
[\medskipamount]
\> $S_3 \equiv$ \> $j:=0;\ y:=\BLANK$;                            \\
\>                   \> $*[\,y \neq \AST;\FILTER\,?\,y \ra c[j]:=y;j:=j+1\,]$.
\end{mytabbing}
is a solution to this problem.

Here the sequence of characters is initially stored in the array
$a[0:M-1]$ of characters in the process \SENDER. The last element of the array
is the special character \AST, i.e., $a[M-1] = \AST$.  The process
\FILTER\ has an array $b$ of characters serving as an intermediate
store for processing the character sequence and the process \RECEIVER\
has an array $c$ of characters to store the result of the filtering
process.  For coordinating its activities the process \FILTER\ uses
two integer variables $in$ and $out$ pointing to elements in the array
$b$.

The process \FILTER\ can communicate with both other processes.  It can
receive characters from process \SENDER\ until \AST\ has been
received and it can transmit all nonblank characters to the
process \RECEIVER.  The Boolean parts of the generalized guards
of the \FILTER\ process can both evaluate to \textbf{true}.  In that
case the next action can be either a communication between \SENDER\ and
\FILTER\ or between \FILTER\ and \RECEIVER.  Consequently this CSP
program exhibits a nondeterministic behaviour.

The process \SENDER\ terminates once it has sent all its $M$
characters to the \FILTER\ process.  The process \FILTER\ terminates
when it has received the character \AST\ and it has transmitted to
\RECEIVER\ all nonblank characters it has received.  Finally, the
process \RECEIVER\ terminates once it has received from \FILTER\ the
character \AST.  Thus the parallel composition of these three
processes terminates if \SENDER\ sends as the last of its $M$
characters the \AST.

The proof system proposed in \cite{AFR80} extends the one for the
guarded commands language by the following axioms and proofs rules
that deal with the communication.
For simplicity we assume that all variables and expressions are of the same type.
\III

\NI
INPUT
\[
\HT{P}{{PR}_i?x}{Q}
\]

\NI
OUTPUT
\[
\HT{P}{{PR}_i!t}{P}
\]

\III

\NI
GENERALIZED ALTERNATIVE COMMAND
\[ \frac{\HT{P \A B_i}{\alpha_i}{R_i}, \HT{R_i}{S_i}{Q}, i \in \C{\LLn}}
        {\HT{P}{[\qed^n_{i=1}\ B_i; \alpha_i \ra S_i]
}{Q}} \]
\III

\NI
GENERALIZED REPETITIVE COMMAND
\[ \frac{\HT{P \A B_i}{\alpha_i}{R_i}, \HT{R_i}{S_i}{P}, i \in \C{\LLn}}
        {\HT{P}{*[\qed^n_{i=1}\ B_i; \alpha_i \ra S_i]
}{P \A \bigwedge_{i=1}^{n} \neg B_i}} \]

The INPUT axiom may look strange since it allows us to conclude an
arbitrary postcondition.  However, the used assertions still have to
pass the cooperation test.  This test refers to the proof outlines
which are defined as in the case of Owicki-Gries approach.  

Suppose now that we established the proof outlines
$\HT{P_i}{S_i^*}{Q_i}$, where $i \in \C{\LLn}$ and $S_1, \LL, S_n$ are
respective programs of the processes ${PR}_1, \LL, {PR}_n$.  We say
that these proof outlines \emph{cooperate} if

\begin{itemize}
\item the assertions used in $\HT{P_i}{S_i^*}{Q_i}$ contain
  no variables subject to change in $S_j$ for $i \neq j$,

\item $\HT{pre_1 \A pre_2}{{PR}_j?x\, \|\, {PR}_i !t}{post_1 \A post_2}$
  holds whenever $\HT{pre_1}{{PR}_j?x}{post_1}$ and \\
  $\HT{pre_2}{{PR}_i!t}{post_2}$ are taken respectively from the
  proof outlines $\HT{P_i}{S_i^*}{Q_i}$ and $\HT{P_j}{S_j^*}{Q_j}$.

\end{itemize}

Intuitively, proof outlines cooperate if they help each other to validate the post
conditions of the i/o commands present in these proofs. 
To establish cooperation the following axiom is needed.
\III

\NI
COMMUNICATION
\[
  \HT{\mathbf{true}}{{PR}_j?x \| {PR}_i!t}{x = t}
\]
provided ${PR}_j?x$ and ${PR}_i!t$ match.
\III

This axiom simply states that, as mentioned before, the effect of the
joint execution of a pair of matching i/o commands
${PR}_j?x$ and ${PR}_i!t$ is that of an assignment $x := t$.
Recall that in CSP, the processes composed by parallel composition
do not share variables, so $x$ cannot occur in $t$. This justifies the
simple postcondition $x=t$ for the assignment $x:=t$.

Then the following proof rule allows one to draw conclusion about the parallel composition
of processes:
\III

\NI
CSP PARALLELISM
\[
\begin{array}{c}
\mbox{The proof outlines\ } \HT{P_i}{S^*_i}{Q_i},  %            \\
i \in \C{\LLn},\ \mbox{for the processes ${PR}_1, \LL, {PR}_n$ cooperate}                                   \\
[-\medskipamount]
\hrulefill                                                          \\
\HT{\bigwedge^n_{i=1}\ P_i}{[{PR}_1 :: S_1 \| \LL \| {PR}_n :: S_n]}{\bigwedge^{n}_{i=1} Q_i}
\end{array}
\]
\vspace*{1mm}

To reason about the CSP programs, as in the case of the Owicki-Gries
approach, the AUXILIARY VARIABLES rule is needed. The presented
reasoning about deadlock freedom is analogous as in \cite{OG76a},
though the reasons for a deadlock can now be different. In particular,
a process can be blocked forever if the control in its program is just
before an i/o command that addresses a process that terminated.  Using
the proposed proof system some example CSP programs were proved
correct in \cite{AFR80}. 
(In \cite{Moi83} it was pointed out that one
of the correctness proofs contained an error and a corrected version
was presented.)  The proof system was subsequently proved in
\cite{Apt83} to be sound and complete in the sense of Cook.  Neither
\cite{AFR80} nor \cite{Apt83} considered termination.

Independently of \cite{AFR80} a very similar proof system to reason
about CSP programs was proposed in \cite{LG81}, in which the
\emph{satisfaction property} corresponds to the cooperation test.
However, the authors used a different semantics of the generalized
repetitive commands than the one stipulated in \cite{Hoa78} and taken
care of in the proof system of \cite{AFR80}.  On the other hand,
in contrast to \cite{AFR80}, program termination was considered.  A
number of different approaches to reason about communicating processes
was proposed in the literature around that time. They are surveyed in
\cite{HR86}.

Subsequently, a simplified proof system for a fragment of CSP was
proposed in \cite{Apt86} and used to establish correctness of a
solution to the so-called \emph{distributed termination problem}.  In this
fragment a program for each process consists of a guarded commands
program followed by a single generalized repetitive command.  The i/o
commands can appear only in the guards of this repetitive command and thus not as
separate statements. An example of a CSP program written in this
fragment is the above-mentioned solution to the transmission
problem. This fragment was studied independently in \cite{ABC87} and
\cite{Zob88}, where it was shown that each CSP program can be
transformed into a program in this subset using some control
variables.

The underlying idea of this approach is that such simpler CSP programs
can be transformed into Dijkstra's guarded commands language without
introducing any additional variables. By absorbing this transformation
into a proof rule one obtains a proof rule that uses a global
invariant and deals directly with the considered CSP program. This way
this approach dispenses with the cooperation test.  Termination is
naturally dealt with by following the approach used for the guarded
commands language.  This proof system was adopted in \cite{ABO09} and
its two previous editions, where its soundness was shown to be a
direct consequence of the correctness of the above-mentioned program
transformation.

In \cite{AFK88} it was explained that in the context of CSP programs
fairness, a notion we discussed in Subsection \ref{subsec:fairness},
can have various interpretations. One of them states that every pair
of matching i/o commands that is infinitely enabled is also infinitely
often selected.  In \cite{GFK84} a proof rule for dealing with this
form of fairness of the CSP programs was proposed.

The approach of \cite{AFR80} and \cite{LG81} was presented only for
CSP programs without nested parallelism.  It was extended in
\cite{AB90a} to CSP programs that allowed nested parallelism and also
dynamic process creation, a feature not present in CSP. Subsequent
work in this direction, \cite{Boe91}, shifted emphasis to reasoning
about objects, a subject that deserves a separate section.

Finally, let us mention that verification of parallel and distributed programs
in the Hoare-like style was systematically presented in book form in 
\cite{RBH01}.

\section{Object-oriented Programs}
\label{sec:oo}

Object-oriented programming, as exemplified by languages like C++ or Java,
builds upon the notion of an object and concepts like inheritance and subtyping.
The difficulty in reasoning about such programs is that their execution
creates dynamic pointer structures that go beyond the static program
structure that has been the backbone of the syntax-directed
Hoare-style proof rules discussed so far.

\subsection{Language characteristics}

As object-oriented programming has been realized in many, often incompatible, ways, we
clarify first the main characteristics of the objects here considered. These are:
\begin{itemize}
\item
objects possess (and {\em encapsulate}) their own  {\em instance}  variables, 
\item
objects interact via {\em method} calls,
\item
objects can be dynamically {\em created}.
\end{itemize}

Each object consists of a set of instance variables and a set of methods.
In contrast to the formal parameters of procedures and the local
variables of the block statements which only exist {\em temporarily}, the
instance variables of an object exist {\em permanently}.  
The \emph{local state} of an object is a mapping that assigns
values to its instance variables.  Each object represents its local
state by a {\em pointer} to it.  {\em Encapsulation} means that the
instance variables of an object cannot be directly accessed from other
objects; they can be accessed only by the method calls of the object.

A method call invokes a procedure which is executed by
the called object.  The execution of a method call thus involves a
temporary {\em transfer} of control from the local state of the caller object to
that of the called object (also referred to as {\em callee}).  
Upon termination of the method call the control returns to the local state of the caller.
The method calls are the \emph{only way} of transferring control from 
one object to another.

The account of verification of object-oriented programs that follows
is based on \cite[Chapter 6]{ABO09}.  We distinguish two kinds of
variables: the set $\mathit{Var}$ of {\em normal} variables, the ones
considered so far, and the set $\mathit{IVar}$ of instance variables,
which are owned by objects.

We consider a set of \emph{methods}, each defined by means of a
declaration
\[
m(\mathbf{u})::S,
\]
where the identifier $m$ denotes a method, $\mathbf{u}$ is the list of
formal parameters of type $\mathit{Var}$, and $S$ is the \emph{method
  body}, which may include recursive calls of $m$.

Methods are invoked by means of the \emph{parameterized method calls}
that are of the form
\[
  s.m(\mathbf{t}).
\]
where $s$ is an expression that denotes the \emph{called object} and
$\mathbf{t}$ is the list of actual parameters of the method~$m$.

A program consists of a main statement and a set of method definitions.
In programs we use a basic (i.e., not compound) type $\object$
that denotes an infinite set of objects.  The constant $\nulll$ of
type $\object$ represents the \emph{void reference}, a special
construct which does not have a local state.  The normal variable
$\this$ of type $\object$ stores at any moment the currently executing
object.  Inside a method body no assignments to the variable
$\this$ are allowed, that is, this variable is read-only.  Values of type
$\object$ can only be tested for \emph{equality}.  
Variables of type $\object$ are called \emph{object variables}, and 
expressions of type $\object$ are called \emph{object expressions}.

\begin{figure}
    \centering
   \includegraphics[scale=0.4,height=0.13\textheight]{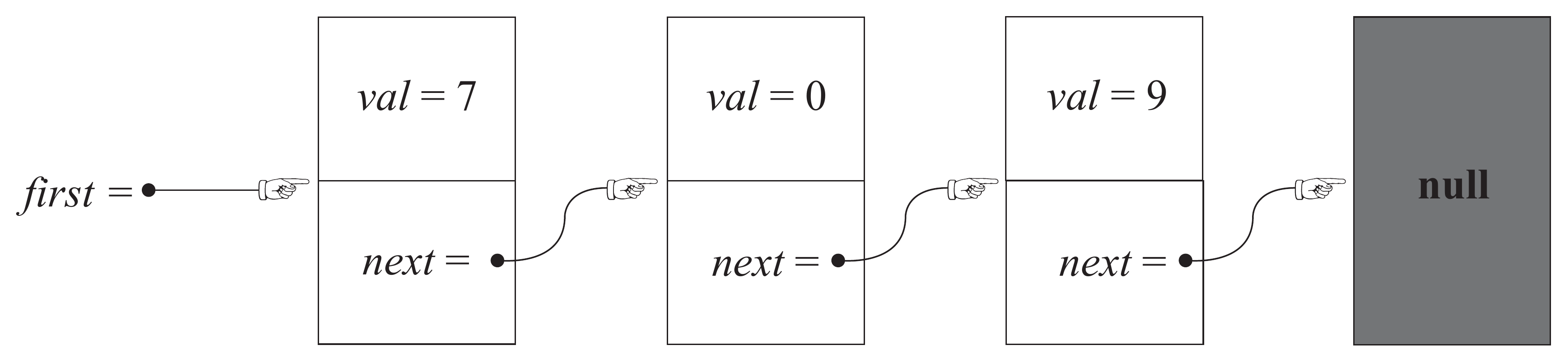}

\caption{\label{fig:list} A linked list. This drawing is taken from \cite[p.~190]{ABO09}.}  
\end{figure}

To illustrate the considered programs consider a recursive method used
to find a zero in a linked list represented by objects.  We represent
such lists using the instance object variable $next$ that links the
objects of the list, and the constant $\nulll$ that allows us to
identify the last element of the list. We assume that each object
stores a value kept in an instance integer variable
$val$. Additionally, we use the normal object variable $\mathit{first}$ to
point to the first object in the list. Figure \ref{fig:list} shows an
example of such a list representation.

The desired method is declared as follows.

\begin{tabbing}
\qquad\qquad 
$\mathit{find}$ :: \= \textbf{if} $val=0$\\
\> \textbf{then} $return:=\this$\\
\> \textbf{else} \= \IF $next\not=\nulll$\\
\>       \> \textbf{then} $next.\mathit{find}$\\
\>       \> \textbf{else} $return:=\nulll$\\
\>       \> \textbf{fi}\\
\> \textbf{fi}
\end{tabbing}

Then upon termination of the call $\this.\mathit{find}$ the object variable $return$
points to the first object in the list that stores zero, if it exists, and otherwise
it returns the void reference, represented by the constant $\nulll$.

\subsection{Reasoning about object-oriented programs}

The methods in object-oriented programs use local updates to
manipulate a global, dynamically changing pointer structure, the
\emph{heap}.  Therefore the assertion language must be able to express
properties of the heap.  To this end, \emph{global expressions} are
introduced by extending the (local) expressions that may appear inside
methods by \emph{navigation expressions} of the form $e.u$, where $e$
is an object expression and $u$ an instance variable.  If $e$ denotes
a certain object (for example \textbf{self}) and $u$ is an instance
variable (for example \emph{next}), then $e.u$ points to the
corresponding object (in the example \textbf{self}.\emph{next}).  This way,
one can navigate from object to object along the pointers in the heap.
Assertions are then constructed from global Boolean expressions by
allowing Boolean combinations and quantification over normal variables
in $\mathit{Var}$.

As an example consider the assertion used in \cite[pp.~226-227]{ABO09} to reason about the above method \emph{find}:
\[
P \equiv \this=a[k]\wedge a[n]=return \wedge \forall\, i\in [k:n-1]: (a[i]\not=\nulll\wedge a[i].val\not=0\wedge a[i].next=a[i+1]).
\]
In addition to the instance object variables $next$ and $return$ and
the instance integer variable $val$ used in the definition of
\emph{find}, one uses here the following normal variables: an array
$a$ of type ${\bf integer}\ra \object$, integer variables $i, k$ and
$n$, and the object variable $\this$.

The assertion $P$ states that the array section $a[k:n]$ stores a linked
list of objects which starts with the object $\this$, ends with the object $return$,
and all of its objects, except possibly the last one, are different
from $\nulll$ and do not store in $val$ zero.
Note the use of the navigation expressions $a[i].val$ and $a[i].next$.

The desired behaviour of the above method \emph{find} can then be
specified by means of the following correctness formula:
\begin{equation}
  \label{equ:oo}
\HT{\mathbf{true}}{\this.\mathit{find}}{Q},  
\end{equation}
where the postcondition $Q$ is defined in terms of the assertion $P$:
\[
Q \equiv (return=\nulll\vee return.val=0)
\wedge
\exists\ a: \exists\ k:\exists\ n\geq k:P.
\]
So the postcondition states that the returned object is $\nulll$ or stores zero and that
for some array section $a[k:n]$ the assertion $P$ holds.

To reason about such correctness formulas we need new axioms and proof rules.

Correctness of the customary assignment $x:=t$ to normal variables $x$
is captured by the ASSIGNMENT axiom that for a given postcondition $P$
calculates the precondition by applying the substitution $[x:=t]$ to $P$.
In object-oriented programs one can additionally use assignments
to instance variables and use the dereferencing $s.m(\mathbf{t})$ in
the method calls. The latter calls for an extension of the assertion
language with the corresponding expressions $e.u$.

F.S.~de Boer proposed in \cite{Boe99} the following textually
identical axiom for such assignments:
\III

\NI
ASSIGNMENT TO INSTANCE VARIABLES 
\[ 
 \HT{P[u:=t]}{u:=t}{P}, 
\]
where $u$ is a (possibly subscripted) instance variable in $\mathit{IVar}$.
\III

So, as in the case of the assignment to subscripted variables, the solution relies on
an extension of the definition of substitution, in this case to the instance variables.

As usual, the definition of the substitution $P[u:=t]$ proceeds 
by induction on the structure of $P$. 
The difference appears at the level of expressions $s$ for which
the definition of $s[u:=t]$ is more elaborate.
If $s$ is of the form $e.u$ for an object expression $e$,
the substitution has to take care of possible aliases of $e.u$.
More precisely, if after applying the substitution inductively to $e$, 
the result refers to the currently active object, i.e., if $e[u:=t] = \this$, 
then the outcome of the substitution is $t$. Otherwise, $u$ is left untouched
and the substitution is applied inductively to $e$.
This is expressed in the following definition:
\[
 e.u[u:=t]\equiv \ITE{e[u:=t]=\this}{t}{e[u:=t].u}.
\]

In case of only one (possibly recursive) method definition, 
the RECURSION rule for procedure calls can be adapted to method calls as follows,
where we use the block statement discussed in Subsection \ref{subsec:local}:
\III

\NI
RECURSION IV
\[ 
  \frac{\HT{P}{s.m(\mathbf{t})}{Q} \vdash 
         \HT{P}{\block{\mathbf{var} \; \this,\mathbf{u} :=s,\mathbf{t};\ S}}{Q}}
        { \HT{P}{s.m(\mathbf{t})}{Q}                          }
\]
where the method $m$ is declared by $m(\mathbf{u})::S$.
\III

So, as for recursive procedures we may use the desired conclusion as a
hypothesis in the correctness proof of a block statement, where
$\this$ and the list $\mathbf{u}$ of formal parameters of the method
are treated as local variables that are respectively initialised by
the called object $s$ and the list $\mathbf{t}$ of actual parameters
and are accessed in the method body $S$.  A rule analogous to the
RECURSION II rule is adopted to deal with total correctness.

To adjust correctness formulas that deal with generic method calls
$y.m(\mathbf{x)}$ to specific objects $s$ and lists of actual
parameters $\mathbf{t}$, we modify the SUBSTITUTION rule as follows,
where we refer to the given set $D$ of method definitions:
\III

\NI
INSTANTIATION 
\[
 \frac{\HT{P}{y.m(\mathbf{x})}{Q}}
  {\HT{P[y,\mathbf{x}:=s,\mathbf{t}]}{s.m(\mathbf{t})}{Q[y,\mathbf{x}:=s,\mathbf{t}]}}
\]
where $y,\mathbf{x}$ is a list of
variables in $\mathit{Var}$ which do not appear in $D$ and
$var(s,\mathbf{t}) \cap change (D) = \ES$.
\III

In \cite{ABO09} these axioms and proof rules were used to establish
total correctness of the above example program expressed by the
correctness formula (\ref{equ:oo}) and of an object-oriented program
that inserts an element into a linked list.

\subsection{Advanced topics in the verification of object-oriented programs}

Various other aspects of object-oriented programming were studied
from the correctness point of view.
One of them is object creation.
An object can be dynamically created by the assignment statement
$u:=\new$, where $u$ is an object variable and $\new$ is a keyword in
the considered programming language.  The execution of this statement
creates a {\em new} object and assigns its identity to the variable
$u$.  This new object comes with a default initialisation of all its
instance variables.  This can be modelled by some bookkeeping of the
set of objects that are currently created, for example by maintaining
a counter referring to an unbounded array.  This allows one to reason
about this assignment using a small program transformation.  A
drawback of this approach is that it refers to an explicit
implementation.

An alternative is to use a substitution of $[x:=\new]$ for object
variables and define its application to a limited class of assertions (we call them
\emph{pure}) that take into account that object variables can only be
compared for equality or be dereferenced in the method calls
$s.m(\mathbf{t})$, and in which one does not quantify over such object variables.
This leads to the following axiom proposed in \cite{Boe99}:

\III

\NI
OBJECT CREATION 
\[
\HT{P[x:=\new]}{x:=\new}{P},
\]
where $x$ is a simple (so not subscripted) object variable and $P$ is a pure assertion.
\III

We omit the details of the definition of the substitution $P[x:=\new]$ and only remark
that one cannot simply replace $x$ in $P$ by the keyword $\new$
because it is {\em not} an expression of the assertion language.
The details are given in \cite[Chapter 6]{ABO09}.  

C.~Pierik and F.S.~de Boer showed in \cite{PB03} how the approach of
\cite{Boe99} can be extended to deal with classes, inheritance and subtyping.
The same authors introduced in \cite{BP03} a general methodology for
obtaining relatively complete Hoare's logics for object-oriented
programs.  A key issue is the extension of Gorelick's most general
formulas \cite{Gor75} (see
Subsection~\ref{subsec:complete-recur-proc}) to deal with the states
of object-oriented programs.

As another contribution to this line of research on verification of
object-oriented programs let us mention \cite{AbrahamBRS05} in which a
Hoare-like proof system for partial correctness and deadlock freedom was
developed for a subset of Java.  The considered subset comprised the
object-oriented core of Java, as well as concurrency via thread
classes, allowing for a multithreaded flow of control.  The Java
concurrency model includes synchronous message passing, dynamic thread
creation, shared-variable concurrency via instance variables, and
coordination via reentrant synchronization monitors.  The verification
method was formulated in terms of proof outlines that were tested both
for interference freedom of shared-variable concurrency, as in
\cite{OG76a}, and for cooperation of synchronous message passing, as in
\cite{AFR80}.  The authors established the soundness and relative
completeness of their proof system.  From an annotated program, a
number of verification conditions were generated and discharged using
the interactive theorem prover PVS \cite{PVS92}.

The verification of Java programs was also in the focus of tool-supported
research projects around the year 2000. The LOOP project led by
B.~Jacobs considered sequential Java programs with specifications
written in the Java Modeling Language (JML)~\cite{JacobsP03}.  JML was
developed by a group led by G.T.~Leavens~\cite{LeavensCCRC05}.  It was
inspired by the idea of \emph{Design-by-Contract} introduced by
B.~Meyer in the context of the object-oriented programming language
Eiffel~\cite{Mey97}.  A contract is a specification in the form of
assertions. The contract is agreed upon before an implementation is
developed that should satisfy this contract.  In its simplest form, a
JML specification of a method comprises a precondition, one
postcondition for normal termination and second one for exceptional
termination, and a set of variables that may be changed.  The LOOP
compiler takes as input a Java program and a JML specification and
outputs files in the syntax of the PVS theorem prover that describe
the semantics of the program and proof obligations generated from the
specification. These are then proven interactively using PVS.

For a subset of sequential Java programs, called NanoJava, a Hoare's
logic was proven sound and relatively complete w.r.t. an operational
semantics of the programs in the interactive theorem prover
Isabelle/HOL \cite{OheimbN02}.  In this approach, as in the paper
  \cite{Ohe99} cited in Subsection \ref{subsec:complete-recur-proc},
a semantic view of assertions was taken, whereby an assertion is a
function from states to Boolean values.  No separate syntactic
representation of assertion was introduced. Thus, subtleties like the
right definition of syntactic substitutions were not considered.

For the object-oriented programming language Eiffel a sound and
relatively complete Hoare's logic was established in~\cite{NordioCMM09}.
The emphasis was on the treatment of features in Eiffel that are not
present in other object-oriented languages. These concern the details
of the exception handling, so-called \emph{once routines}, and
multiple inheritance.

More recently B. Engelmann \cite{EngelmannO16,Engelmann17} considered
object-oriented programs in the context of \emph{dynamic typing}. 
This means that variables do not have an a priori declared
static type but may assume any value during the execution of the program.
As an example consider the following method from \cite{Engelmann17},
where it is assumed that $b$ is a Boolean variable:

\begin{tabbing}
\qquad\qquad 
$num\_or\_string(b) ::$  \textbf{if} $b$ \textbf{then} $x:=y:=5$
                         \textbf{else} $x := \texttt{``foo''};\, y:= \texttt{``bar''}$ 
                         \textbf{fi};
                          $z:= x + y$
\end{tabbing}

\NI
If $b$ is true, the method call $num\_or\_string(b)$ yields the
numeric value $z=10$, but if $b$ is false, it yields the string value
$z = \texttt{``foobar''}$.  So the type of $x$ and $y$ is determined
dynamically during runtime, and it is either numeric or string, with
the operation + being either addition or string concatenation.  Such
dynamically typed programs are present in the list processing language
LISP and in widespread programming languages like Python and
JavaScript.

Engelmann developed in \cite{Engelmann17} a Hoare-like proof system for a model
language \textbf{dyn} of object-oriented programs with dynamic typing. He
proved soundness and relative completeness of his system, thereby
extending the arguments of \cite{Coo78,Gor75} and \cite{BP03}
to the object-oriented part.

\section{Alternative Approaches}
\label{sec:alternative}

\subsection{Weakest precondition semantics and systematic program development}
\label{subsec:wp}

Dijkstra suggested an alternative approach to program
verification, called \emph{weakest precondition semantics}~ \cite{Dij75}. The idea
is that given a program $S$ and a desired postcondition $P$ we would like
to find the weakest precondition $wp(S,Q)$ such that
$\HT{wp(S,Q)}{S}{Q}$ holds in the sense of total
correctness. `Weakest' means here that for any precondition $P$ such
that $\HT{P}{S}{Q}$ holds in the sense of total correctness, the
implication $P \to wp(S,Q)$ holds.

This approach differs from Hoare's original approach by

\begin{itemize}
\item insisting on total correctness instead of on partial correctness,
  
\item assuming that initially only the postcondition is given.
\end{itemize}

Additionally, Dijkstra insisted that the program should be developed
\emph{together} with its correctness proof. In \cite{Dij75} he
advanced this approach for his guarded command language that we
briefly discussed in Subsection \ref{subsec:nondeterminism}. Its
notable feature was the use of programming constructs that support
nondeterminism. Another interesting feature of the language was
\emph{parallel assignment} $\mathbf{x}:=\mathbf{t}$, where
$\mathbf{x}$ is a list of different variables and $\mathbf{t}$ is a
list of expressions of the same length. This construct is for example
useful to swap the values of variables without additional variables:
\[
  x, y := y,x.
\]

The weakest precondition $wp(S,Q)$ is defined by induction on the structure
of the program, with
\begin{itemize}
\item $wp(\mathbf{x}:=\mathbf{t},Q) \equiv Q[\mathbf{x}:=\mathbf{t}]$,
\item $wp(S_1;\ S_2,Q) \equiv wp(S_1,wp(S_2,Q))$,
\end{itemize}
as typical clauses.

The main problem is how to deal with loops.  In \cite{Dij75} the
weakest precondition for guarded commands was defined as follows.
Denote the alternative command $\IFP$ by $\mathit{IF}$, the
repetitive command $\DOP$ by $\mathit{DO}$ and abbreviate
$\bigvee_{i=1}^{n} B_i$ to $\mathit{BB}$.  Then
\[
  wp(\mathit{DO}, Q) \equiv \te k: k \geq 0: H_k(Q),
\]
where
\[
  H_0(Q) \equiv (Q \land \neg BB)
\]
and for $k> 0$
\[
H_k(Q) \equiv wp(\mathit{IF}, H_{k-1}(Q)) \lor H_0(Q).
\]
Intuitively,
$H_k(Q)$ is the weakest precondition guaranteeing proper termination
in a state satisfying $Q$, after at most $k$ guard selections.

Since $k$ is used as a subscript of $H_k$, `$\te k: k \geq 0$' is here
not a customary quantification but a shorthand for an infinite
disjunction, with additionally the formulas $H_k(Q)$ defined by
induction. In other words, so defined weakest precondition of a
repetitive command is not an assertion in a first-order language.
The same is the case for the weakest liberal
preconditions mentioned in Subsection~\ref{subsec:completeness}, that
can be defined for \textbf{while} programs in an analogous way.

On the other hand, if one confines one's attention to the language of
Peano arithmetic, then the weakest (liberal) preconditions can be
encoded as first-order formulas, see for example \cite[Chapter
7]{Win93}.  However, the resulting formulas are not natural and using
them to express loop invariants (recall from
Subsection~\ref{subsec:completeness} that weakest liberal
preconditions can be used in a simple way to express loop invariants)
is an overkill.  This underlies the difficulty of finding simple loop
invariants, the problem we already mentioned in
Subsection~\ref{subsec:hoare1}, and clarifies why finding them is an
important problem in developing correct programs or in establishing
their correctness.

These complications disappear if one identifies assertions with the
sets of states that satisfy them and views the weakest precondition as
a way of assigning inductively semantics to programs. The above
clauses defining $wp$ show then how a program can be viewed as a
\emph{predicate transformer}, a concept further discussed in
Subsection~\ref{subsec:specifications}.  So the weakest precondition
approach can be viewed both as an approach to program verification and
as a way of defining program semantics.

The concept of weakest precondition was used a number of times in this
survey, which shows how this notion permeated research on Hoare's
logic and program verification.
The idea of using it to develop programs together with their correctness proofs
was subsequently presented in book form in \cite{Dij76}, where the
weakest precondition semantics was extended to blocks and
procedures. A systematic development of provably correct programs was
further advanced by David Gries in his book \cite{Gri81}, notably by
proposing a number of heuristics for finding loop invariants.  
Other books devoted to this subject are \cite{Bac86,DF88,Kal90,BW08}.
Various aspects of the weakest precondition semantics were further
discussed in book form in \cite{DS90}.  The problem of finding loop
invariants became central in the subsequent study of program
verification and development.  \cite{FMV14} surveyed various ways of
constructing loop invariants, and provides their classification and
analysis.

Another matter relevant for a systematic program development is
termination.  Both the WHILE II and RECURSION II rules provide only a
kind of template for an actual termination proof, without explaining
how the termination functions are to be found. Some sophisticated
techniques were developed to establish termination. They go beyond the
framework of Hoare's logic or the weakest precondition semantics and
rely on various methods developed in other areas, notably term
rewriting systems. The authors of \cite{CPR11} provided an accessible
account of the recent developments.  In particular, they explain a
recent alternative approach to proving termination called
\emph{disjunctive termination argument}: only one of a disjunction of
the termination functions needs to decrease, but this has to be the
case after any number of iterations of the loop. It is argued that
disjunctive termination arguments are easier to find than the
classical termination argument dating back to Turing \cite{Tur49} (see
Subsection~\ref{subsec:Turing}), where a single termination function has to
decrease its values taken from a well-founded set with each iteration
of the loop.

One difficulty of this approach for program verification is that it is
not clear how to extend the weakest precondition semantics to deal
with advanced programming constructs, for instance arbitrary
procedures as considered in Section \ref{sec:procedures}.  In
contrast, Hoare's approach is more flexible. In particular, as we have
seen in Subsection \ref{subsec:adaptation}, the reasoning can be
supported by various adaptation rules. This cannot be done in the
framework of the weakest precondition semantics which requires
computing a single assertion.  On the other hand, as we saw in
Subsection \ref{subsec:prob}, the weakest precondition semantics
turned out to be helpful for the verification of probabilistic
programs.

\subsection{Specifying in Hoare's logic}

Until now we used Hoare's logic exclusively to reason about 
programs.  But the formalism of correctness formulas can be also used
to specify programs, though some care needs to be exercised.
For example, the correctness formula
$\HT{\mathbf{true}}{S}{z=x+y}$ does not properly specify that a still to
be constructed program $S$ computes in $z$ the sum of $x$ and $y$ because also
$S \equiv x:=0; y:=0; z:=0$ satisfies this correctness formula.

To overcome this problem, one specifies the set $var(S)$ of variables
that $S$ may access. Then every variable outside $var(S)$ is known
to keep its value during the computation of $S$. These variables
can be used to freeze values of the initial state to use them
in the postcondition evaluated in the final state.
For example, by postulating $var(S) = \{x,y,z\}$, the correctness formula
\[
  \HT{x=x_0 \land y=y_0}{S}{x=x_0 \land y=y_0 \land z=x_0 + y_0}
\]
specifies that $S$ computes in $z$ the sum of $x$ and $y$ while
leaving $x$ and $y$ unchanged.
This specification technique is also at the heart of the most general
formulas introduced by Gorelick in his completeness proof
for recursive procedures~\cite{Gor75} and discussed in
Subsection~\ref{subsec:complete-recur-proc}.

In a similar vein the adaptation rules keep track of the set $var(S)$
of the variables accessed by a program~$S$, sometimes 
refined into the variables that are changed and those that are only
read by $S$.

To support the specification of programs, variants of Hoare's logic
have been proposed that introduce distinguished variables that may
never be accessed by programs. By different authors, these variables
are called logical variables (in \cite{OheimbN02}), 
logical constants (in \cite{Mor94},
further discussed in the next subsection), specification-only
variables, or auxiliary variables (in \cite{Kle99}).  The last name,
however, is in conflict with the concept of auxiliary variable
introduced by Owicki and Gries in the context of parallel
programs~\cite{OG76a} and discussed in
Subsection~\ref{subsec:parallel}.

T.~Kleymann developed in \cite{Kle99} a variant of Hoare's logic where
such explicit logical variables appear in the assertions and a new
consequence rule is proposed that takes care of these variables. As a
result, he showed that this consequence rule is powerful enough
to achieve adaptation completeness (a notion discussed in Subsection
  \ref{subsec:adaptation}) without further adaptation rules.

We mentioned auxiliary variables first in the context of the
Owicki-Gries method, when discussing verification of parallel programs
in Subsection~\ref{subsec:parallel}.  The need for auxiliary variables
in correctness proofs was already observed in \cite{Cli73}.  In
\cite{ABO09} it was noted that the AUXILIARY VARIABLES rule is already
needed to reason about disjoint parallel programs.  Indeed, the
correctness formula
\[ 
\HT{x=y}{[x:=x+1 \| y:=y+1]}{x=y} 
\]
cannot be proved using the DISJOINT PARALLELISM rule and the rules of
the proof system ${\cal H}$.

The most extensive analysis of these matters was provided in
\cite{GR16} that clarified and extended initial results of
\cite{Kle98} and \cite{Kle99}.  The authors of \cite{GR16} showed that
the AUXILIARY VARIABLES rule is a \emph{derived} rule both in the
proof system ${\cal H}$ and in a proof system for parameterless
recursive procedures, which means that it can be eliminated from any
proof that uses it.  In the case of disjoint parallel programs the
authors showed that this rule can be replaced by the simpler
$\te$-INTRODUCTION rule discussed in Subsection
\ref{subsec:adaptation}.

They also showed that for parallel programs with shared variables the
AUXILIARY VARIABLES rule turns out to be essential. On the other hand,
no other proof rules are needed.  Indeed, Owicki proved in
\cite{Owi76c} that the above presented proof system for parallel
programs is complete in the sense of Cook.

\subsection{Programming from specifications}
\label{subsec:specifications}

R.-J.~Back \cite{Back80} and C.~Morgan \cite{Mor94}
extended Dijkstra's approach to a methodology for ``programming from specifications''
by a rule-based step-by-step development of specifications into programs.
To this end, they introduced a \emph{specification} statement. 
In \cite{Mor94} it takes the form $\mathbf{x}:[P,Q]$, with the meaning
\begin{quote}
``\emph{If} the initial state satisfies the precondition $P$ 
\emph{then} change only the variables
listed in $\mathbf{x}$ so that the resulting final state
satisfies the postcondition $Q$.''
\end{quote}
While the variables that are changed by a specification
$S \equiv \mathbf{x}:[P,Q]$ are explicitly mentioned, namely
$change(S) = \{\mathbf{x}\}$, there is no information which variables
are accessed, i.e., $var(S)$ is not defined.  Recall that various
Hoare-style proof rules for programs $S$ have application conditions
concerning $var(S)$. Also, Gorelicks's most general formulas require
fresh variables outside of $var(S)$ to freeze their initial values.
To overcome this weakness, Morgan considers what he calls
\emph{logical constants} (occasionally also called
\emph{specification-only variables}), that by definition appear only in
assertions and are thus never changed by any program or
specification. With a logical constant $X$ the initial values of a
variable $x$ can be frozen. For example, $x:[x=X, x>X]$ specifies that
$x$ should be increased. The amount of the increment is left
unspecified.

The idea is that specifications and programs are handled on the same footing,
so that programming operators like sequential composition or loops can
be applied to specifications, as well. Constructs $S_1$ and $S_2$ in
this extended syntax can be compared by a \emph{refinement relation}:
$S_1 \sqsubseteq S_2$ denotes that $S_1$ is refined by $S_2$.
Semantically, specifications and programs are considered as 
\emph{predicate transformers} that transform given postconditions
into the corresponding weakest preconditions. 
Refinement $S_1 \sqsubseteq S_2$ means that for all postconditions $Q$
the implication
\[
  wp(S_1,Q) \ra wp(S_2,Q)
\]
holds, i.e., $S_2$ establishes the postcondition $Q$ in at least all states
where $S_1$ establishes $Q$.
For example, $x:[x=X, x>X] \sqsubseteq x:=x+42$.
In this approach specifications are stepwise \emph{refined} to programs by 
the applications of refinement rules, described in \cite{Mor94}.

Morgan \cite{Mor94} also considers a number of advanced programming
concepts like recursive procedures with parameters, modules with local
declarations of variables and procedures, and data refinement.

Subtly different from Morgan's specification statement is the 
\emph{generic command} of J.~Schwarz \cite{Sch77}. 
It is written as $[P \Rightarrow Q][X]$,
where $P$ is a precondition, $Q$ a postcondition, and $X$ a set of variables. 
Semantically, a generic command denotes a certain state transformer.
Assume an interpretation $I$. A \emph{state transformer $T$ based on}
a finite set $X$ of variables is a binary relation on the set $\Sigma$
of states that has only read or write access to the states via the
variables in the set $X$.  For each program $S$ its meaning is a state
transformer based on $var(S)$.

Let $\mathcal{T}(X)$ be the set of all state transformers based on
$X$.  State transformers are ordered by the set inclusion:
$T_1 \subseteq T_2$ means that $T_1$ produces less output than
$T_2$. The least element is the empty state transformer $\emptyset$
producing no output, corresponding to a program that never
terminates. Now, the semantics $\mathcal{M}_I$ relative to $I$ of a
generic command $[P \Rightarrow Q][X]$ is given by
\[
  \MSI{[P \Rightarrow Q][X]} = \bigcup \{S \mid S \in \mathcal{T}(X) \text{ and $\HT{P}{S}{Q}$ is true in $I$ in the sense of partial correctness} \}.
\]

So given an interpretation $I$, $[P \Rightarrow Q][X]$ denotes the
largest state transformer $S$ based on $X$ that satisfies
$\HT{P}{S}{Q}$ in the sense of partial correctness.  `Largest' refers
here to the ordering of state transformers by the set inclusion.  In
\cite{Old83} it was shown that for $\{\mathbf{x}\} = X$ the formula
$W$ of the ADAPTATION II rule, introduced in Subsection~\ref{subsec:adaptation},
expresses the weakest liberal precondition
$wlp_I([P \Rightarrow Q][X],R)$ of the generic command
$[P \Rightarrow Q][X]$ w.r.t. a given postcondition $R$.  This
connects generic commands via the weakest liberal preconditions to the
ADAPTATION II rule.

Schwarz used generic commands for stating an alternative version of
a RECURSION rule, but did not embark on program development.
Morgan's specification statement $\mathbf{x}:[P,Q]$ is similar to the generic command
$[P \Rightarrow Q][X]$, except that
the list $\mathbf{x}$ does not record the variables that are only read, which are
also covered by $X$.

\subsection{Algorithmic logic and dynamic logic}

Hoare's logic is geared towards establishing program
correctness. However, from the point of view of mathematical logic it
has a very rigid syntax: the correctness formulas cannot be negated or
combined, for example by disjunction.  As a result one cannot view
Hoare's logic as an extension or a modification of some existing
logics, even though it crucially relies on first-order logic and its
extensions, for example by allowing subscripted variables.  In some
alternative approaches one could view reasoning about programs as an
extension of reasoning within existing logics.  We discuss now briefly
two most prominent examples.

\emph{Algorithmic logic} was originally proposed in \cite{Sal70} and
presented in book form in \cite{MS87}.  It extends first-order
language by expressions that can be interpreted as programs and
constructs that allow one to mix formulas and programs.
Interestingly, substitutions for a sequence of variables are viewed
as atomic programs. This is equivalent to the aforementioned 
parallel assignment of Dijkstra, and also shows a close connection
with the ASSIGNMENT axiom.  Programs are built by allowing formulas as
tests and using program composition, conditionals, and loops, all
written in a compact notation. For example,
$\mathbf{if} \ B \ \mathbf{then} \ S \ \mathbf{else} \ T$ is written
as \underline{v}$[BST]$. In turn, the construct $*[BST]$ corresponds
to $\mathbf{while} \ B \ \mathbf{do} \ S \ \mathbf{od}; T$.

Expressions of the form $S \phi$, where $S$ is a program and $\phi$ a
formula, correspond to the strongest postcondition introduced in
Subsection \ref{subsec:completeness}.  Further, it is shown how
termination of the considered programs can be expressed as a formula
that admits countable disjunction.  This is analogous to the
definition of $wp(\mathit{DO}, Q)$ given earlier.

The relation between Hoare's logic and algorithmic logic becomes clear
when one realizes that the correctness formula $\HT{P}{S}{Q}$ can be
expressed as the implication $S P \to Q$.  Consequently, rules used in
Hoare's logic can be readily reproduced as rules in algorithmic logic,
in particular various forms of adaptation rules. This straightforward
modelling, however, does not yield new insights concerning program
verification.

Research on algorithmic logic focused mostly on such matters as studies
of consistency and the infinitary completeness of selected theories,
derivation of the normal forms of programs, and axiomatization of
various data structures, rather than on (relatively) complete
axiomatizations of fragments concerned with specific features of
programming languages, a direction Hoare's logic took.

\emph{Dynamic logic} was originally proposed in \cite{Pra76} and
presented in book form in \cite{Har79} and more extensively in
\cite{HKT00}.  It is very similar to algorithmic logic introduced six
years earlier, though it was developed independently.  It enriches
first-order logic by constructs reminiscent of modal logic. First,
programs are defined starting from atomic actions and tests, using the
sequential composition ($;$), nondeterministic composition ($\cup$)
(absent in algorithmic logic), and iteration ($*$) that corresponds to
Kleene's star. In the dynamic logic syntax the $\WDD{B}{S}$ statement
can be expressed as $(B;S)^* \cup \neg B$.

Further, one admits formulas of the form $[S] \phi$, where
$S$ is a program and $\phi$ is a formula, with the intended
interpretation ``every execution of the program $S$ from the
current state leads to a state in which $\phi$ is true''.  The
formulas and programs are defined by simultaneous induction, allowing
the usual Boolean connectives. A dual formula to $[S] \phi$ is
$\langle S \rangle \phi$, defined by:
\[
  \langle S \rangle \phi \equiv \neg [S] \neg \phi.
\]
So its intended interpretation is ``some execution of the program
$S$ from the current state leads to a state in which $\phi$ is
true''.  The $[S]$ and $\langle S \rangle$ operators can
thus be viewed as the counterparts of the $\LTLsquare$ and $\LTLdiamond$
operators in propositional modal logic, but
parameterized with a program $S$.

Research on dynamic logic mainly focused on a study of various
fragments or extensions, with the corresponding sound and complete
axiomatizations, and the corresponding decidability and computational
complexity results.  In particular, dynamic logic was extended in
\cite{Har79} to deal with recursive procedures. 

Typical axioms are:
\[
  [S; T] \phi \lra [S] [ T] \phi,
\]
which corresponds to one of the mentioned clauses that define the weakest precondition,
and
\[
  [S^*] \phi \lra \phi \land [S][S^*] \phi,
\]
that captures the idea that $^*$ stands for the infinite iteration.

The relation between Hoare's logic and dynamic logic is easily
established by noticing that the correctness formula $\HT{P}{S}{Q}$
can be expressed as the implication $P \to [S]Q$.  So $[S]Q$ models
what is called the weakest liberal precondition introduced in
Subsection \ref{subsec:completeness}.  Consequently, as in the case of
algorithmic logic, proof rules of the proof system ${\cal H}$ of
Subsection \ref{subsec:hoare1} can be translated into proof rules of
dynamic logic.  Moreover, these translated rules are derivable from
the adopted axioms and proof rules of dynamic logic.

Further, thanks to the richer syntax,
it is possible to express other program properties and discuss such
properties like program equivalence.  For example, the following
formula states that the program $S$ is deterministic:
\[
\langle S \rangle \mathbf{true} \ra  [S] \mathbf{true}.
\]

Dynamic logic was generalized in a number of ways, for example to
\emph{epistemic dynamic logic} (see, e.g., \cite{BR16}), to allow
reasoning about modalities and change.  Also it was extended to
specify and reason about \emph{hybrid systems}, i.e., systems where a
discrete control interacts with a continuous dynamics
\cite{Platzer08}. Further, the KeY system (discussed Section
\ref{sec:concluding}) is based on dynamic logic.

\subsection{Temporal logic and model checking}

Research on verification of parallel programs carried out in the
seventies showed limitations of Hoare's logic in reasoning about
concurrent programs.  We discussed three properties of such programs:
partial correctness, termination, and absence of deadlock. However, in
contrast to the sequential programs, concurrent programs are often
supposed to operate repeatedly, in a cyclic fashion. For instance, a
solution to the mutual exclusion problem deals with infinite
executions of the program components operating in parallel.  This
calls for a study of properties (such as an \emph{eventual access})
that cannot be expressed in Hoare's logic.

To express such concepts and to systematically reason about them
A.~Pnueli proposed in \cite{Pnu77} to use \emph{temporal logic}.
Using it one can express in a natural way various program properties
that do not necessarily deal with the input/output behaviour of a
program.  In contrast to Hoare's logic the reasoning about programs is
not syntax-directed. Instead, one usually reasons about specific
control points in a program and a relation between them.

For example, the following formula states that a process $P$
infinitely often enters its critical section \emph{CS}:
\[
  \LTLsquare \LTLdiamond \: in \, \emph{CS}
\]
where $in \, \emph{CS}$ is a formula that states that
the control in the process $P$ is within \emph{CS}.

In turn, to express the strong fairness assumption for the repetitive
command
\[
S \equiv \DOP
\]
we can use the following formula:
\[
\bigwedge_{i=1}^{n} (\LTLsquare \LTLdiamond \: (at \, S \land B_i) \to \LTLsquare \LTLdiamond \: at \, S_i),
\]
where $at \, T$ holds when the control in the considered program is just in front of $T$.

Appropriate axioms and proof rules were then developed to reason about such formulas.
Temporal logic, applied to concurrent programs, grew into an impressive
research area, see in particular the books \cite{MP91,MP95} of Z.~Manna
and A.~Pnueli. This line of research should be viewed as complementary to
Hoare's logic, that is why we do not devote more space to it.

In that context a distinction, first advanced in \cite{Lam77}, is
useful. A \emph{safety property} states `nothing bad will happen'
during a program execution, while a \emph{liveness property} states
that eventually `something good will happen' during a program
execution. According to this distinction partial correctness, absence
of errors, and deadlock freedom are safety properties. In contrast,
program termination, fairness, eventual access, infinite access,
etc.~are liveness properties. In temporal logic these properties can
be formulated by means of invariants that become formulas of the form
$\LTLsquare \phi$, while liveness properties are formulas of the form
$\LTLdiamond \phi$.  Hoare's logic is a convenient vehicle to prove
safety properties.  Its modification to deal with termination is also
quite natural.  As we saw in Section \ref{subsec:fairness} one can
also investigate within this logic fairness of nondeterministic
programs. In fact, even more advanced liveness properties, such as
fairness of parallel programs and eventual access properties can be
studied, as well (see \cite{OA88}). However, the treatment becomes
awkward, as one has to reason then about appropriately transformed
programs. This is in contrast to temporal logic that allows one to
reason about liveness properties in a simple way, without any need for
program transformations.

Temporal logic led in turn to \emph{model checking}.  Independently,
E.M.~Clarke and A.~Emerson \cite{EC82}, as well as J.P.~Queille and
J.~Sifakis \cite{QS81}, discovered that the problem of checking
whether a finite-state system satisfies (is a \emph{model} of) a
propositional temporal logic formula is decidable, providing efficient
algorithms for it.  This was the start of enormous research activities
extending the scope of model checking so that even industrial-size
problems could be tackled \cite{ClarkeGHJLMN93}.  Interestingly, model
checking is often used to debug a system because in case the system
does not satisfy the temporal logic specification, model checkers can
provide a counterexample that is helpful for understanding the
mismatch between a system and its specification.  In the recent years
model checking has been extended to infinite-state systems, mostly by
automatically constructing and refining abstractions of the system, a
method known as \emph{counterexample-guided abstraction refinement}, see
\cite{GEGAR-JACM-2003}.  The state of the art of model checking is
represented in the handbook \cite{ModelChecking2018}.

Another approach based on temporal logic is \emph{Temporal Logic of
Actions} (TLA) introduced by L.~Lamport in~\cite{Lamport94}.  It is
a method of specifying and reasoning about concurrent systems in
which systems and their properties are represented in the same
temporal logic. A system is specified by a temporal formula
representing a single loop that in each iteration nondeterministically
chooses one action for execution.  An action describes a basic step of
the considered system and is specified by a predicate using normal and
primed variables as we have seen it already in Turing's example in
Subsection~\ref{subsec:Turing}.  The standard form of a TLA
specification is
\[
 \mathit{Init} \land \LTLsquare\,[\mathit{Next}]_{\mathbf{x}} \land \mathit{Live},
\]
where $\mathit{Init}$ is the initial predicate, $\mathit{Next}$ is the
next-state relation expressed as a disjunction of actions,
$\mathbf{x}$ is the list of all variables occurring in these actions,
and $\mathit{Live}$ is a temporal formula that specifies a liveness
condition in terms of fairness assumptions.  To allow for refinement
of concurrent systems in the presence of different granularities of
the basic steps, \emph{stuttering steps} are considered.  To hide
parts of the state, quantification over variables is possible.  The
extension TLA+ comes with facilities to structure system
specifications into modules, see ~\cite{Lamport2002}.  Properties of a
subset of TLA+ specifications can be verified with the model checker
TLC discussed in~\cite{Lamport2002}.

\subsection{Separation logic}

This approach to program verification was originally developed as an
extension of Hoare's logic to reason locally about about pointer
structures, see \cite{OHearnRY01,Rey02,OHearnYR04,OHearn19}.  To cope with
pointers, separation logic builds upon a semantic model, in which a
state is a pair $(s,h)$ consisting of a \emph{store} $s$ and a
\emph{heap} $h$.  A store $s$ is a mapping from variables to values
(so a state in the sense of Subsection \ref{subsec:soundness}), which
may be data values such as integers, or pointer values such as
addresses.  A heap $h$ is a finite partial mapping from addresses (or
cells) to values, which again can be data or pointers. It is assumed
that the addresses are integers, which in turn are values.

Several low-level statements for manipulating the heap were considered
in~\cite{OHearnRY01,OHearn19}.  Let $x$ stand for a variable and $e$ for
an integer expression, which can denote an address in the heap.  One
can explicitly distinguish between an address and its contents: $[e]$
denotes the contents of the heap at address $e$.  \emph{Assignments}
affect only the store.  Besides the normal assignments $x:=e$ there
are \emph{lookups} $x:=[e]$, where $e$ is interpreted as an address in
the heap and the contents of $e$ is assigned to the variable $x$ in
the store.  \emph{Mutations} affect the heap.  An \emph{update}
$[e]:=e'$ expresses that the contents of the address $e$ in the heap
becomes value of the expression $e'$.  Further,
$x := \mathbf{alloc()}$ expresses that the address $x$ is newly
allocated to the heap, and $\mathbf{free}(x)$ expresses that the
address $x$ is deallocated from the heap.

Separation logic extends the usual assertion language of Hoare's logic
to specify properties of the heap: $\mathbf{emp}$ asserts that the
heap is empty, $e \mapsto e'$ asserts that the heap consists of one
cell, with address $e$ and contents $e'$, and $e \mapsto -$ asserts
that the heap consists of one cell, with address $e$ but unknown
content.  The main new operator in separation logic is the
\emph{separation conjunction}, written * and pronounced ``and
separately''.  The assertion $P*Q$ expresses that the heap can be
split into two disjoint parts in which $P$ and $Q$ hold, respectively.
Using separation conjunction, one can specify larger parts of the
heap.  For instance, the assertion $(x \mapsto 21) * (y \mapsto 42)$
describes two separate cells with addresses $x$ and $y$ and contents
21 and 42, respectively.  As abbreviation one uses
$e \mapsto e_1, \dots, e_n$ to stand for
$e \mapsto e_1 * e+1 \mapsto e_2 * \dots * e+n-1 \mapsto e_n$ and thus
asserting that the heap consists of $n$ adjacent cells with addresses
$e, \dots, e+n-1$ and contents $e_1, \dots, e_n$, respectively.  For
example, the assertion $(x \mapsto 21,y) * (y \mapsto 42,x)$ concisely
specifies a heap with a cyclic pointer structure, in which
$x \mapsto 21,y$ stands for $x \mapsto 21 * x+1 \mapsto y$ and
similarly with $y \mapsto 42,x$.  It consists of two
separate parts of the heap at the addresses $x$ and $y$ that contain
21 and 42, respectively, and a pointer to the other address.

For mutations and lookups the following axioms were stated in
\cite{Brookes07,OHearn19}:
\III

\NI
ALLOCATION
\[
 \HT{\mathbf{emp}}{x:=\mathbf{alloc()}}{x \mapsto -}       
\]
DE-ALLOCATION
\[
 \HT{x \mapsto -}{\mathbf{free}(x)}{\mathbf{emp}}       
\]
UPDATE
\[
 \HT{e \mapsto -}{[e] := e'}{e \mapsto e'}       
\]
LOOKUP
\[
 \HT{P[x:=e'] \land e \mapsto e'}{x := [e]}{P \land e \mapsto e'}       
\]
where $x$ does not occur in $e$ or $e'$.
\III

Separation conjunction enables the formulation of proof rules for local reasoning
about components of parallel programs. 
Crucial is the following rule, which is essentially the INVARIANCE rule
in a semantic setting of heaps:
\III

\NI
FRAME
\[ \frac{ \HT{P}{S}{Q}           }
        { \HT{P * R}{S}{Q * R} } 
\]
where  $\mathit{free}(R) \cap change(S)=\ES$.  
\III

In separation logic, this rule serves to extend a local specification
involving only the variables and parts 
of the heap that is used
by $S$ by adding assertions about variables and
other parts of the heap not modified by $S$.
Thus the FRAME rule is considered as the key to
local reasoning about the heap \cite{Rey02}.
\III

Separation logic was originally used for the verification of sequential
programs manipulationg pointer structures. Exploiting its ability to reason
explicitly about the heap, the approach was later extended in  \cite{OHearn07} to reason
in a modular way about concurrent programs.
The idea is that two threads that operate on disjoint parts of the heap do not 
interfere, and thus can be verified in isolation.
This is captured by the following rule, which is the 
DISJOINT PARALELISM rule in a semantic setting with heaps:
\III

\NI
CONCURRENCY
\[ \frac{ \HT{P_1}{S_1}{Q_2}, \ \HT{P_2}{S_2}{Q_2}           }
        { \HT{P_1 * P_2}{[S_1 \| S_2]}{Q_1 * Q_2} } 
\]
where $\mathit{free}(P_1,Q_1) \cap change(S_2) = \mathit{free}(P_2,Q_2) \cap change(S_1) = \ES$.
\III

In this rule, the separation conjunction $P_1*P_2$ in the precondition
of the parallel composition $[S_1 \|S_2]$ is true if the heap can be
partitioned into sub-heaps making the local preconditions of the
components $S_1$ and $S_2$ true.  
The premises of the rule state that the components $S_1$ and $S_2$ establish
local postconditions $Q_1$ and $Q_2$, respectively, 
which in the conclusion are combined into the global postcondition $Q_1*Q_2$. 

A major issue was developing a coherent semantical model for this
\emph{concurrent separation logic}.  This was solved by 
S.~Brookes in \cite{Brookes07}.  His semantics evaluates resource-sensitive
partial correctness formulas of the form $\Gamma \vdash \HT{P}{S}{Q}$,
where $\Gamma$ is a \emph{resource context} that specifies for each
resource name occurring in the program $S$ a finite set of variables, a
protection list, and a resource invariant, and a proof system allows
one to reason about these formulas.

\subsection{Relational Hoare logic}
\label{subsec:relational}

We conclude this overview of alternative approaches by discussing a
line of research that began with the work of N.~Benton,
\cite{Ben04}. In it Hoare's logic was modified to verify correctness
of various optimizing program transformations.  This was achieved by
proposing a proof system in which one reasons at the same time about a
pair of programs the variables of which are related by some
precondition and postcondition.  To this end the customary correctness
formulas were replaced by \emph{judgments} concerning two
\textbf{while} programs, $S_1$ and $S_2$. These are statements of the
form
\[
S_1 \sim S_2 : \Psi \Rightarrow \Phi,
\]
where $\Psi$ and $\Phi$ are relations on program states.  Informally,
such a judgment states that if initially the relation $\Psi$ between
the variables of the programs $S_1$ and $S_2$ holds, then after their
independent executions the relation $\Phi$ between
their variables holds.  (So a notation $\HT{\Psi}{S_1 \sim S_2}{\Phi}$
would have been more intuitive.)  Typically both programs use the same
variables, so to indicate from which program the variable is taken the
indices $\langle 1 \rangle$ and $\langle 2 \rangle$ are used in $\Psi$
and $\Phi$.

As an example consider the technique of \emph{invariant hoisting}.  The
program
\[
  S_1 \equiv \WDD{i<n}{x:=y+1; \ i:=i+x}
\]
can be optimized to
\[
  S_2 \equiv x:=y+1; \ \WDD{i<n}{i:=i+x}.
\]
The resulting programs compute the same values for the variables $i$ and $n$.
This can be expressed as the judgment
\[
  S_1 \sim S_2 : \Phi \Rightarrow \Phi,
\]
with
$\Phi \equiv i\langle 1 \rangle = i\langle 2 \rangle \land n\langle 1
\rangle = n\langle 2 \rangle \land y\langle 1 \rangle = y\langle 2
\rangle$. So the first occurrence of $\Phi$ expresses the information
that prior to the programs executions the values of their variables
$i, n, y$ are respectively equal, while the second occurrence states that
the values of these variables are respectively equal after the executions
of both programs.

In general, if we rename $S_1$ to $S'_1$ by appending to each of its variables
the index $\langle 1 \rangle$, and similarly with $S_2$ for which we use
the index $\langle 2 \rangle$, then the
judgment $S_1 \sim S_2 : \Psi \Rightarrow \Phi$ can be interpreted as
the correctness formula $\HT{\Psi}{S'_1;\ S'_2}{\Phi}$.

The resulting logic was called \emph{Relational Hoare Logic} (RHL),
where the qualification `Relational' referred to the fact that instead
of assertions relations were used. An example rule is the following
one dealing with the conditional statement:
\[
  \frac{S_1 \sim S_2 : \Psi \Rightarrow \Phi}
  {\ITE{B}{S_1}{S_2} \sim S_1 : \Psi \Rightarrow \Phi}
\]
The premise of this rule states that in the context 
determined by the $\Psi$ and $\Phi$ relations, the statement $S_1$ can be replaced
by $S_2$. Thus the second branch of the conditional statement can be
replaced by $S_1$, and consequently this conditional statement
can be simplified to $S_1$.

Appropriate proof rules formulated in this framework made it possible
to verify various known compiler optimization techniques, including the above
example of invariant hoisting.

In \cite{BGB09} this approach was extended to study equivalence of
probabilistic programs. The considered programs extend \textbf{while}
programs by allowing a standard feature of probabilistic programs,
called \emph{random sampling}, which is an assignment of the form
$x:= \cal D$, where $\cal D$ is a probability distribution over the
values of the type of $x$.  The semantics of such programs, as in
\cite{HartogV02}, is then a mapping from the set of states to the set of
distributions over the states.

The semantics of the judgments is defined in such a way
that it respects the fact that in both considered programs the same
probability distributions are used. In particular the judgment
$S \sim S : \mathbf{true} \Rightarrow x\langle 1\rangle = x\langle
2\rangle$, where $S \equiv x:= \cal D$, is true.

The resulting logic was called \emph{probabilistic Relational
Hoare Logic} (pRHL). 
pRHL is not a simple extension of RHL because in presence
of probabilities some rules of RHL become unsound. For example, the rule
\[
  \frac{S_1 \sim S_2 : \Psi \Rightarrow \Phi_1, \ S_1 \sim S_2 : \Psi \Rightarrow \Phi_2}
  {S_1 \sim S_2 : \Psi \Rightarrow \Phi_1 \land \Phi_2}
\]
is sound in RHL (because the programs are deterministic) but not in pRHL.

This work was carried out in the context of \texttt{CertiCrypt}, a
system built on top of the interactive theorem prover \texttt{Coq}
(see Section \ref{sec:concluding}), that makes it possible to provide
machine-checked correctness proofs of cryptographic algorithms.

In \cite{BKOB13} this framework was further generalized to allow for a
probabilistic reasoning about \emph{differential privacy}, 
a notion of privacy that guarantees that
the behaviour of an algorithm taking values from a database hardly
changes when the database is slightly modified.  Differential privacy
in particular provides a formal guarantee that information about
specific participants in a database is not revealed by the algorithm.
The resulting formalism is called \emph{approximate probabilistic Relational
Hoare Logic} (apRHL).  

In apRHL, the judgments of \cite{BGB09} were generalized to
parameterized judgments concerning two probabilistic programs.  The
appropriate proof rules generalize those of \cite{BGB09}.  This
framework was implemented in a system called \texttt{CertiPriv} built
on top of \texttt{Coq}, which was in particular used to provide
machine-checked proofs of soundness of the considered rules.  This
work was further pursued in \cite{BGAHKS14}, where differential
privacy was dealt with by means of a transformation of a probabilistic
program into a non-probabilistic one that simulates two executions of
the original program.  As a result, differential privacy of a single
program could be established using the original proof system
${\cal H}$ presented in Subsection \ref{subsec:hoare1}.

\section{Final Remarks: a Summary and an Assessment}
\label{sec:concluding}

Hoare's logic had a huge impact on program verification, notably by
allowing one to approach it in a systematic way, using the logical
apparatus of formal proofs.  Combining it with the research on program
semantics made it possible to argue about the soundness and relative
completeness of the underlying proof systems. The syntax-directed form
of Hoare's logic suggested a natural research agenda, which ---as we
have seen--- allowed one to deal with several programming constructs
and forms of program construction, including higher-order procedures,
nondeterminism, concurrency, and object-orientation.

This survey aimed at providing a systematic exposition of these developments.
Because of space considerations we had to omit an account of various
Hoare-style proof rules for such concepts as program jumps
(\cite{CH72}), go-to statement (\cite[Chapter 10]{Bak80}, written by
A. de Bruijn), or several forms of abrupt loop termination present in
Java (\cite{HJ00}).

In the seventies Hoare's logic was used to define programming
languages.  In \cite{Hoa72} and \cite{HW73} an axiomatic definition of
the programming language {\sc Pascal} was given. These papers provided
proof rules for simple constructs such as the \textbf{case} statement
and the \textbf{repeat} statement.  However, the presentation was
incomplete. For example, no account of reasoning about recursive
procedures or pointers was given. This work was pursued in
\cite{LondonGHLMP78} where axioms and proof rules in Hoare's logic for
the programming language {\sc Euclid} were presented.

From the current perspective one can see that such axiomatic
presentations were not rigorous since no soundness proofs were
provided to justify the introduced axioms and rules, notably the
recursion rule. To see that such soundness proofs are not superfluous
recall from Subsection \ref{subsec:adaptation} the observation of
\cite{Old83} that the adaptation rule for {\sc Euclid} is not
sound. Also, termination was not dealt with in these papers and the
reasoning about it can be another source of possible, subtle, errors
(see for example the discussion at the end of Subsection
\ref{subsec:complete-recur-proc}). In fact, as we saw, reasoning about
soundness within the framework of Hoare's logic started only after
these two papers were published.

But even if such soundness proofs were presented, given the size of
the considered programming languages, there would be a non-trivial
chance of errors.  In fact, we mentioned a number of times that
arguments about correctness of various proof rules in Hoare's logic or
about soundness or relative completeness of some specific proof
systems have led to various, occasionally, pretty subtle errors.

A logical remedy is to use automated reasoning to argue about various
Hoare’s logics.  This brings us to \emph{computer aided
  verification}, one of the important research directions that emerged
in the eighties.  Since 1990 it is a subject of an annual conference
with the same name (abbreviated to CAV). In relation
to Hoare's logic this work naturally divides into two categories:

\begin{itemize}

\item mechanical verification by means of interactive theorem provers
  of properties of various Hoare logics, such as soundness and
  relative completeness proofs,
  
\item computer aided verification of selected programs.
  
\end{itemize}

In what follows we refer to a number of systems that are the outcome
of several years of research and associated ongoing implementations.
Given that mechanical verification is not the subject of this survey,
we do not discuss these systems in detail.  In each system, to
appropriately formalize the assertion-based reasoning about programs
and reason about the corresponding proof systems, several important
design decisions have to be taken concerning the choice of the
underlying assertion language, proof representation, choice of proof
tactics, form of the interaction with the user, etc.  A separate
survey on this subject would be a most welcome addition to this one.

An early contribution in the first category was \cite{Sok87}, where
soundness of the original proof system of Hoare from \cite{Hoa69} was
established in LCF (that stands for Logic for Computable Functions),
an early interactive automated theorem prover developed in the
seventies, see, e.g, \cite{GMW79}.

Next, in \cite{Kle98}, \cite{Nip02} and \cite{Nip02-CSL} soundness and
relative completeness of Hoare's logics for partial and total
correctness of \textbf{while} programs and programs with recursive
procedures was established in the \textsc{Lego} and Isabelle/HOL
interactive theorem provers, the latter system described initially in
\cite{NPW02}.  In Subsection~\ref{subsec:complete-recur-proc}, we
mentioned already that in \cite{Ohe99} soundness and relative
completeness of a proof system for a programming language with the
mutually recursive procedures with the call-by-value mechanism was
established in Isabelle.  In all these works assertions are identified
with sets of states that satisfy them.

Further, in \cite{NN99} the semantics and proof system of \cite{OG76a} for
partial correctness of parallel programs, that we discussed in
Subsection \ref{subsec:parallel}, was formalized in the Isabelle/HoL
system.  Subsequently the authors proved soundness of this proof
system and verified a number of correctness proof examples.  An
analogous formalization was carried out in \cite{Bal06} on the basis
of dynamic logic in the KIV system, described in \cite{KIV00}.
This approach combined symbolic execution of the operational semantics
of the programs with induction.

The works in the second category concern specific programs. They rely
on various systems and tools that support mechanical verification of
programs based on Hoare's logic or some of the alternative approaches
discussed in the previous section.  An early example is \cite{Gor88},
where a simple program verifier implemented in LISP is described. It
consists of a generator of verification conditions and a theorem
prover. It was used in particular to justify Hoare's correctness proof
given in Figure \ref{fig:1} in Subsection~\ref{subsec:hoare1}.  The
following more recent selective contributions underscore the
importance of mechanical verification.  Several other recent systems
and tools for program verification are described in \cite{PR18}.

\begin{itemize}

\item In his paper \cite{Hoa71a} concerned with the correctness of 
  the \texttt{FIND} program, Hoare expressed the desire for computer
  support in ``formulating the lemmas, and perhaps even checking the
  proofs.'' Only much later, Filli{\^a}tre \cite{Filliatre07}
  published a mechanized proof of \texttt{FIND} using the interactive theorem
  prover \texttt{Coq} and following Hoare's proof from \cite{Hoa71a}
  as closely as possible. He noticed that Hoare's informal termination
  proof does not meet the requirements of a termination function in
  the sense that the additional invariants used by Hoare are not real
  invariants. 

  \texttt{Coq} is based on a formal
  language called the \emph{Calculus of Inductive Constructions} that
  extends typed lambda calculus, see, e.g., \cite{Chl13}. It
  was used to verify formally proofs of a number of famous
  mathematical results, including the Four Color theorem.

\item
  In \cite{CDELSW16} a tool-supported correctness proofs of
  \texttt{Quicksort} and its variants were reported.  The authors used
  the Hoare's logic based verification tool \texttt{Dafny}~\cite{Leino10}.
  
  \texttt{Dafny} comprises (1) an imperative, class-based language for
  programs annotated with the features needed for verification (such
  as assertions, \emph{framing} clauses expressing the objects that
  are allowed to be changed, \emph{ghost variables}, i.e., auxiliary
  variables in the sense of Subsection~\ref{subsec:parallel}, and
  termination functions) and (2) a verifier that translates annotated
  \texttt{Dafny} programs to an intermediate language input by the
  tool \texttt{Boogie}~\cite{BCDJL05}.  The latter is a
  fully-automatic deductive verification system that generates
  verification conditions which are then passed to an SMT
  (satisfiability modulo theories) solver, where the default SMT
  solver is \texttt{Z3} \cite{MouraBjo08}.  \texttt{Boogie} uses the
  weakest precondition semantics discussed in
  Subsection~\ref{subsec:wp}.

\item \texttt{TimSort} is a standard sorting algorithm provided by
  several programming framework, including the Java standard
  library. While trying to verify it the authors of \cite{GBBHRS19}
  discovered a bug.  They formally specified a revised version and
  mechanically verified its correctness, including termination, in the
  KeY system.

  KeY is an extensive software development system that supports in
  particular specification and formal verification of object-oriented
  software, see \cite{ABBHSU16}. As mentioned earlier, it is based on
  dynamic logic.  The KeY-Hoare tool (see \cite{BH16}), built on top
  of the KeY system, allows one to reason about partial and total
  correctness of \textbf{while} programs in an extension of Hoare's
  logic with explicit state updates.
\end{itemize}

In recent years research on Hoare's logic visibly slowed down,
probably due to the fact that through hundreds of publications it
achieved its main goal of creating a comprehensive formal framework to
reason about various classes of programs.  On the other hand, its
versatility results in new applications in various contexts.  We
provide here three recent examples.

The first, due to \cite{AMO13}, concerns reasoning about linear
systems (for example linear differential equations) that are expressed
graphically as block diagrams. The authors, building upon the
framework of abstract Hoare's logic developed in \cite{AMMO09}
mentioned in Subsection \ref{subsec:completeness}, provided a sound
Hoare-like proof system that allowed them to reason about various
examples of linear systems, including linear filters and the steam
boiler problem, a classic example of a hybrid system.

The second one concerns reasoning about time complexity of the
programs.  In \cite{HN18} three Hoare's logics for reasoning about
time bounds, including the original logic due to \cite{Nie87}, were
formalized and shown to be sound and relatively complete.

The final one is an application to quantum programs, a research
direction that originated with the work of M.~Ying \cite{Yin12}.  For
a recent overview of the developments on this subject, that
successfully parallel the developments of the customary Hoare's logic,
see \cite{Yin19}. A related approach of \cite{Unr19} follows the line
of research started with \cite{Ben04} and \cite{BGB09} and introduces
\emph{quantum Relational Hoare Logic} (qRHL) that allows one to reason
about how the outputs of two quantum programs relate given a relation
between their inputs.

One should also mention here recent work aimed at supporting teaching
of Hoare's logic.  In \cite{SS14} an account was given of a tool called
HAHA (Hoare Advanced Homework Assistant) that was specifically
designed to teach Hoare's logic.  The tool supports reasoning about
\textbf{while} programs with integer variables and arrays.

Hoare's logic is an example of an \emph{assertion-based} method of
program verification, in the sense that it relies on the use of
assertions.  This is also a common feature of the approaches to
program verification discussed in the previous section.  In this sense
we can say that all these approaches are direct or less direct
successors of Turing's approach from \cite{Tur49}.

However, there are other theories of program verification, notably of
parallel and distributed programs, that take a different approach.  A
prominent example is CCS (Calculus for Communicating Systems) of Robin
Milner that focuses on reasoning about
equivalence of abstract distributed programs~ \cite{Mil80,Mil89}.
It is at the origin of a number of related theories 
and led to the field of \emph{process algebra}~\cite{BPS2001}.
A comparative survey of these
approaches would form a natural complement to this one.

\subsection*{Acknowledgements}

We would like to thank all three reviewers for exceptionally detailed
and helpful referee reports that led us to improve and expand the presentation.

\bibliographystyle{myalpha}
\bibliography{ao,new,abo}

\appendix

\section{Turing's example}

We present here in Figure \ref{Turing-while} a proof outline for a
\textbf{while} program corresponding to Turing's example given in
Figure \ref{Turing-annotated} of Subsection \ref{subsec:Turing}.  The
program without the assertions was shown in~\cite{OlderogW12}.  The
qualification ${\bf inv}$ is used to annotate the loop invariants and
${\bf bd}$ to annotate the termination functions. So this is a proof
outline establishing total correctness of the program w.r.t.~the
precondition $n \ge 1$ and postcondition $v=n!$.  The assertions used
for partial correctness are similar to the ones used in the
correctness proof of~\cite{deBakker75} of a program corresponding to
Turing's example that uses both a \textbf{while} and a \textbf{repeat}
statement.

\begin{figure}

\begin{tabbing}
\qquad\qquad
\= $\{n \ge 1\}$  \\
\> \ $r:=1;$ \\
\> \ $u:=1;$ \\
\> \ $v:=u;$ \\                                      
\> $\{{\bf inv :}\ P_1 \equiv v=r!  \land u=r! \land 1 \le r \le n\}$ \\
\> $ \{{\bf bd :}\ t_1 \equiv n-r\}$ \\
\> \WHILE $r < n$  \DO  \\
\> $\{P_1 \land r<n\}$ \\
\> $\{v=r! \land u+v = 2 \cdot v \land 1 \le 2 \le r+1 \le n\}$ \\
\> \quad  \      $s:=1$;  \\
\> \quad  \      $u:=u+v$;  \\
\> \quad  \       $s:=s+1;$   \\
\> \quad $\{{\bf inv :}\ P_2 \equiv v=r! \land u = s \cdot v
                                       \land 1 \le s \le r+1 \le n\}$ \\
\> \quad $ \{{\bf bd :}\ t_2 \equiv r+1-s\}$ \\
\> \quad \WHILE $s \le r$ \DO  \\
\> \quad \quad $\{P_2 \land s \le r\}$ \\
\> \quad \quad $\{v=r! \land u +v = (s+1) \cdot v
                                       \land 1 \le s+1 \le r+1 \le n\}$ \\
\> \quad \quad  \     $u:=u+v;$  \\
\> \quad \quad  \    $s:=s+1$  \\
\> \quad \quad $\{P_2\}$ \\
\> \quad \OD \\
\> \quad $\{P_2 \land s>r\}$ \\
\> \quad $\{ v=r! \land u = s \cdot v 
                                       \land 1 \le s =r+1 \le n \}$ \\
\> \quad $\{ u = (r+1)\cdot r! \land 1 \le r+1 \le n\}$ \\
\> \quad $\{ u=(r+1)! \land 1 \le r+1 \le n \}$ \\

\> \quad \ $r:=r+1;$ \\
\> \quad \ $v:=u$ \\
\> \quad $\{ P_1 \}$ \\
\> \OD \\
\> $\{P_1 \land r \ge n\}$ \\ 
\> $\{v=r! \land r \le n \land r \ge n\}$ \\                                                       
\> $\{v=n!\}$  
\end{tabbing}

\caption{Turing's example as a proof outline of a corresponding  \textbf{while} program.}
\label{Turing-while}

\end{figure}

\label{lastpage}

\end{document}